\documentclass[modern]{aastex62}

\usepackage{amsmath}
\usepackage{amssymb}
\usepackage{rotating}

\newcommand\tE{t_{\rm E}}
\newcommand\thetaE{\theta_{\rm E}}

\newcommand{\ra}[4]{${#1}^{\rm h}{#2}^{\rm m}{#3}\fs{#4}$}
\newcommand{\dec}[4]{${#1}\arcdeg{#2}\arcmin{#3}\farcs{#4}$}
\newcommand{\rashort}[3]{${#1}^{\rm h}{#2}^{\rm m}{#3}^{\rm s}$}
\newcommand{\decshort}[3]{${#1}\arcdeg{#2}\arcmin{#3}\arcsec$}
\newcommand\Fs{F_{\rm s}}
\newcommand\Fb{F_{\rm b}}

\received{January 1, 2018}
\revised{January 1, 2018}
\accepted{January 1, 2018}

\shorttitle{Microlensing optical depth and event rate maps from OGLE-IV}
\shortauthors{Mr\'oz et al.}

\begin{document}

\title{Microlensing optical depth and event rate toward the Galactic bulge from  8 yr of OGLE-IV observations}

\correspondingauthor{Przemek Mr\'oz}
\email{pmroz@astrouw.edu.pl}

\author[0000-0001-7016-1692]{Przemek Mr\'oz}
\affil{Astronomical Observatory, University of Warsaw, Al. Ujazdowskie 4, 00-478 Warszawa, Poland}

\author[0000-0001-5207-5619]{Andrzej Udalski}
\affil{Astronomical Observatory, University of Warsaw, Al. Ujazdowskie 4, 00-478 Warszawa, Poland}

\author[0000-0002-2335-1730]{Jan Skowron}
\affil{Astronomical Observatory, University of Warsaw, Al. Ujazdowskie 4, 00-478 Warszawa, Poland}

\author[0000-0002-0548-8995]{Micha\l{} K. Szyma\'nski}
\affil{Astronomical Observatory, University of Warsaw, Al. Ujazdowskie 4, 00-478 Warszawa, Poland}

\author[0000-0002-7777-0842]{Igor Soszy\'nski}
\affil{Astronomical Observatory, University of Warsaw, Al. Ujazdowskie 4, 00-478 Warszawa, Poland}

\author[0000-0002-9658-6151]{\L{}ukasz Wyrzykowski}
\affil{Astronomical Observatory, University of Warsaw, Al. Ujazdowskie 4, 00-478 Warszawa, Poland}

\author[0000-0002-2339-5899]{Pawe\l{} Pietrukowicz}
\affil{Astronomical Observatory, University of Warsaw, Al. Ujazdowskie 4, 00-478 Warszawa, Poland}

\author[0000-0003-4084-880X]{Szymon Koz\l{}owski}
\affil{Astronomical Observatory, University of Warsaw, Al. Ujazdowskie 4, 00-478 Warszawa, Poland}

\author[0000-0002-9245-6368]{Rados\l{}aw Poleski}
\affil{Department of Astronomy, Ohio State University, 140 W. 18th Ave., Columbus, OH~43210, USA}
\affil{Astronomical Observatory, University of Warsaw, Al. Ujazdowskie 4, 00-478 Warszawa, Poland}

\author[0000-0001-6364-408X]{Krzysztof Ulaczyk}
\affil{Department of Physics, University of Warwick, Coventry CV4 7 AL, UK}
\affil{Astronomical Observatory, University of Warsaw, Al. Ujazdowskie 4, 00-478 Warszawa, Poland}

\author[0000-0002-9326-9329]{Krzysztof Rybicki}
\affil{Astronomical Observatory, University of Warsaw, Al. Ujazdowskie 4, 00-478 Warszawa, Poland}

\author[0000-0002-6212-7221]{Patryk Iwanek}
\affil{Astronomical Observatory, University of Warsaw, Al. Ujazdowskie 4, 00-478 Warszawa, Poland}

\begin{abstract}
The number and properties of observed gravitational microlensing events depend on the distribution and kinematics of stars and other compact objects along the line of sight. In particular, precise measurements of the microlensing optical depth and event rate toward the Galactic bulge enable strict tests of competing models of the Milky Way. Previous estimates, based on samples of up to a few hundred events, gave larger values than expected from the Galactic models and were difficult to reconcile with other constraints on the Galactic structure. 

Here we used long-term photometric observations of the Galactic bulge by the Optical Gravitational Lensing Experiment (OGLE) to select a homogeneous sample of 8000 gravitational microlensing events. We created the largest and most accurate microlensing optical depth 
and event rate maps of the Galactic bulge. The new maps ease the tension between the previous measurements and Galactic models. They are consistent with some earlier calculations based on bright stars and are systematically $\sim 30\%$ smaller than the other estimates based on ``all-source'' samples of microlensing events. The difference is caused by the careful estimation of the source star population.

The new maps agree well with predictions based on the Besan\c{c}on model of the Galaxy. Apart from testing the Milky Way models, our maps may have numerous other applications, such as the measurement of the initial mass function or constraining the dark matter content in the Milky Way center. The new maps will also inform the planning of future space-based microlensing experiments by revising the expected number of events.
\end{abstract}


\section{Introduction} \label{sec:intro}

Gravitational microlensing is detectable when an angular separation between a lens and a source is of the order of or smaller than an angular Einstein ring radius,
\begin{equation}
\thetaE = \sqrt{\kappa M \pi_{\rm rel}},
\end{equation}
where $M$ is the mass of the lens, $\pi_{\rm rel}=\mathrm{1\,au}\,(1/D_l-1/D_s)$ is the relative lens--source parallax ($D_l$ and $D_s$ are distances to the lens and source, respectively), and $\kappa= 8.144\,\mathrm{mas}\,M_{\odot}^{-1}$. The microlensing optical depth toward a given source describes the probability that the source falls into the Einstein radius of some lensing foreground object. 

The microlensing optical depth toward one source at distance $D_s$ depends only on the distribution of matter along the line of sight,
\begin{equation}
\tau (D_s) = \frac{4\pi G}{c^2}\int_0^{D_s} \rho(D_l) \frac{D_l (D_s-D_l)}{D_s} dD_l,
\end{equation}
where $\rho(D_l)$ is the mass density of lenses. As the optical depth is independent of the mass function and kinematics of lenses, its measurements allow us to study the distribution of stars and other compact objects toward the Galactic bulge. In practice, however, it is only viable to observe the integrated optical depth, which is averaged over all detectable sources in a given patch of sky, so it may weakly depend on their mass function and star formation history, as well as interstellar extinction,
\begin{equation}
\tau = \frac{1}{N_{\rm s}} \int_0^{\infty} \tau(D_s) dn(D_s),
\end{equation}
where $dn(D_s)$ is the number of detectable sources in the range $[D_s,D_s+dD_s]$ and $N_{\rm s}=\int_0^{\infty}dn(D_s)$ \citep{kiraga1994}. 

The differential microlensing event rate toward a given source is
\begin{equation}
\frac{d^4\Gamma}{dD_l dM d^2\boldsymbol{\mu_{\rm rel}}} = 2 D_l^2 \thetaE |\boldsymbol{\mu_{\rm rel}}| n(D_l) f(\boldsymbol{\mu_{\rm rel}}) g(M),
\end{equation}
where $M$ is the lens mass, $n(D_l)$ is the local number density of lenses, $f(\boldsymbol{\mu_{\rm rel}})$ is the two-dimensional probability density for a given lens--source relative proper motion $\boldsymbol{\mu_{\rm rel}}$, and $g(M)$ is the mass function of lenses \citep{batista2011}. Contrary to the optical depth, the event rate explicitly depends on the mass function of lenses and their kinematics.

The microlensing optical depth can be also interpreted as the fraction of sky covered by the angular Einstein rings of all lenses. Thus, from the observational point of view, the optical depth is proportional to the fraction of time sources spend inside the Einstein ring, and it can be estimated using the following formula that was derived by \citet{udalski1994c}:
\begin{equation}
\label{eq:tau_obs}
\tau = \frac{\pi}{2N_{\rm s}\Delta T}\sum_i\frac{t_{\mathrm{E},i}}{\varepsilon(t_{\mathrm{E},i})},
\end{equation}
where $N_{\rm s}$ is the total number of monitored source stars, $\Delta T$ is the duration of the survey, $t_{\mathrm{E},i}$~is the Einstein timescale of the $i$th event (which is defined as \mbox{$\tE = \thetaE/|\boldsymbol{\mu_{\rm rel}}|$}), and $\varepsilon(t_{\mathrm{E},i})$ is the detection efficiency (probability of finding an event) at that timescale. The event rate is given by
\begin{equation}
\label{eq:gamma_obs}
\Gamma = \frac{1}{N_{\rm s}\Delta T}\sum_i\frac{1}{\varepsilon(t_{\mathrm{E},i})}.
\end{equation}

Direct studies of the central regions of the Milky Way are difficult because of high interstellar extinction and crowding. Precise measurements of the microlensing optical depth and event rate toward the Galactic bulge, although difficult, provide strong constraints on theoretical models of the Galactic structure and kinematics \citep[e.g.,][]{han_gould2003,wood2005,kerins2009,awiphan2016,wegg2016,binney2018}. 

\begin{table}
\caption{Previous Measurements of the Microlensing Optical Depth toward the Galactic Bulge.}
\label{tab:tau1}
\centering
\begin{footnotesize}
\begin{tabular}{lccrrrl}
\hline
\hline
\multicolumn{1}{c}{Collaboration} & \multicolumn{1}{c}{Location} & \multicolumn{1}{c}{Optical Depth} & \multicolumn{1}{c}{$N_{\rm stars}$} & \multicolumn{1}{c}{$N_{\rm events}$} & \multicolumn{1}{c}{$\Delta T$} & \multicolumn{1}{c}{Source} \\
& $(l,b)$ & $(\times 10^{-6})$ & $(\times 10^{6})$ & & (yr) & \\
\hline
OGLE-I & $(1^{\circ},-4^{\circ})$ & $3.3 \pm 1.2$ & 0.95 & 9 & 2 & \citet{udalski1994c}\\
MACHO & $(2.30^{\circ},-2.65^{\circ})$ & $>1.3$ & 0.43 & 4 & 1 & \citet{alcock1995} \\
MACHO & $(2.55^{\circ},-3.64^{\circ})$ & $3.9^{+1.8}_{-1.2}$ & 1.3 & 45 & 1 & \citet{alcock1997b} \\
MACHO & $(2.68^{\circ},-3.35^{\circ})$ & $2.43^{+0.39}_{-0.38}$ & 17 & 99 & 3  & \citet{alcock2000} \\
MACHO & $(3.9^{\circ},-3.8^{\circ})$ & $2.0 \pm 0.4$ & 2.1 & 52 & 5 &  \citet{popowski2001} \\
MACHO & $(2.22^{\circ},-3.18^{\circ})$ & $2.01^{+0.34}_{-0.32}$ & 17 & 99 & 2 & \citet{popowski2002} \\
MOA & $(3.0^{\circ},-3.8^{\circ})$ & $2.59^{+0.84}_{-0.64}$ & 230 & 28 & 2 & \citet{sumi2003} \\
EROS-2 & $(2.5^{\circ},-4.0^{\circ})$ & $0.94 \pm 0.29$ & 1.42 & 16 & 3 & \citet{afonso2003} \\
MACHO & $(1.50^{\circ},-2.68^{\circ})$ & $2.17^{+0.47}_{-0.38}$ & 6 & 62 & 7 & \citet{popowski2005} \\
OGLE-II & $(1.16^{\circ},-2.75^{\circ})$ & $2.55^{+0.57}_{-0.46}$ & 1.5 & 32 & 4 &  \citet{sumi2006} \\
EROS-2 & EROS-2 fields & $1.68 \pm 0.22$ & 5.6 & 120 & 6 & \citet{hamadache2006}\\
MOA-II & MOA-II fields & $1.87^{+0.15}_{-0.13}$ & 90.4 & 474 & 2 & \citet{sumi2013} \\
MOA-II & MOA-II fields & $1.53 ^{+0.12}_{-0.11}$ & 110.3 & 474 & 2 & \citet{sumi_penny2016} \\
\hline
\end{tabular}
\end{footnotesize}
\end{table}

The first measurement of the microlensing optical depth toward the Galactic bulge was carried out by \citet{udalski1994c} and based on data from the first phase of the Optical Gravitational Lensing Experiment (OGLE) from 1992--1993 (see Table~\ref{tab:tau1} for a compilation of previous measurements). They found nine microlensing events in a systematic search of $\sim 10^6$ light curves, and they calculated $\tau = (3.3 \pm 1.2) \times 10^{-6}$, which was greater than contemporary theoretical estimates ($(0.4-1)\times 10^{-6}$; \citealt{paczynski1991,griest1991,kiraga1994}). A similar conclusion was reached by \citet{alcock1995,alcock1997b} based on MACHO project observations of the Galactic bulge. 
These seminal papers boosted the development of the field, but as we now know, the calculated optical depths are prone to systematic errors, especially due to miscalculation of number of monitored sources. The early photometry was done using the point-spread function (PSF) fitting method, which in crowded fields faces more challenges than the difference image analysis (DIA) that is normally used in modern microlensing surveys. 

These first measurements of the optical depth led to the realization that most of the observed microlensing events are caused by lenses located in the Galactic bulge \citep{paczynski1994,zhao1995} and reinforced the idea that the inner regions of the Milky Way have a bar-like structure elongated along the line of sight \citep[e.g.,][]{blitz1991,stanek1994}. The first measurements stimulated the development of improved models of the Galactic bulge \citep[e.g.,][]{zhao1996,fux1997,nikolaev1997,peale1998,gyuk1999,sevenster1999,grenacher1999}. Nonetheless, all of these models predicted the optical depth in the direction of MACHO fields in the range $(1.1-2.2)\times 10^{-6}$, a factor of 2--4 lower than the reported values.

The implementation of the DIA technique \citep{alard1998} led to the improvement of the quality of the photometry in very dense stellar fields toward the Galactic bulge. This enabled the surveys to detect more microlensing events and precisely measure their parameters. The optical depth measurements based on MACHO \citep[$2.43^{+0.39}_{-0.38} \times 10^{-6}$;][]{alcock2000} and MOA-I \citep[$2.59^{+0.84}_{-0.64} \times 10^{-6}$;][]{sumi2003} data were still higher than the theoretical predictions. \citet{binney2000} and \citet{bissantz2002} argued that such high optical depths cannot be easily reconciled with other constraints, such as the Galactic rotation curve and the mass density near the Sun. Nearly two decades later, \citet{sumi_penny2016} suggested that these measurements suffer from biased source star counts and are overestimated.

In addition, \citet{popowski2001} and \citet{popowski2002} noticed that previous microlensing optical depth measurements underestimated (or completely ignored) the influence of blending on the estimation of event parameters from the light curves. The Galactic bulge fields are extremely crowded, and there should be many faint unresolved stars within the seeing disk of any bright star. The omission of blending results in underestimated Einstein timescales. In highly blended events, as demonstrated by \citet{wozniak1997}, the event timescale, impact parameter, and blending parameter may be severely correlated, which renders robust timescale measurements difficult.

\citet{popowski2001} proposed determining the microlensing optical depth using exclusively red clump giants as sources because they are subject to little blending and it is easy to estimate their total number. Several measurements of the microlensing optical depth toward the Galactic bulge based on red clump giants were published by the EROS ($0.94 \pm 0.29 \times 10^{-6}$; \citealt{afonso2003} ; $1.68 \pm 0.22 \times 10^{-6}$; \citealt{hamadache2006}), MACHO \citep[$2.17^{+0.47}_{-0.38} \times 10^{-6}$;][]{popowski2005}, and OGLE-II \citep[$2.55^{+0.57}_{-0.46} \times 10^{-6}$;][]{sumi2006} groups. These estimates were lower than those based on all-star samples of events \citep{alcock2000,sumi2003}.

The current largest microlensing optical depth and event rate maps are based on 2~yr (2006--2007) of observations of the Galactic bulge by the MOA-II survey \citep{sumi2013}. \citet{sumi2011} and \citet{sumi2013} found over 1000 microlensing events in that data set, but only 474 events were used for the construction of event rate maps. All events are located in 22 bulge fields covering about 42~deg$^2$ in the ranges $-5^{\circ}<l<+10^{\circ}$ and $-7^{\circ}<b<-1^{\circ}$. Three years after the MOA-II publication, \citet{sumi_penny2016} realized that the sample of red clump giants, which was used to scale the number of observed sources and thus optical depths and event rates, was incomplete, most likely due to crowding and high interstellar extinction. The completeness increased with the Galactic latitude -- from 70\% at $b=-1.5^{\circ}$ to 100\% in fields located far from the Galactic plane ($b=-6^{\circ}$). This affected the measured optical depth and event rates, which were systematically overestimated at low Galactic latitudes. The revised all-source optical depth measurements were much lower than those published by \citet{sumi2013}, which alleviated (but did not completely remove) the tension with the previous measurements based on red clump giant stars \citep{popowski2005,hamadache2006,sumi2006}. A similar bias may have affected the early MACHO and MOA measurements \citep{alcock2000,sumi2003}.

Large samples of microlensing events were also recently reported by \citet[][OGLE-III]{wyrzykowski2015,wyrzykowski2016}, \citet[][VVV]{navarro2017,navarro2018}, and \citet[][KMTNet]{kim2018,kim2018_2}, but these authors did not attempt to calculate optical depths and event rates.

The original MOA-II optical depth maps \citep{sumi2013} were used by \citet{awiphan2016} to modify the Besan\c{c}on Galactic model \citep{robin2014}. For example, they needed to include M dwarfs and brown dwarfs in the mass function of lenses to match the timescale distribution of microlensing events. \citet{awiphan2016} noticed that the predicted optical depths at low Galactic latitudes were about 50\% lower than those reported by \citet{sumi2013}. This discrepancy can only be partially explained by the \citet{sumi_penny2016} findings; the theoretical optical depth is a factor of $\sim 1.6$ lower than the revised MOA-II measurements. The revised MOA-II data \citep{sumi_penny2016} were also used by \citet{wegg2016} to constrain the dark matter fraction in the inner Galaxy. 

The accurate microlensing event rates are also of interest for the astronomical community, for example, for the preparation of future space-based microlensing surveys like the \textit{Wide Field Infrared Survey Telescope} (\textit{WFIRST}; \citealt{spergel2015}) or \textit{Euclid} \citep{penny2013}. The current Galactic models seemed to not be precise enough to predict reliable event rates, and they had to be scaled to match the observations \citep{penny2013,penny2019}. For example, \citet{penny2019} had to multiply the predicted rates by a factor of 2.11 to match the Sumi \& Penny (2016) results.

All of these model constraints and predictions are still based on a relatively small sample of microlensing events, and many authors have raised the need for optical depths from the larger OGLE sample \citep[e.g.,][]{wegg2016,penny2019}. In this paper, we aim to address these needs.

The basic information about the OGLE-IV survey and the data set used in the analysis is included in Section~\ref{sec:data}. Section~\ref{sec:selection} presents the selection of microlensing events. In Section~\ref{sec:counts}, we estimate the completeness of the OGLE star catalogs and the number of observable sources. The calculations of the microlensing event detection efficiency are described in Sections~\ref{sec:blending}--\ref{sec:im_sim}. The main scientific results and their implications are discussed in Section~\ref{sec:results}.

\section{Data}
\label{sec:data}

The photometric data analyzed in this paper were collected as part of the OGLE sky survey, which is one of the largest long-term photometric sky surveys worldwide. All analyzed observations were collected during the fourth phase of the project \citep[OGLE-IV;][]{udalski2015} during 2010--2017. The survey uses the dedicated 1.3\,m Warsaw Telescope, located at Las Campanas Observatory, Chile. (The observatory is operated by the Carnegie Institution for Science.) The telescope is equipped with a mosaic camera that consists of 32 CCD detectors, each of $2048\times 4102$ pixels. The OGLE-IV camera covers a field of view of 1.4~deg$^2$ with a pixel scale of $0.26''$ per pixel. 

\begin{figure}
\centering
\includegraphics[width=.95\textwidth]{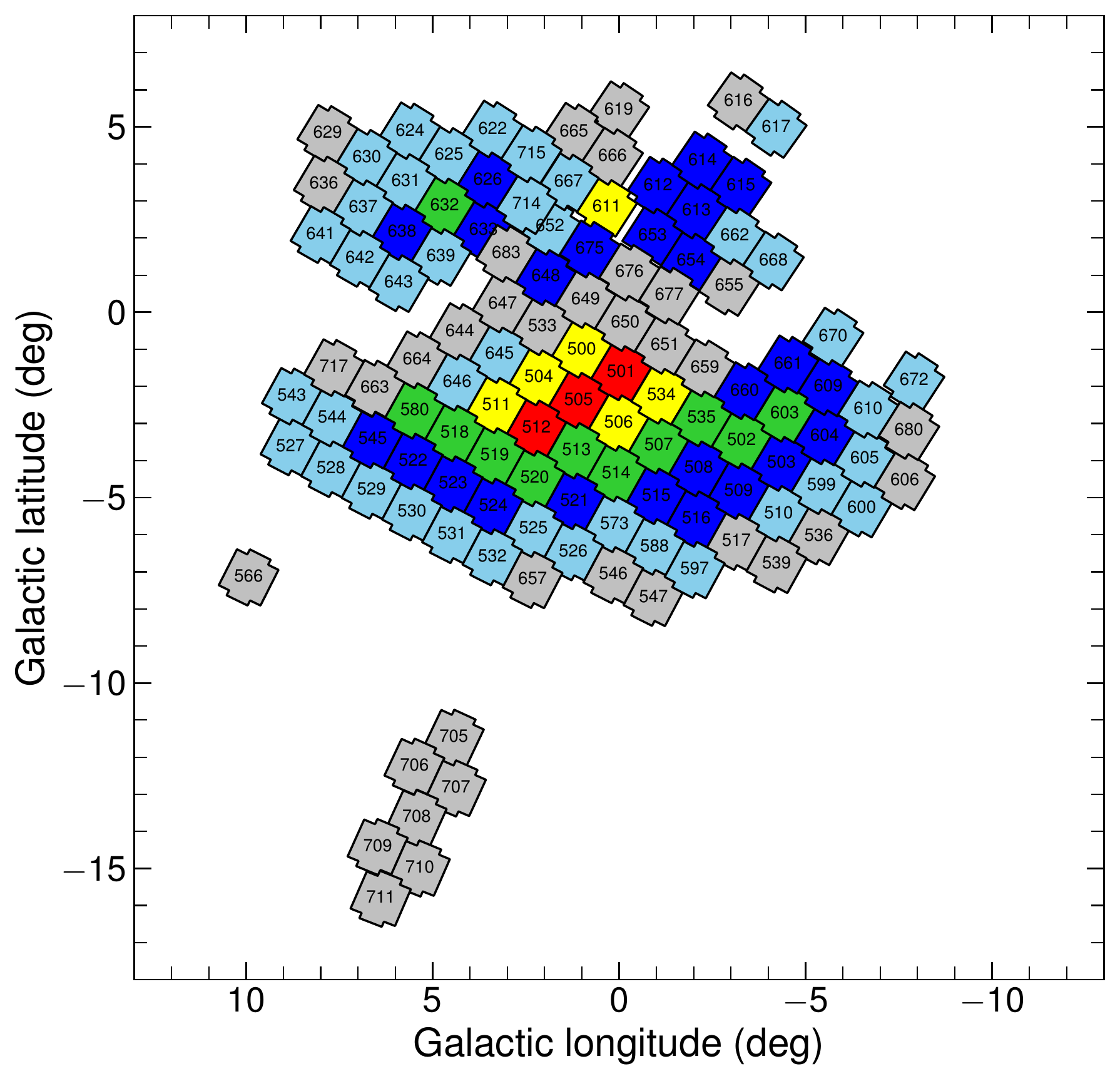}
\caption{OGLE-IV fields toward the Galactic bulge. Colors mark the typical cadence of observations: red -- one observation every 20 minutes, yellow -- one observation every 60 minutes, green -- two to three observations per night, blue -- one observation per night, cyan -- one observation per 2 nights. Silver fields were regularly observed during the years 2010--2013, usually once every 2--3 days.}
\label{fig:fields}
\end{figure}

We searched for microlensing events in 121 fields located toward the Galactic bulge that have been observed for at least two observing seasons (filled polygons in Figure~\ref{fig:fields}). These fields cover an area of over 160~deg$^2$ and contain over 400 million sources in OGLE databases. Typical exposure times are 100--120~s, and the vast majority of observations are taken through the $I$-band filter, closely resembling that of a standard Cousins system. The magnitude range of the survey is $12<I<21$, but the limiting magnitude depends on the crowding of a given field (as shown in Section~\ref{sec:counts}). Fields are grouped and scheduled for observations with one of several cadences. Some fields switch groups or are paused for the next season.

Nine fields that are observed with the highest cadence (BLG500, BLG501, BLG504, BLG505, BLG506, BLG511, BLG512, BLG534, and BLG611) have already been analyzed by \citet{mroz2017} with the aim of measuring the frequency of free-floating planets in the Milky Way. Here we use the sample of microlensing events presented in that paper to calculate optical depths and event rates in the subset of high-cadence fields. We also make use of image-level simulations that have been carried out by \citet{mroz2017} to measure the detection efficiency of microlensing events. The data were collected between 2010 June 29 and 2015 November 8. Each light curve consists of 4500--12,000 single photometric measurements, depending on the field.

For the remaining 112 fields, which are the main focus of this paper, we used data collected during a longer period, between 2010 June 29 and 2017 November 1, whenever available. Because of the changes in the observing strategy of the survey, some of these fields were observed for a shorter period of time (from two to five Galactic bulge seasons). Most of these fields (76; i.e., 68\%), however, were monitored for nearly 8~yr. The majority of light curves consist of 100--2000 data points. 

Basic information about all analyzed fields (equatorial and Galactic coordinates, number of monitored sources, number of epochs) is presented in Table~\ref{tab:allfields} in Appendix~\ref{ch:app1}.

The OGLE photometric pipeline is based on the DIA method \citep{alard1998,wozniak2000}, which allows for obtaining very accurate photometry in dense stellar fields. A reference image of each field is constructed by stacking the three to six highest-quality frames. This reference image is then subtracted from incoming frames, and the photometry is performed on the subtracted images. Variable and transient objects that are detected on the subtracted images are then assigned and stored in either of the two databases. The ``standard'' database holds the light curves of all stellar-like objects previously identified on the reference frame, while ``new'' objects (those that are not registered as stellar on the reference images) are stored separately. The detailed descriptions of the image reductions, calibrations, and OGLE photometric pipeline are included in \citet{wozniak2000}, \citet{udalski2003}, and \citet{udalski2015}.

\section{Selection of Events in Low-cadence Fields}
\label{sec:selection}

The selection algorithms of the microlensing events and final selection cuts were similar to those used by \citet{mroz2017}, although with some small differences. Because the contamination from instrumental artifacts (such as reflections within the telescope optics) in the analyzed fields is much less severe than in high-cadence fields, we were able to relax the selection criteria compared to the earlier work \citep{mroz2017}. All criteria are summarized in Table~\ref{tab:crit_2}.

It is known that the photometric uncertainties returned by DIA are underestimated and do not reflect the actual observed scatter in the data. Thus, we began the analysis by correcting the reported uncertainties using the procedure proposed by \citet{skowron2016}. For stars fainter than approximately $I=15$, the error bars were corrected using the formula $\delta m_{i,\mathrm{new}}=\sqrt{(\gamma\delta m_i)^2+\varepsilon^2}$, where $\gamma$ and $\varepsilon$ are the parameters determined for each field separately. They were measured based on the scatter of constant stars (typically, $\gamma=1.2-1.6$ and $\varepsilon=0.002-0.004$). For the brightest stars, there is an additional correction resulting from the non-linear response of the detector. The error bar correction coefficients were not available for 11 fields and we closely followed \citet{skowron2016} to calculate the missing values. Subsequently, we transformed magnitudes into flux. The search procedure consisted of three steps.

\begin{table}[b]
\caption{Selection Criteria for High-quality Microlensing Events in Low-cadence OGLE-IV Fields.}
\label{tab:crit_2}
\centering
\footnotesize
\begin{tabular}{p{0.3\textwidth}p{0.5\textwidth}p{0.1\textwidth}}
\hline
Criteria & Remarks & Number \\
\hline
All Stars in Databases & & 353,789,948\\
\hline
$\chi^2_{\rm out}/{\rm dof} \leq 2.0$ & No variability outside the 720 day (or 360 day) window centered \mbox{on the event} \\
$n_{\rm DIA} \geq 3$ & Centroid of the additional flux coincides with the source star centroid\\
$\chi_{3+}=\sum_i(F_i-F_{\rm base})/\sigma_i\geq 32$ & Significance of the bump & 23,618\\
\hline
$s<0.4$ & Rejecting photometry artifacts \\
$A \geq 0.1$ mag & Rejecting low-amplitude variables \\
$n_{\rm bump}=1$ & Rejecting objects with multiple bumps & 18,397 \\
\hline
& Fit quality: & \\
$\chi^2_{\rm fit}/\rm{dof} \leq 2.0$ & $\chi^2$ for all data \\
$\chi^2_{\rm fit,\tE}/\rm{dof} \leq 2.0$ & $\chi^2$ for $|t-t_0|<\tE$ \\
$2,455,377\leq t_0\leq 2,458,118$ & Event peaked between 2010 June 29 and 2017 December 31 \\
$u_0 \leq 1$ & Maximum impact parameter \\
$\tE \leq 300$\,d & Maximum timescale \\
$I_{\rm s} \leq 21.0$ & Maximum $I$-band source magnitude \\
$F_{\rm b} > -0.1$ &  Maximum negative blend flux, corresponding \mbox{to $I=20.5$ mag star} \\
$f_{\rm s} > 0.01$ &  Rejecting highly blended events & 5790\\
\hline
\end{tabular}
\end{table}

\textit{Step 1:} We began the analysis with over 350~million objects in the ``standard'' databases.  First, we searched for any kind of brightening in the light curves. We searched for at least three consecutive data points that are at least $3\sigma_{\rm base}$ above the baseline flux $F_{\rm base}$. The baseline flux and its dispersion were calculated using data points outside a 720 day window centered on the event after removing $5\sigma$ outliers (if the light curve was shorter than 6~yr, we used a 360 day window instead). We required the light curve outside the window to be flat ($\chi^2_{\rm out}/\mathrm{d.o.f.} \leq 2.0$), which allowed us to remove the majority of variable stars and image artifacts. We also required at least three magnified data points to be detected on the subtracted images during the candidate event ($n_{\rm DIA}\geq 3$), meaning that the centroid of the additional flux coincided with the source star centroid (within $0.5''$). That selection cut enabled us to remove any contamination from asteroids, as well as contamination from spurious events and photometric artifacts. For each candidate event, we calculated $\chi_{3+}=\sum_i(F_i-F_{\rm base})/\sigma_i$; the summation is performed over all consecutive data points at least $3\sigma_{\rm base}$ above the baseline. We required $\chi_{3+}\geq 32$. These simple selection criteria allowed us to reduce the number of candidate microlensing events to 23,618.

\textit{Step 2:} Subsequent cuts were devised to remove any additional obvious nonmicrolensing light curves. We removed all objects with two or more brightenings in the light curve -- mostly dwarf novae and other erupting variable stars. We discarded all candidate events with amplitudes smaller than 0.1~mag to minimize the contamination from pulsating red giants. The real microlensing events with such a small amplitude typically yield an inaccurate estimation of the event timescale; hence, they are not essential for the current analysis. As in \citet{mroz2017}, we also removed all candidates that were located close to each other and magnified in the same images; these are spurious detections caused by reflections within the telescope or nonuniform background. In this step, we removed 5221 objects from the sample.

\textit{Step 3:} Finally, we fitted the microlensing point-source point-lens model to the light curves of the remaining 18,397 candidates. The microlensing magnification depends on three parameters -- the time $t_0$ and projected separation $u_0$ (in Einstein radius units) between the lens and the source during the closest approach and the Einstein timescale $\tE$ -- and is given by
\begin{equation}
A = \frac{u^2+2}{u\sqrt{u^2+4}},
\end{equation}
where $u=\sqrt{u_0^2+(t-t_0)^2/\tE^2}$. The observed flux is $F_{\rm model}(t_i)=\Fs A(t_i) + \Fb$, where $\Fs$ and $\Fb$ describe the source flux and the unmagnified blended flux, respectively. As the observed flux depends linearly on $\Fs$ and $\Fb$, they were calculated analytically using the least-squares method for each set of $(t_0,u_0,\tE)$.

The best-fit parameters were found by minimizing the function $\chi^2 = \sum_i (F_i-F_{\rm model}(t_i))^2/\sigma^2_i$ using the Nelder--Mead algorithm\footnote{We used the C implementation of the algorithm by John Burkardt, which is distributed under the GNU LGPL license (https://people.sc.fsu.edu/$\sim$jburkardt/c\_src/asa047/asa047.html).}. During the modeling, we iteratively removed any $4\sigma$ outliers, provided that the adjacent data points were within $3\sigma$ of the model. To quantify the quality of the fit, we calculated $\chi^2_{\rm fit}$ for the entire data set and $\chi^2_{\rm fit,t_E}$ for data points within $\tE$ of the maximum of the event (i.e., $|t-t_0|<\tE$). We calculated five- and four-parameter models (with the blend flux set to zero). We allowed for some amount of the negative blending in the five-parameter fits ($F_{\rm b} \geq -F_{\rm min}$, where $F_{\rm min}=0.1$ is the absolute flux corresponding to an 20.5 mag star). If $\Fb < -F_{\rm min}$ and the four-parameter model was marginally worse ($\Delta\chi^2<9$) than the five-parameter model, we chose the former. We required $\chi^2/\mathrm{dof}\leq 2$ in the chosen fit.

We were left with 5790 objects, which will constitute our final sample of microlensing events used for the construction of optical depth and event rate maps in low-cadence fields. The uncertainties of the model parameters were estimated using the Markov chain Monte Carlo technique using the \textsc{Emcee} sampler coded by \citet{foreman2013}. To take into account our limits on negative blending, we added the following prior on the blend flux:
\begin{equation}
\mathcal{L}_{\rm prior} = 
\begin{cases}
1 & \mathrm{if}\ F_{\rm b} \geq 0, \\
\exp{\left(-\frac{F_{\rm b}^2}{2\sigma^2}\right)} & \mathrm{if}\ F_{\rm b} < 0,
\end{cases}
\end{equation}
where $\sigma = F_{\rm min}/3$ ($F_{\rm min}=0.1$ is the absolute flux corresponding to $I=20.5$). The best-fit parameters and their uncertainties are reported in Table \ref{tab:params}. The uncertainties represent the 68\% confidence range of the marginalized posterior distributions.

\begin{figure}[h]
\centering
\includegraphics[width=0.65\textwidth]{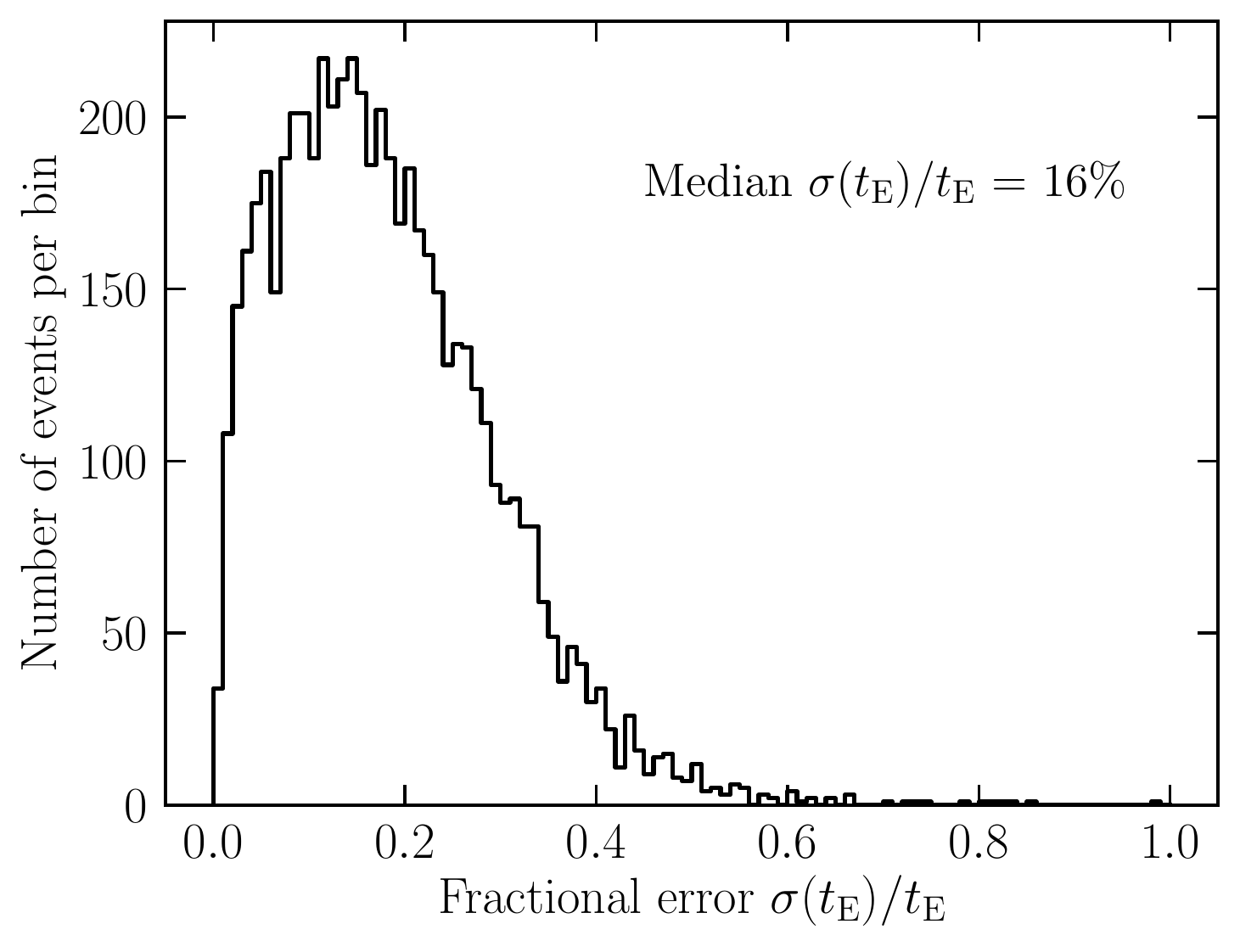}
\caption{Distribution of fractional uncertainties of Einstein timescales of microlensing events in low-cadence OGLE-IV fields. The median uncertainty is 16\%, while $\sigma(\tE)/\tE\leq 0.5$ for 98\% of events in the analyzed sample.}
\label{fig:tE_errorbars}
\end{figure}

Light curves of selected events were inspected by a human expert, from which we estimate the purity of our sample of microlensing events to be very high ($\sim99.5\%$). Figure~\ref{fig:tE_errorbars} shows the distribution of fractional uncertainties of Einstein timescales. The median uncertainty is 16\%, and for 98\% of events in the analyzed sample ,$\sigma(\tE)/\tE\leq 0.5$. There are two main factors influencing our measurements of $\tE$: the source brightness and impact parameter (which corresponds to maximal magnification). The fainter the source and the larger the impact parameter, the larger the uncertainties. 

Of the 5790 events from our sample, 3958 (68\%) have been announced in real time by the OGLE Early Warning System (EWS) \citep{udalski2003}; the remaining 1832 events (32\%) are new discoveries. For comparison, from 2011 to 2017, 6959 microlensing alerts in low-cadence fields were announced by the EWS, and about 10\% of these are anomalous or binary. The EWS also contains some lower-amplitude events or events on sources fainter than $I=21$.

We calculated more detailed statistics for the field BLG660 as an example. Our step~1 criteria selected 180 candidate microlensing events; the visual inspection of light curves showed that 138 were indeed microlensing events,  while 125 were found by the EWS. Two objects reported in the EWS were not microlensing events (variable stars), and three were detected in ``new'' databases. Nine genuine EWS events were not identified by our search algorithm (mostly because of the variability in the baseline, the low significance of the event, or the small number of magnified data points), and 27 events were detected only by our search algorithm. Furthermore, 111 objects are common. Thus, our search algorithm was able to find $(111+27)/(111+27+9)=94\%$ of events in that field.
However, only 94 events from the field BLG660 (i.e., 64\%) satisfied all of our selection criteria and were included in the final sample of events. Half of the rejected events have very faint sources ($I_{\rm s}\geq 21$) and, as a consequence, their parameters are not well measured. The remaining events are anomalous, do not fulfill the constraints on $t_0$ or $u_0$, or have light curves that are noisy; thus, they do not meet the $\chi^2$ fit quality criterion.

\begin{deluxetable*}{@{}lrrrrrrrl@{}}[h]
\tablecaption{Best-fitting Parameters of the Analyzed Microlensing Events in Low-cadence OGLE Fields (First 40 Objects) \label{tab:params}}
\tabletypesize{\scriptsize}
\tablehead{
\colhead{Star} & \colhead{R.A.} & \colhead{Decl.} & \colhead{$t_0$ (HJD)} &
\colhead{$t_{\rm E}$ (days)} & \colhead{$u_0$} & \colhead{$I_{\rm s}$} & \colhead{$f_{\rm s}$} &
\colhead{EWS ID}
}
\startdata
BLG617.16.73378 & \ra{17}{13}{08}{00} & \dec{-29}{48}{13}{0} & $2455434.91 ^{+0.23}_{-0.22}$ & $ 24.41 ^{+4.55}_{-2.85}$ & $0.206 ^{+0.041}_{-0.045}$ & $20.11 ^{+0.31}_{-0.23}$ & $0.91 ^{+0.22}_{-0.22}$ &  \\
BLG617.24.41328 & \ra{17}{13}{54}{30} & \dec{-29}{36}{35}{5} & $2457462.79 ^{+0.57}_{-0.66}$ & $199.66 ^{+18.49}_{-17.42}$ & $0.139 ^{+0.019}_{-0.016}$ & $20.29 ^{+0.14}_{-0.16}$ & $0.25 ^{+0.04}_{-0.03}$ & OB160231 \\
BLG616.32.9872 & \ra{17}{13}{56}{45} & \dec{-28}{08}{40}{8} & $2455840.42 ^{+0.28}_{-0.30}$ & $ 38.60 ^{+15.65}_{-9.95}$ & $0.021 ^{+0.010}_{-0.007}$ & $20.86 ^{+0.44}_{-0.40}$ & $0.87 ^{+0.39}_{-0.29}$ &  \\
BLG617.24.47717 & \ra{17}{13}{56}{92} & \dec{-29}{32}{34}{2} & $2457594.85 ^{+0.24}_{-0.24}$ & $ 21.55 ^{+4.75}_{-2.77}$ & $0.211 ^{+0.048}_{-0.052}$ & $20.34 ^{+0.34}_{-0.25}$ & $1.07 ^{+0.28}_{-0.29}$ & OB161459 \\
BLG617.07.43804 & \ra{17}{14}{00}{26} & \dec{-30}{08}{42}{2} & $2457842.43 ^{+1.62}_{-1.62}$ & $ 38.77 ^{+18.74}_{-8.75}$ & $0.691 ^{+0.336}_{-0.317}$ & $19.74 ^{+1.01}_{-0.83}$ & $0.36 ^{+0.41}_{-0.22}$ & OB170395 \\
BLG617.07.22676 & \ra{17}{14}{06}{09} & \dec{-30}{04}{59}{8} & $2456434.71 ^{+0.52}_{-0.51}$ & $ 39.08 ^{+6.76}_{-4.41}$ & $0.349 ^{+0.068}_{-0.074}$ & $20.16 ^{+0.33}_{-0.26}$ & $0.95 ^{+0.26}_{-0.25}$ & OB130673 \\
BLG617.15.7193 & \ra{17}{14}{06}{52} & \dec{-29}{56}{21}{5} & $2456038.68 ^{+0.44}_{-0.43}$ & $ 26.22 ^{+5.42}_{-3.22}$ & $0.438 ^{+0.099}_{-0.111}$ & $19.60 ^{+0.43}_{-0.33}$ & $0.72 ^{+0.26}_{-0.24}$ & OB120420 \\
BLG617.14.111445 & \ra{17}{14}{18}{39} & \dec{-29}{48}{49}{5} & $2455652.64 ^{+0.03}_{-0.03}$ & $ 15.53 ^{+1.07}_{-1.00}$ & $0.087 ^{+0.008}_{-0.007}$ & $19.14 ^{+0.10}_{-0.10}$ & $0.40 ^{+0.04}_{-0.04}$ &  \\
BLG617.31.97087 & \ra{17}{14}{18}{45} & \dec{-29}{20}{03}{2} & $2456479.74 ^{+0.01}_{-0.01}$ & $ 28.27 ^{+0.72}_{-0.71}$ & $0.034 ^{+0.001}_{-0.001}$ & $18.75 ^{+0.03}_{-0.03}$ & $0.98 ^{+0.03}_{-0.03}$ & OB130992 \\
BLG617.14.111841 & \ra{17}{14}{21}{39} & \dec{-29}{50}{01}{3} & $2457834.40 ^{+0.15}_{-0.13}$ & $  7.97 ^{+2.18}_{-1.46}$ & $0.056 ^{+0.031}_{-0.024}$ & $19.76 ^{+0.43}_{-0.38}$ & $0.61 ^{+0.26}_{-0.20}$ & OB170396 \\
BLG616.14.35260 & \ra{17}{14}{40}{27} & \dec{-28}{37}{58}{6} & $2456059.85 ^{+0.52}_{-0.54}$ & $ 54.45 ^{+8.89}_{-5.39}$ & $0.356 ^{+0.060}_{-0.072}$ & $19.53 ^{+0.32}_{-0.23}$ & $0.85 ^{+0.20}_{-0.21}$ &  \\
BLG617.31.40734 & \ra{17}{14}{41}{78} & \dec{-29}{17}{31}{7} & $2457174.55 ^{+0.62}_{-0.61}$ & $ 10.26 ^{+3.23}_{-1.46}$ & $0.616 ^{+0.144}_{-0.211}$ & $19.59 ^{+0.69}_{-0.39}$ & $0.69 ^{+0.30}_{-0.33}$ &  \\
BLG617.14.6899 & \ra{17}{14}{47}{57} & \dec{-29}{56}{44}{3} & $2456829.48 ^{+0.81}_{-0.80}$ & $ 29.53 ^{+9.78}_{-5.38}$ & $0.590 ^{+0.209}_{-0.211}$ & $19.18 ^{+0.72}_{-0.57}$ & $0.48 ^{+0.33}_{-0.23}$ & OB141193 \\
BLG617.31.20057 & \ra{17}{14}{49}{02} & \dec{-29}{12}{40}{2} & $2456785.64 ^{+0.06}_{-0.07}$ & $ 34.09 ^{+6.88}_{-5.22}$ & $0.052 ^{+0.013}_{-0.011}$ & $20.70 ^{+0.27}_{-0.25}$ & $0.29 ^{+0.08}_{-0.06}$ & OB140835 \\
BLG617.05.69384 & \ra{17}{15}{09}{42} & \dec{-30}{12}{28}{7} & $2456104.16 ^{+0.05}_{-0.06}$ & $  3.34 ^{+0.75}_{-0.38}$ & $0.267 ^{+0.051}_{-0.071}$ & $19.22 ^{+0.39}_{-0.23}$ & $0.82 ^{+0.20}_{-0.24}$ & OB121027 \\
BLG617.05.57777 & \ra{17}{15}{10}{18} & \dec{-30}{20}{07}{5} & $2457830.83 ^{+1.51}_{-1.63}$ & $270.00 ^{+50.51}_{-40.90}$ & $0.186 ^{+0.048}_{-0.040}$ & $20.37 ^{+0.30}_{-0.30}$ & $0.07 ^{+0.02}_{-0.02}$ & OB170165 \\
BLG616.05.90565 & \ra{17}{15}{10}{89} & \dec{-28}{54}{27}{7} & $2455598.78 ^{+0.38}_{-0.30}$ & $ 17.17 ^{+0.63}_{-0.43}$ & $0.095 ^{+0.017}_{-0.025}$ & $17.00 ^{+0.11}_{-0.06}$ & $0.94 ^{+0.05}_{-0.09}$ &  \\
BLG617.22.46738 & \ra{17}{15}{19}{23} & \dec{-29}{36}{58}{2} & $2456574.28 ^{+0.13}_{-0.13}$ & $ 21.19 ^{+2.52}_{-2.00}$ & $0.128 ^{+0.018}_{-0.017}$ & $19.89 ^{+0.17}_{-0.15}$ & $1.38 ^{+0.20}_{-0.20}$ &  \\
BLG617.22.38934 & \ra{17}{15}{19}{76} & \dec{-29}{38}{32}{3} & $2457132.60 ^{+0.02}_{-0.02}$ & $ 25.79 ^{+0.66}_{-0.60}$ & $0.118 ^{+0.004}_{-0.004}$ & $17.88 ^{+0.04}_{-0.04}$ & $0.93 ^{+0.03}_{-0.03}$ & OB150455 \\
BLG617.05.8156 & \ra{17}{15}{31}{67} & \dec{-30}{16}{59}{5} & $2456436.94 ^{+0.07}_{-0.08}$ & $ 27.86 ^{+2.30}_{-1.77}$ & $0.068 ^{+0.007}_{-0.007}$ & $19.85 ^{+0.12}_{-0.10}$ & $1.15 ^{+0.11}_{-0.12}$ & OB130699 \\
BLG617.12.109866 & \ra{17}{15}{40}{52} & \dec{-29}{49}{51}{4} & $2457845.65 ^{+0.54}_{-0.52}$ & $ 81.52 ^{+2.88}_{-2.57}$ & $0.414 ^{+0.018}_{-0.020}$ & $18.65 ^{+0.08}_{-0.07}$ & $1.04 ^{+0.06}_{-0.07}$ & OB170208 \\
BLG617.29.115121 & \ra{17}{15}{42}{04} & \dec{-29}{12}{06}{9} & $2456143.38 ^{+0.70}_{-0.75}$ & $ 75.17 ^{+5.11}_{-3.59}$ & $0.494 ^{+0.034}_{-0.047}$ & $19.08 ^{+0.16}_{-0.11}$ & $1.02 ^{+0.11}_{-0.14}$ & OB120851 \\
BLG617.29.102385 & \ra{17}{15}{43}{18} & \dec{-29}{18}{44}{1} & $2457484.80 ^{+0.63}_{-0.58}$ & $ 21.29 ^{+5.36}_{-2.67}$ & $0.506 ^{+0.122}_{-0.158}$ & $19.35 ^{+0.56}_{-0.36}$ & $0.70 ^{+0.28}_{-0.28}$ & OB160624 \\
BLG617.04.62618 & \ra{17}{15}{50}{74} & \dec{-30}{18}{46}{9} & $2457275.58 ^{+1.69}_{-1.84}$ & $ 22.81 ^{+5.47}_{-3.27}$ & $0.566 ^{+0.107}_{-0.151}$ & $19.55 ^{+0.51}_{-0.29}$ & $0.78 ^{+0.24}_{-0.29}$ &  \\
BLG617.29.88791 & \ra{17}{15}{55}{68} & \dec{-29}{08}{20}{3} & $2457941.92 ^{+0.35}_{-0.36}$ & $ 28.37 ^{+4.20}_{-2.23}$ & $0.451 ^{+0.063}_{-0.089}$ & $19.29 ^{+0.33}_{-0.20}$ & $0.85 ^{+0.18}_{-0.22}$ & OB171155 \\
BLG617.21.47343 & \ra{17}{16}{00}{70} & \dec{-29}{32}{38}{2} & $2457107.33 ^{+0.71}_{-0.67}$ & $ 33.90 ^{+19.29}_{-9.26}$ & $0.344 ^{+0.235}_{-0.172}$ & $20.11 ^{+0.90}_{-0.81}$ & $0.37 ^{+0.41}_{-0.21}$ & OB150454 \\
BLG617.29.41603 & \ra{17}{16}{01}{07} & \dec{-29}{17}{42}{0} & $2455857.41 ^{+0.42}_{-0.40}$ & $ 35.73 ^{+1.23}_{-0.92}$ & $0.492 ^{+0.016}_{-0.029}$ & $16.93 ^{+0.10}_{-0.05}$ & $0.95 ^{+0.05}_{-0.08}$ & OB111302 \\
BLG617.04.54041 & \ra{17}{16}{09}{35} & \dec{-30}{06}{06}{4} & $2457963.81 ^{+0.42}_{-0.43}$ & $ 10.90 ^{+2.83}_{-1.74}$ & $0.357 ^{+0.085}_{-0.102}$ & $20.35 ^{+0.46}_{-0.30}$ & $0.98 ^{+0.31}_{-0.34}$ & OB171467 \\
BLG617.28.101826 & \ra{17}{16}{28}{47} & \dec{-29}{20}{36}{4} & $2456462.98 ^{+0.37}_{-0.37}$ & $ 20.63 ^{+3.27}_{-1.65}$ & $0.563 ^{+0.068}_{-0.114}$ & $19.21 ^{+0.37}_{-0.20}$ & $0.86 ^{+0.17}_{-0.25}$ & OB130950 \\
BLG617.11.107345 & \ra{17}{16}{31}{47} & \dec{-29}{46}{55}{6} & $2457155.46 ^{+1.70}_{-1.68}$ & $ 38.30 ^{+19.59}_{-7.74}$ & $0.557 ^{+0.227}_{-0.266}$ & $20.37 ^{+0.98}_{-0.63}$ & $0.55 ^{+0.44}_{-0.33}$ & OB151021 \\
BLG617.03.87926 & \ra{17}{16}{31}{48} & \dec{-30}{18}{45}{1} & $2455984.69 ^{+0.04}_{-0.04}$ & $ 51.71 ^{+3.17}_{-2.68}$ & $0.006 ^{+0.001}_{-0.001}$ & $20.04 ^{+0.08}_{-0.07}$ & $1.03 ^{+0.07}_{-0.07}$ & OB120030 \\
BLG617.20.63992 & \ra{17}{16}{39}{97} & \dec{-29}{39}{56}{8} & $2457835.07 ^{+0.04}_{-0.03}$ & $ 18.48 ^{+0.87}_{-0.73}$ & $0.052 ^{+0.003}_{-0.003}$ & $18.78 ^{+0.07}_{-0.06}$ & $0.98 ^{+0.06}_{-0.06}$ & OB170284 \\
BLG617.03.55959 & \ra{17}{16}{40}{95} & \dec{-30}{20}{47}{5} & $2457988.33 ^{+0.28}_{-0.28}$ & $ 11.19 ^{+3.29}_{-1.58}$ & $0.224 ^{+0.057}_{-0.081}$ & $19.59 ^{+0.52}_{-0.29}$ & $0.80 ^{+0.24}_{-0.30}$ &  \\
BLG617.03.26929 & \ra{17}{16}{44}{41} & \dec{-30}{19}{46}{3} & $2456135.78 ^{+0.04}_{-0.05}$ & $ 54.31 ^{+2.95}_{-2.70}$ & $0.049 ^{+0.003}_{-0.003}$ & $19.78 ^{+0.07}_{-0.07}$ & $0.38 ^{+0.03}_{-0.02}$ & OB121056 \\
BLG617.28.43862 & \ra{17}{16}{44}{54} & \dec{-29}{17}{37}{6} & $2456230.88 ^{+3.97}_{-4.26}$ & $141.95 ^{+14.56}_{-8.21}$ & $0.838 ^{+0.070}_{-0.116}$ & $19.22 ^{+0.30}_{-0.16}$ & $0.93 ^{+0.15}_{-0.22}$ & OB121403 \\
BLG616.20.2812 & \ra{17}{16}{46}{34} & \dec{-28}{25}{36}{4} & $2456034.77 ^{+0.02}_{-0.02}$ & $ 13.76 ^{+1.44}_{-0.97}$ & $0.073 ^{+0.022}_{-0.029}$ & $19.05 ^{+0.19}_{-0.14}$ & $0.93 ^{+0.13}_{-0.15}$ &  \\
BLG616.20.31331 & \ra{17}{16}{52}{76} & \dec{-28}{20}{29}{3} & $2456417.81 ^{+0.04}_{-0.04}$ & $ 52.37 ^{+0.53}_{-0.37}$ & $0.218 ^{+0.002}_{-0.003}$ & $16.48 ^{+0.02}_{-0.01}$ & $0.99 ^{+0.01}_{-0.02}$ & OB131368 \\
BLG617.28.21305 & \ra{17}{16}{55}{48} & \dec{-29}{11}{04}{1} & $2457167.08 ^{+0.46}_{-0.47}$ & $ 98.73 ^{+15.58}_{-13.30}$ & $0.238 ^{+0.054}_{-0.043}$ & $19.89 ^{+0.26}_{-0.28}$ & $0.42 ^{+0.12}_{-0.09}$ &  \\
BLG617.28.17554 & \ra{17}{16}{56}{59} & \dec{-29}{15}{02}{8} & $2455992.87 ^{+0.05}_{-0.05}$ & $ 18.17 ^{+2.05}_{-1.77}$ & $0.045 ^{+0.006}_{-0.006}$ & $20.09 ^{+0.15}_{-0.14}$ & $0.84 ^{+0.12}_{-0.11}$ & OB120156 \\
BLG616.19.79418 & \ra{17}{17}{24}{57} & \dec{-28}{23}{07}{7} & $2456102.24 ^{+0.15}_{-0.16}$ & $ 10.87 ^{+5.35}_{-3.63}$ & $0.223 ^{+0.175}_{-0.094}$ & $20.03 ^{+0.67}_{-0.80}$ & $0.20 ^{+0.22}_{-0.09}$ &  \\
\enddata
\tabletypesize{\small}
\tablecomments{For each parameter, we provide the median and $1\sigma$ confidence interval derived from the marginalized posterior distribution from the Monte Carlo chain. Here $I_{\rm s}$ is the source brightness and $f_{\rm s} = F_{\rm s}/(F_{\rm s}+F_{\rm b})$ is the blending parameter. Equatorial coordinates are given for the epoch J2000, and OBNNMMMM stands for OGLE-20NN-BLG-MMMM. (This table is available in its entirety in machine-readable form.)}
\end{deluxetable*}

\begin{deluxetable*}{llrrrrrrr}
\tablecaption{Star Surface Density (arcmin$^{-2}$) in selected \textit{HST} Fields toward the Galactic Bulge \citep{holtzman2006,brown2010}. \label{tab:hst_density}}
\tablecolumns{9}
\tablewidth{0pt}
\tablehead{
\colhead{\textit{HST} Field} &
\colhead{OGLE Field} &
\colhead{R.A.} &
\colhead{Decl.} &
\colhead{$\Sigma^{HST}_{20}$} & 
\colhead{$\Sigma^{\rm LF}_{20}$} &
\colhead{$\Sigma^{HST}_{21}$} & 
\colhead{$\Sigma^{\rm LF}_{21}$} & 
\colhead{$\Sigma^{\rm simul}_{21}$}
}
\startdata
mw\_bulge\_u2c901 & BLG505.19 & \rashort{17}{59}{00} & \decshort{-29}{12}{13} & 2429.2 & 2395.9 & 4592.9  & 4390.0 & 3860.0 \\
mw\_bulge\_u2cl02 & BLG513.12 & \rashort{18}{03}{09} & \decshort{-29}{51}{45} & 1603.8 & 1476.1 & 2825.3 & 2636.5 & 2769.4 \\
mw\_bulge\_u2tw01 & BLG513.12 & \rashort{18}{03}{09} & \decshort{-29}{51}{44} & 1610.8 & 1476.1 & 2841.2 & 2636.5 & 2769.4 \\
mw\_bulge\_u66h01 & BLG513.12 & \rashort{18}{03}{10} & \decshort{-29}{51}{44} & 1636.6 & 1476.1 & 2848.4 & 2636.5 & 2769.4 \\
mw\_bulge\_u6ls02 & BLG511.01 & \rashort{18}{05}{04} & \decshort{-27}{42}{29} & 1521.3 & 1610.2 & 2760.3 & 2912.0 & 2950.3 \\
mw\_bulge\_u2oq03 & BLG532.08 & \rashort{18}{22}{15} & \decshort{-29}{19}{33} &  --    & -- &  524.0\tablenotemark{a} & -- & 533.4\tablenotemark{a} \\
ogle29            & BLG600.25 & \rashort{17}{48}{15} & \decshort{-37}{09}{01} &  575.0 & -- & 1070.6 & -- & 1251.6\\
stanek            & BLG501.17 & \rashort{17}{54}{42} & \decshort{-29}{49}{30} & 2416.3 & 2389.8 & 5009.3 & 4802.7 & 4124.6 \\
sweeps            & BLG505.19 & \rashort{17}{58}{59} & \decshort{-29}{12}{18} & 2295.0 & 2395.9 & 4377.4 & 4390.0 & 3860.0 \\
baade             & BLG513.12 & \rashort{18}{03}{10} & \decshort{-29}{56}{34} & 1537.0 & 1476.1 & 2841.7 & 2636.5 & 2769.4 \\
\enddata
\tablenotetext{a}{Surface density in the range $19.5 < I < 21$.}
\tablecomments{Here $\Sigma_{20}$ and $\Sigma_{21}$ are the surface densities in the ranges $14<I<20$ and $14<I<21$, respectively. The star surface density was calculated using the \textit{HST} images, matching the template LF, and image-level simulations.}
\end{deluxetable*}

\begin{figure}
\centering
\includegraphics[width=0.8\textwidth]{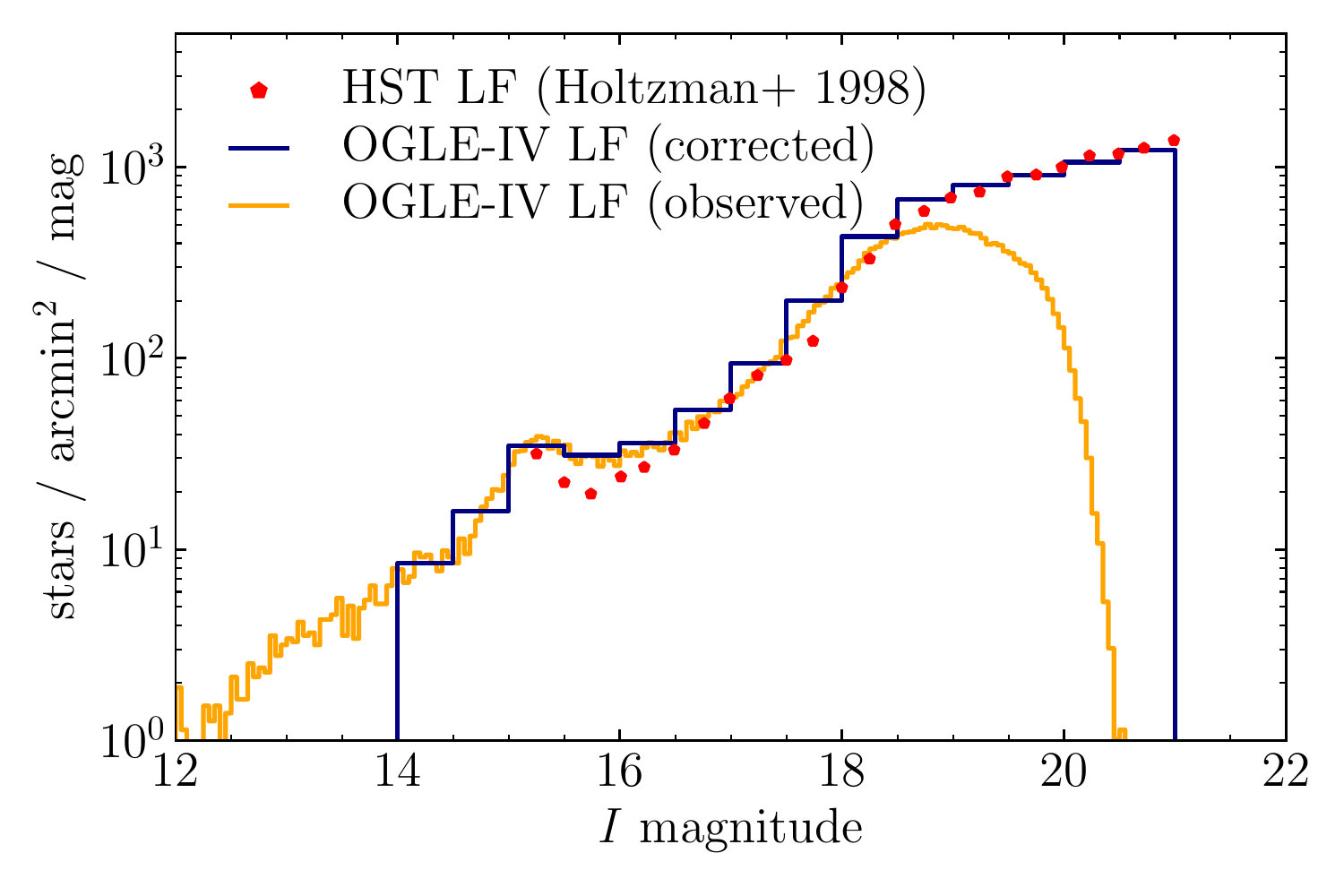}
\caption{Comparison between the OGLE-IV LF of the field BLG513.12 and the \textit{HST} LF of the same region \citep{holtzman1998}. The detection efficiency simulations are sufficient for correcting the LF (blue histogram) so that it becomes consistent with deep \textit{HST} observations (red points).}
\label{fig:hst_lf}
\end{figure}

\begin{figure}
\includegraphics[width=0.32\textwidth]{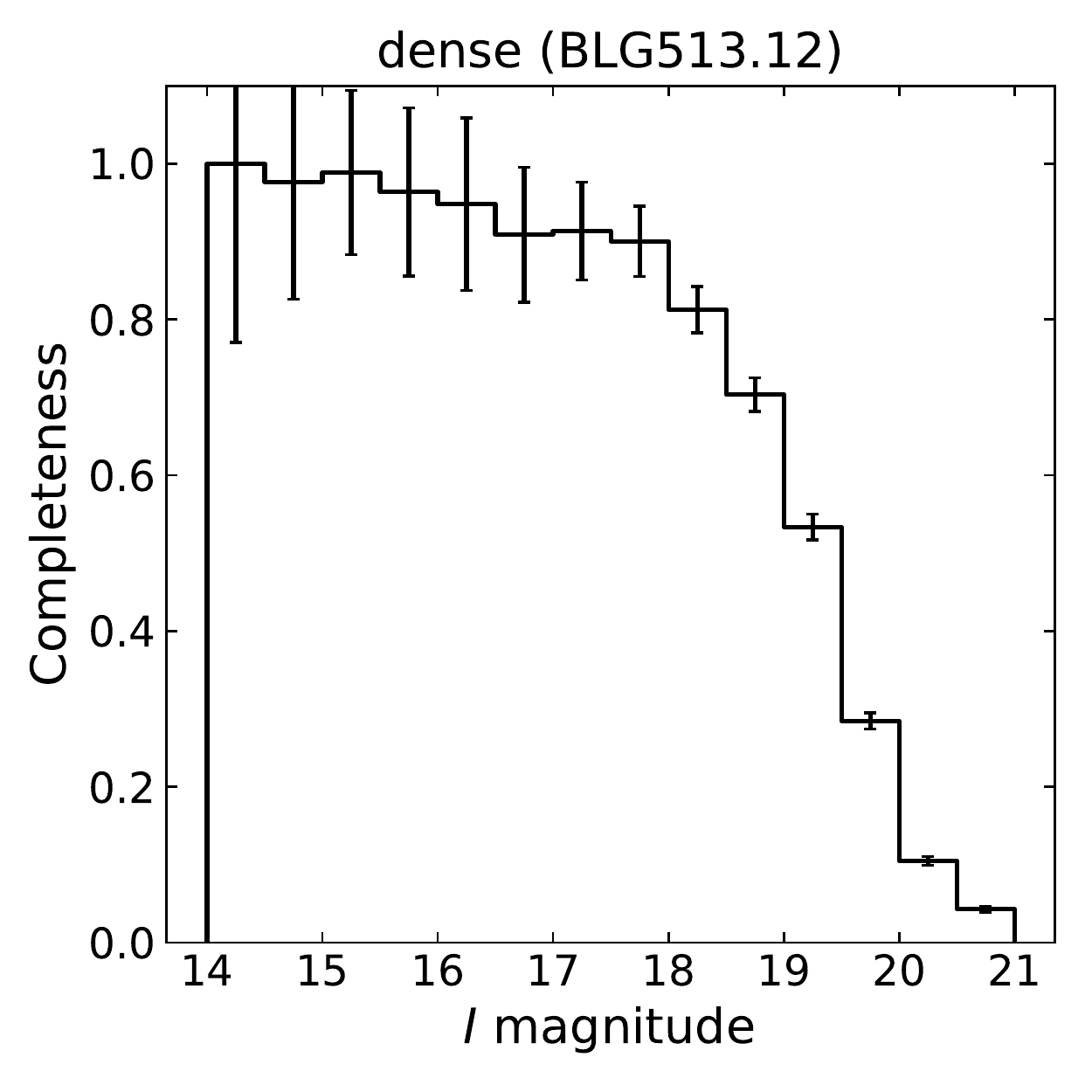}
\includegraphics[width=0.32\textwidth]{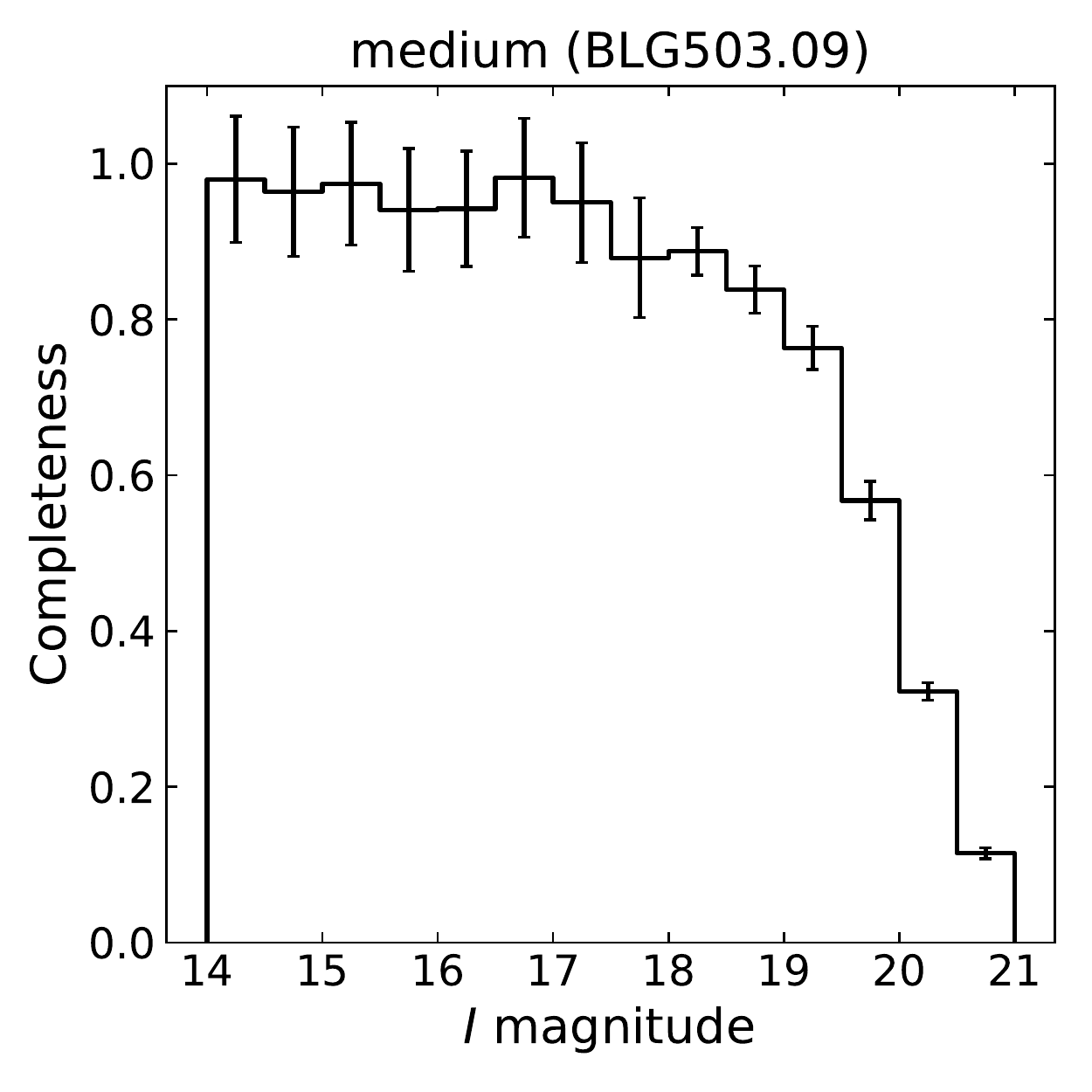}
\includegraphics[width=0.32\textwidth]{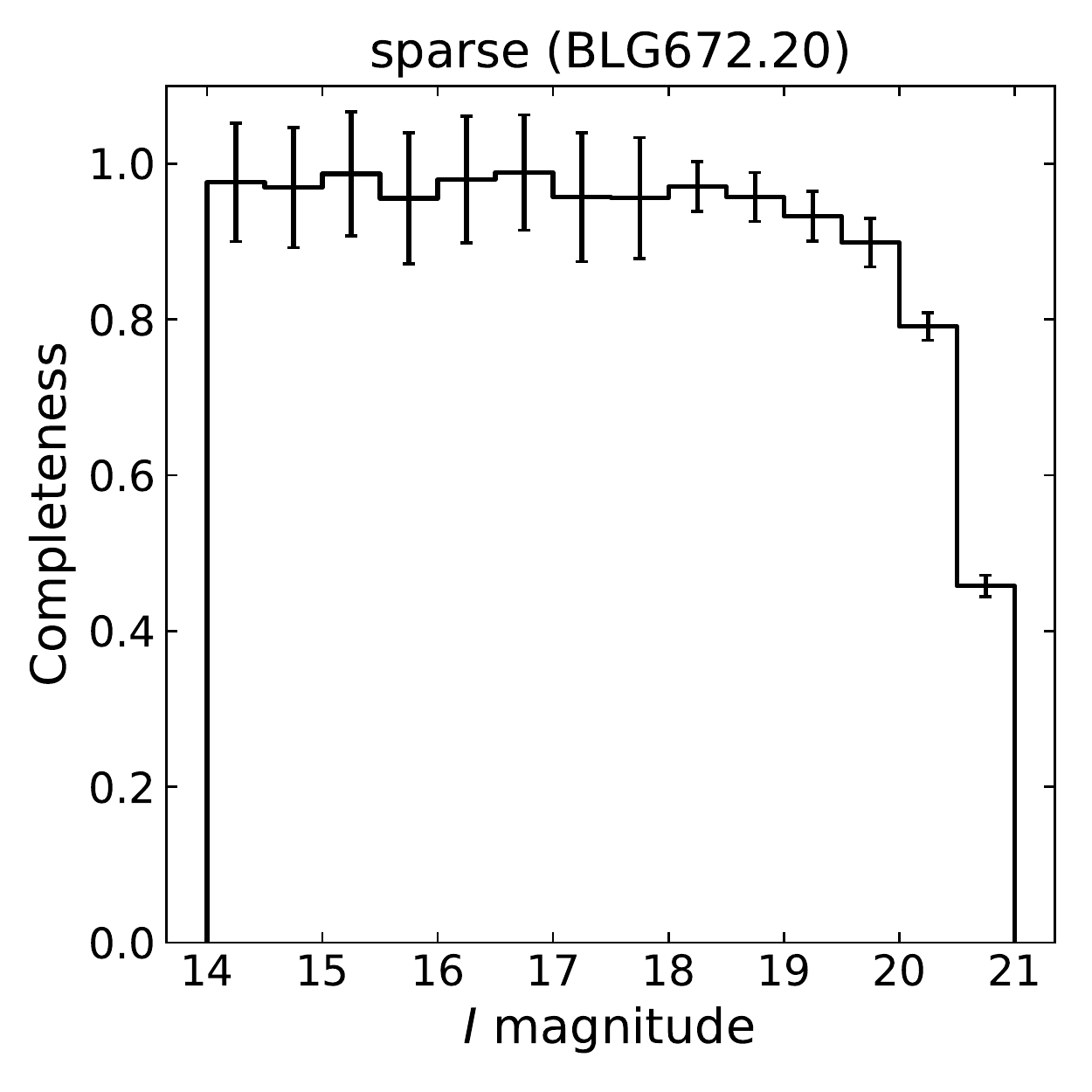}
\caption{Completeness of star counts toward three OGLE fields as a function of brightness. The selected three fields represent regions of high (BLG513.12: $2800$~arcmin$^{-2}$), medium (BLG503.09: $1800$~arcmin$^{-2}$), and low (BLG672.20: $500$~arcmin$^{-2}$) star density. The OGLE star counts are nearly complete down to $I\approx 18$ (BLG513.12), $I\approx 19$ (BLG503.09), and $I\approx 20$ (BLG672.20), respectively.}
\label{fig:cts_completeness}
\end{figure}

\section{Star Counts}
\label{sec:counts}

The number of monitored sources is an essential quantity in microlensing optical depth calculations. While it is usually presumed that star catalogs are nearly complete at the bright end of the luminosity function (LF), this is not true for faint sources. (In fact, the incompleteness in red clump giant counts may have led to the discrepancy between optical depths based on bright and faint sources; \citealt{sumi_penny2016}). The density of stars brighter than $I=21$ in the most crowded regions of the Galactic bulge exceeds 4000 stars per arcmin$^2$, which corresponds to about 0.7 unresolved blends in a typical seeing disk ($\mathrm{FWHM}=0.8-0.9''$) of a star. A faint star can be hidden in the glow of a bright neighbor, or two faint stars cannot be resolved, and the total brightness of the blend is higher than the brightness of either of the sources. Star catalogs might therefore be highly incomplete, especially in crowded fields. We calculated the number of monitored sources using three independent methods, all of which yielded consistent results. 

The most robust approach to counting the source stars is to use deep, high-resolution images of a given field taken with the \textit{Hubble Space Telescope} (\textit{HST}). This method is, however, impractical in our study, because sufficiently deep \textit{HST} pointings are available for only a few sight lines toward the Galactic bulge \citep[e.g.,][]{holtzman1998,holtzman2006}. Moreover, the observed number density of stars may vary on small angular scales due to variable and patchy interstellar extinction. 

We used several \textit{HST} pointings as a ``ground truth'' to test the accuracy of other methods of counting source stars. We used the database of stellar photometry of several Galactic bulge fields obtained using the Wide Field Planetary Camera 2 (WFPC2) on board \textit{HST} \citep{holtzman2006}. The WFPC2 camera has a field of view of 4.97~arcmin$^2$ and a pixel scale of $\sim 0.0455''$ or $\sim0.1''$ per pixel, depending on the detector \citep{holtzman1995}. The obser\-va\-tions were taken through the F814W filter and transformed to the Cousins $I$ magnitudes. \citet{holtzman2006} also provided information on the completeness of the photometry as a function of brightness based on image-level simulations, which allowed us to correct the observed LFs. Additionally, we used the archival \textit{HST} observations of four Galactic bulge fields \citep{brown2009,brown2010} taken with the Wide Field Camera 3 \citep{mackenty2010}. Our results for 10~\textit{HST} fields are reported in Table~\ref{tab:hst_density}.

The most common approach to assess the number of sources monitored in the microlensing experiment is to use one deep LF of a single field as a template \citep[e.g.,][]{alcock2000,sumi2003}. The template LF is shifted in brightness and rescaled so that the brightness and number of red clump stars match the observed bright end of the LF in a given direction. We used this method to calculate the number of monitored sources in 452 selected subfields. The template LF was constructed using deep \textit{HST} observations \citep{holtzman1998} for faint stars ($I\geq 17$) and the OGLE-IV LF of the field BLG513.12 for bright stars ($I<17$). While the presented method can work well for neighboring regions, it may fail for fields located at or above the Galactic equator (as well as in the Galactic plane, far from the bulge), where the shape of the LF may be different.  

We therefore tried a novel approach. The pixel size ($\sim 0.26''$) of the OGLE-IV camera and typical PSF size of stars on reference images ($0.8-0.9''$) are much better than in previous experiments (although still inferior to the \textit{HST} images). We carried out a series of image-level simulations to estimate the completeness of our star catalogs. We injected artificial stars ($14<I<21$) into random locations on real OGLE images, stacked the images into the deep reference image, and ran our star detection pipeline \citep{wozniak2000} in exactly the same way as real star catalogs were created. We injected 5000 stars per frame so that the density of the stars tended to increase by less than $5\%$. We consider the artificial star as detected if (1) the measured centroid is consistent (within 1.5~pixels) with the location where the star was placed and the closest star from the original catalog is at least 2.1~pixels away (the artificial star is in an ``empty'' field) or if (2) the measured brightness of the artificial star $I_{\rm sim}$ is closer to the input brightness $I_{\rm in}$ than to the brightness of the real neighboring star $I_{\rm nei}$ (i.e., $|I_{\rm sim}-I_{\rm in}|<|I_{\rm sim}-I_{\rm nei}|$) if such a neighbor was detected within 2.1~pixels on the original frame (in other words, the real star from the original catalog becomes a blend). The star detection algorithm can separate neighboring objects as close as less than 2~pixels away, but its effectiveness depends on the flux and flux ratio. 

\begin{figure}
\centering
\includegraphics[width=0.8\textwidth]{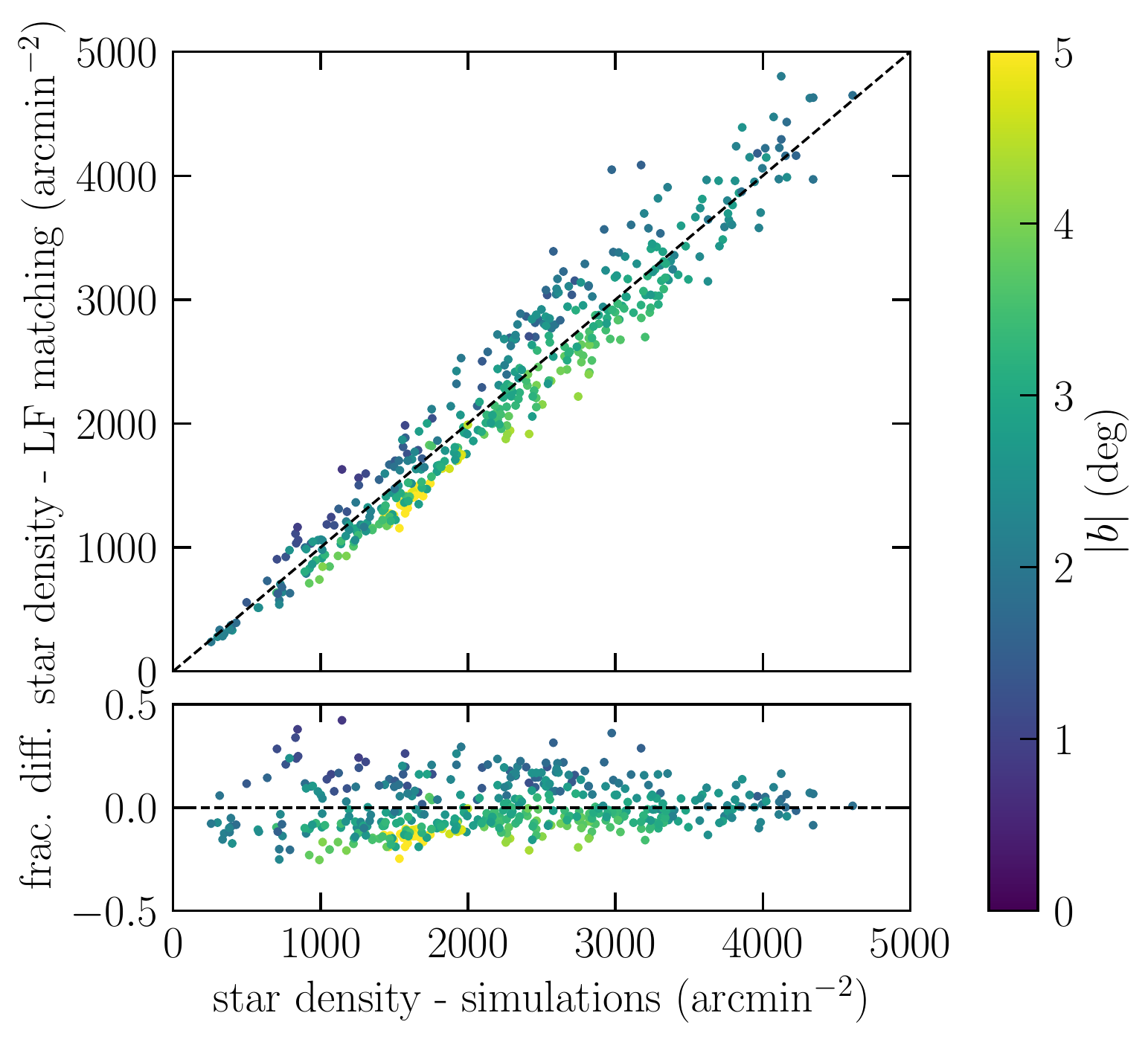}
\caption{Comparison between the star surface density calculated using image-level simulations and that measured by matching LFs (upper panel). The lower panel presents the fractional difference. Data points are color-coded by the absolute value of the Galactic latitude.}
\label{fig:comparing_cts}
\end{figure}

\begin{figure}
\centering
\includegraphics[height=0.7\textwidth]{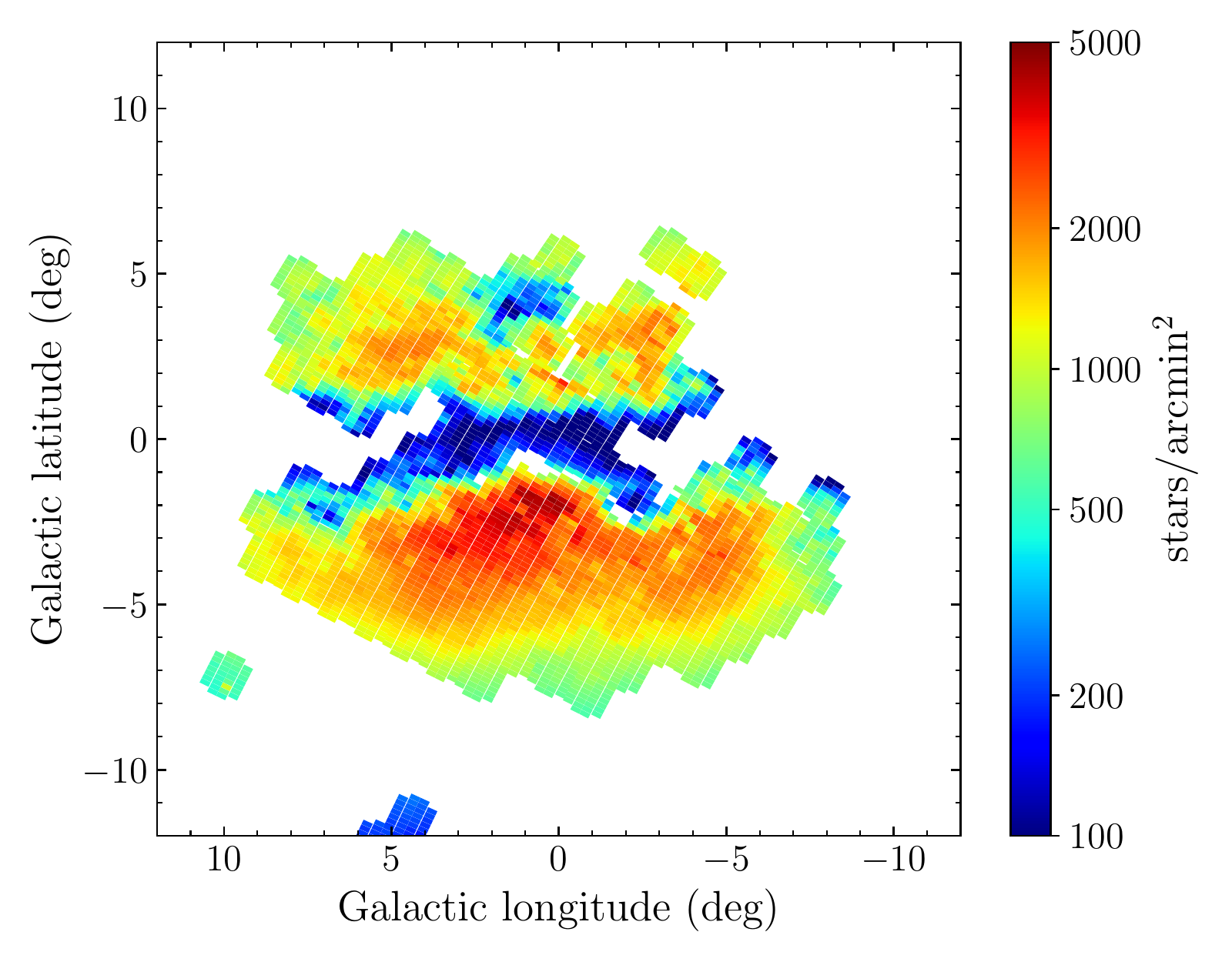}
\caption{Density of stars down to $I=21$ in OGLE-IV fields toward the Galactic bulge.}
\label{fig:density}
\end{figure}

We calculated the completeness of our star catalogs in 14 0.5~mag wide bins and corrected the observed LFs for each subfield. This approach works well for $I<20.5$. In a few cases of the most crowded fields, we needed to extrapolate the LF for the faintest sources ($I>20.5$) based on two or three earlier bins. Our LF of the field BLG513.12 is consistent with the \textit{HST} results \citep{holtzman1998} (Figure \ref{fig:hst_lf}). Star catalogs are nearly complete down to $I=18$ in the most crowded fields and even to $I=20$ in relatively empty fields (Figure~\ref{fig:cts_completeness}). The overall completeness (down to $I=21$) typically varies from 30\% to 80\%, depending on the field. 

Figure~\ref{fig:comparing_cts} shows the comparison between the star surface density (down to $I=21$) measured using image-level simulations ($\Sigma_{21}^{\rm simul}$) and that estimated from matching the template LF ($\Sigma_{21}^{\rm LF}$). On average, $\Sigma_{21}^{\rm LF}/\Sigma_{21}^{\rm simul}=1.01 \pm 0.01$, although we noticed a small bias. In fields located close to the Galactic plane, $\Sigma_{21}^{\rm LF}>\Sigma_{21}^{\rm simul}$, while far from the plane, $\Sigma_{21}^{\rm LF}<\Sigma_{21}^{\rm simul}$. This bias may be related to the fact that Galactic disk sources modify the shape of the LF at low latitudes, and our assumption that its shape is identical across all fields is invalid. The residuals presented in the lower panel of Figure~\ref{fig:comparing_cts} are correlated with Galactic latitude, i.e.,  $\Sigma_{21}^{\rm LF}/\Sigma_{21}^{\rm simul}-1=p(|b|)$, where $p$ is a second-order polynomial. The scatter about this function (0.08) is a good measure of the systematic error in the source counts.

Our comparison of measured star densities to those inferred directly from \textit{HST} images (Table~\ref{tab:hst_density}) indicates that both proposed methods of counting sources (template matching and image-level simulations) are accurate to about 10--15\%. These tests demonstrate that we are presently unable to measure the number of monitored sources with accuracy better than 10\%. In turn, optical depths and event rates may suffer from systematic errors at the \textbf{10\% level}. Because our sample of microlensing events is large, the accuracy of inferred optical depths and event rates will be mostly limited by the accuracy of the determination of the number of sources, not by small numbers of events as in the previous studies.

The number and surface density of stars in the analyzed subfields calculated using image-level simulations are presented in Table \ref{tab:stars} and Figure \ref{fig:density}.

\begin{deluxetable*}{lrrrrrrrr}
\tablecaption{Surface Density of Stars in OGLE-IV Subfields Calculated Using Image-level Simulations \label{tab:stars}}
\tablehead{
\colhead{Subfield} & \colhead{R.A.} & \colhead{Decl.} & \colhead{$l$} & 
\colhead{$b$} & \colhead{$\Sigma_{18}$} & \colhead{$\Sigma_{21}$} & \colhead{$N_{18}$} & \colhead{$N_{21}$} \\
\colhead{} & \colhead{(J2000.0)} & \colhead{(J2000.0)} & \colhead{(deg)} & 
\colhead{(deg)} & \colhead{(arcmin$^{-2}$)} & \colhead{(arcmin$^{-2}$)} & \colhead{} & \colhead{}
}
\startdata
BLG500.01 & 268.526 & --29.081 & 0.829 & --1.666 & 356.6 & 4333.5 & 59437 & 722294 \\
BLG500.02 & 268.351 & --29.081 & 0.751 & --1.534 & 276.5 & 3174.1 & 46075 & 528914 \\
BLG500.03 & 268.175 & --29.081 & 0.673 & --1.401 & 225.8 & 2704.4 & 37609 & 450453 \\
BLG500.04 & 267.999 & --29.081 & 0.595 & --1.269 & 166.7 & 1588.2 & 27758 & 264465 \\
BLG500.05 & 267.823 & --29.081 & 0.516 & --1.137 &  85.2 &  829.9 & 14184 & 138160 \\
BLG500.08 & 268.694 & --28.759 & 1.180 & --1.630 & 270.6 & 4225.4 & 45092 & 704120 \\
BLG500.09 & 268.518 & --28.759 & 1.102 & --1.497 & 277.4 & 3963.2 & 46193 & 659959 \\ 
BLG500.10 & 268.342 & --28.759 & 1.024 & --1.365 & 231.3 & 2724.2 & 38507 & 453530 \\
BLG500.11 & 268.167 & --28.759 & 0.946 & --1.232 & 185.8 & 1757.4 & 30932 & 292575 \\
\enddata
\tablecomments{Here $\Sigma_{18}$ and $\Sigma_{21}$ are the surface densities of stars brighter than $I=18$ and $I=21$, respectively, and $N_{18}$ and $N_{21}$ are the numbers of stars brighter than $I=18$ and $I=21$, respectively. We note that the subfield (reference image) area may be slightly larger than the area covered by a single CCD detector because the reference image is the sum of a few frames that may be somewhat offset. (This table is available in its entirety in machine-readable form.)}
\end{deluxetable*}

\section{Distribution of the Blending Parameter}
\label{sec:blending}

Due to the high density of stars toward the Galactic bulge and the PSF size of objects on reference images, some sources cannot be resolved on OGLE template images (this phenomenon is called ``blending''). A faint star can be hidden in the glow of a bright neighbor, or two faint stars cannot be resolved, and the total brightness of the blend is higher than the brightness of either of the sources. We used the image-level simulations that were described in the previous section to derive the distribution of the blending parameter $f_{\rm s}$ as a function of the brightness of the baseline star. These distributions will be necessary for catalog-level simulations of detection efficiency of microlensing events in our experiment. The blending parameter is defined as the ratio between the source flux and the total flux of the detected object (i.e., the sum of the fluxes of the source and unrelated blends). 

Previously, \citet{wyrzykowski2015} used archival \textit{HST} observations of the \mbox{OGLE-III} field BLG206 to obtain the distribution of the blending parameter in that field. They matched OGLE stars to individual stars present on the \textit{HST} image and calculated the ratio of their flux to the total brightness of the object detected on the OGLE template image. Then, they assumed that the distribution of blending is the same across all analyzed OGLE-III fields.

We used image-level simulations to construct distributions of the blending parameter in all analyzed fields. We matched stars injected into images with those detected on reference images (we used a matching radius of 1.5~pixels). The blending parameter is simply $f_{\rm s} = F_{\rm in} / F_{\rm out}$, where $F_{\rm in}$ is the input flux and $F_{\rm out}$ is the flux measured on the template image. Sources were drawn from the completeness-corrected OGLE LFs of corresponding fields.

\begin{figure}
\centering
\includegraphics[width=0.49\textwidth]{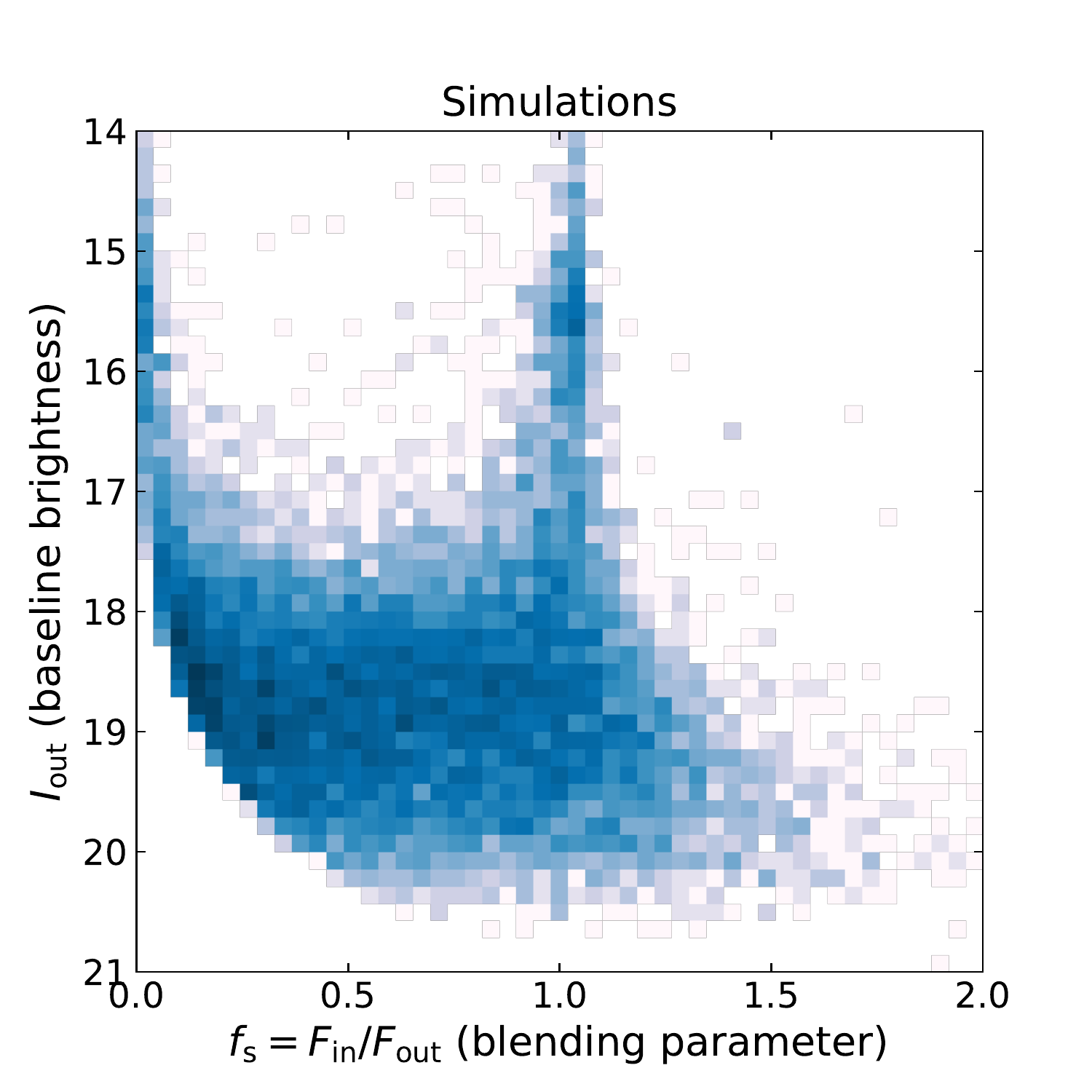}
\includegraphics[width=0.49\textwidth]{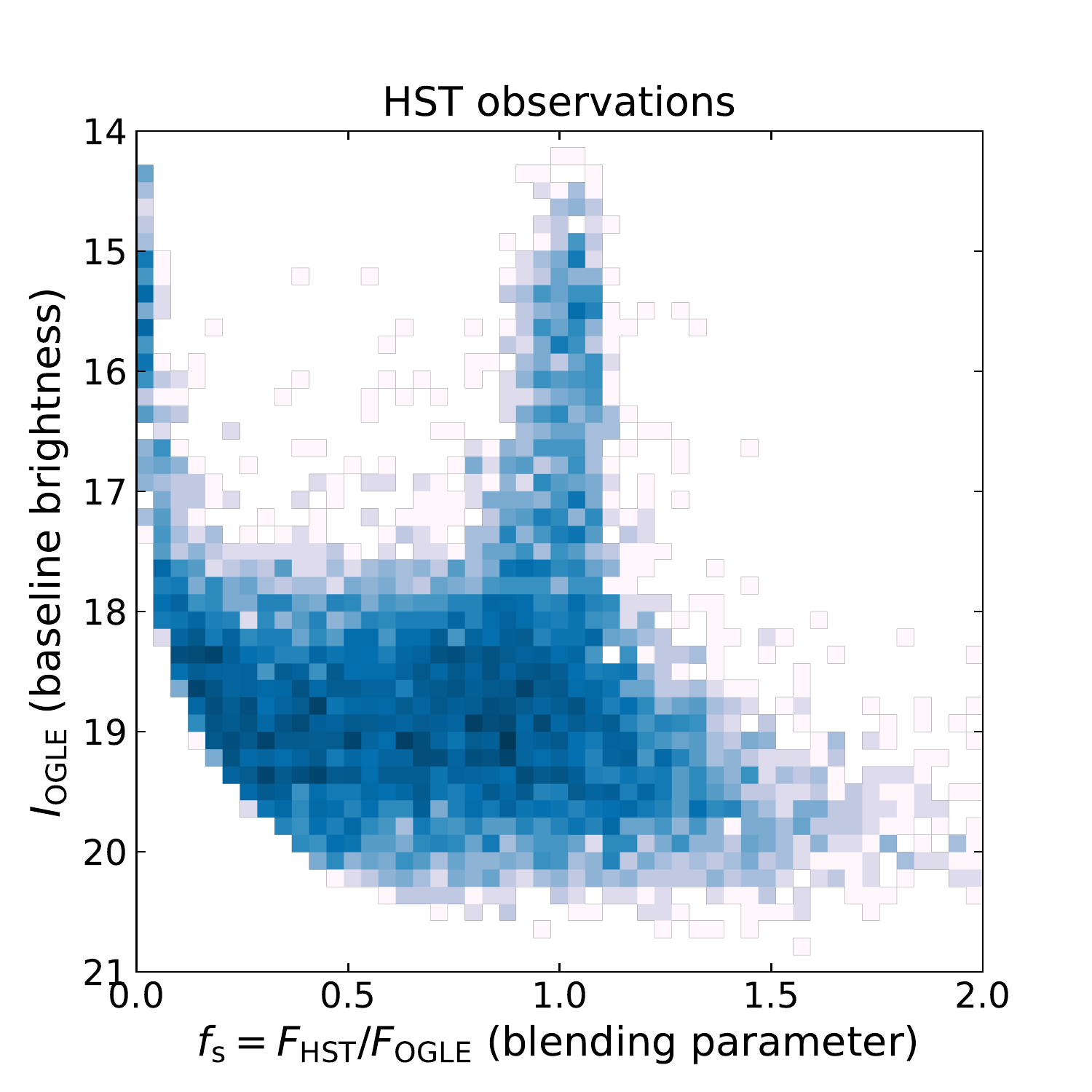}\\
\includegraphics[width=0.49\textwidth]{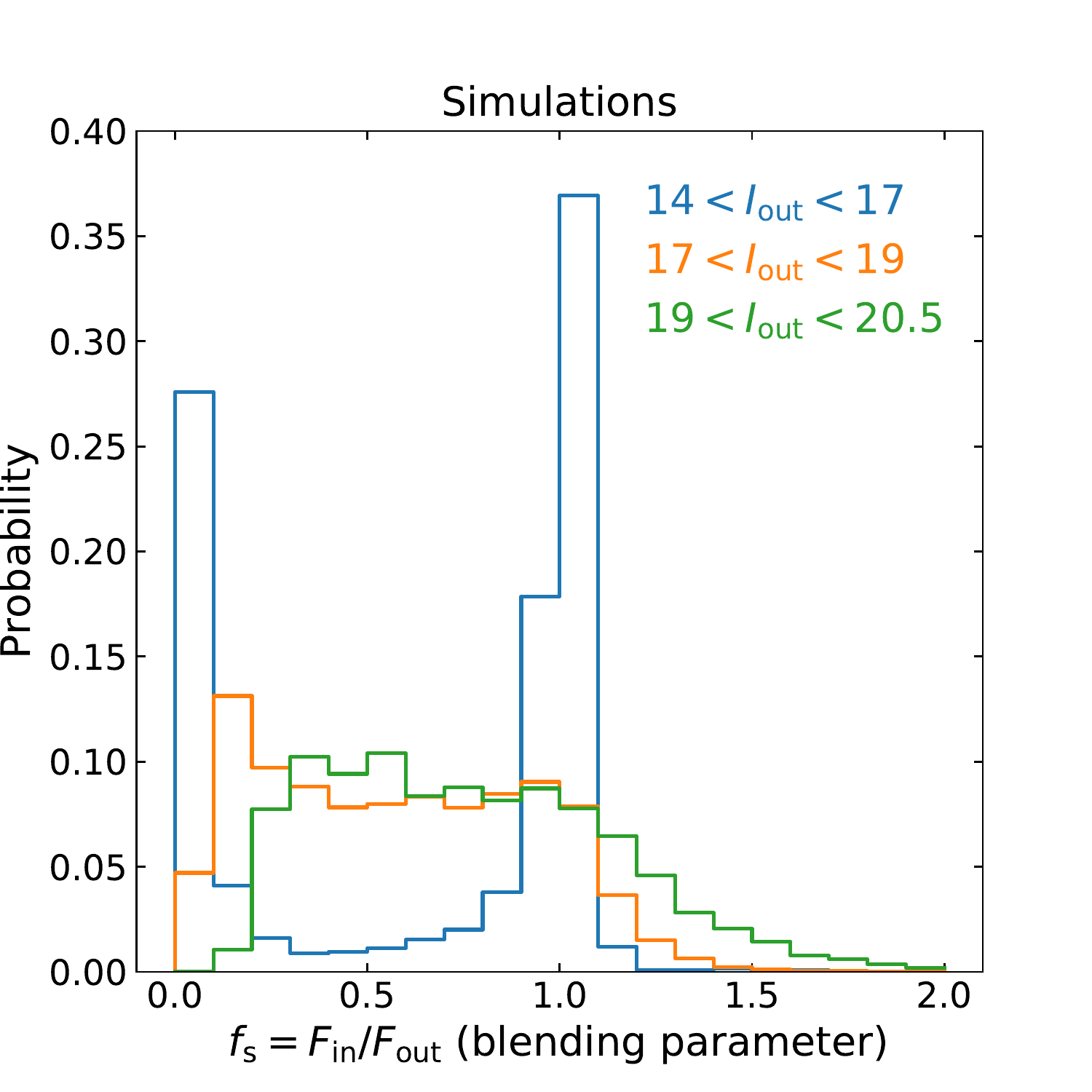}
\includegraphics[width=0.49\textwidth]{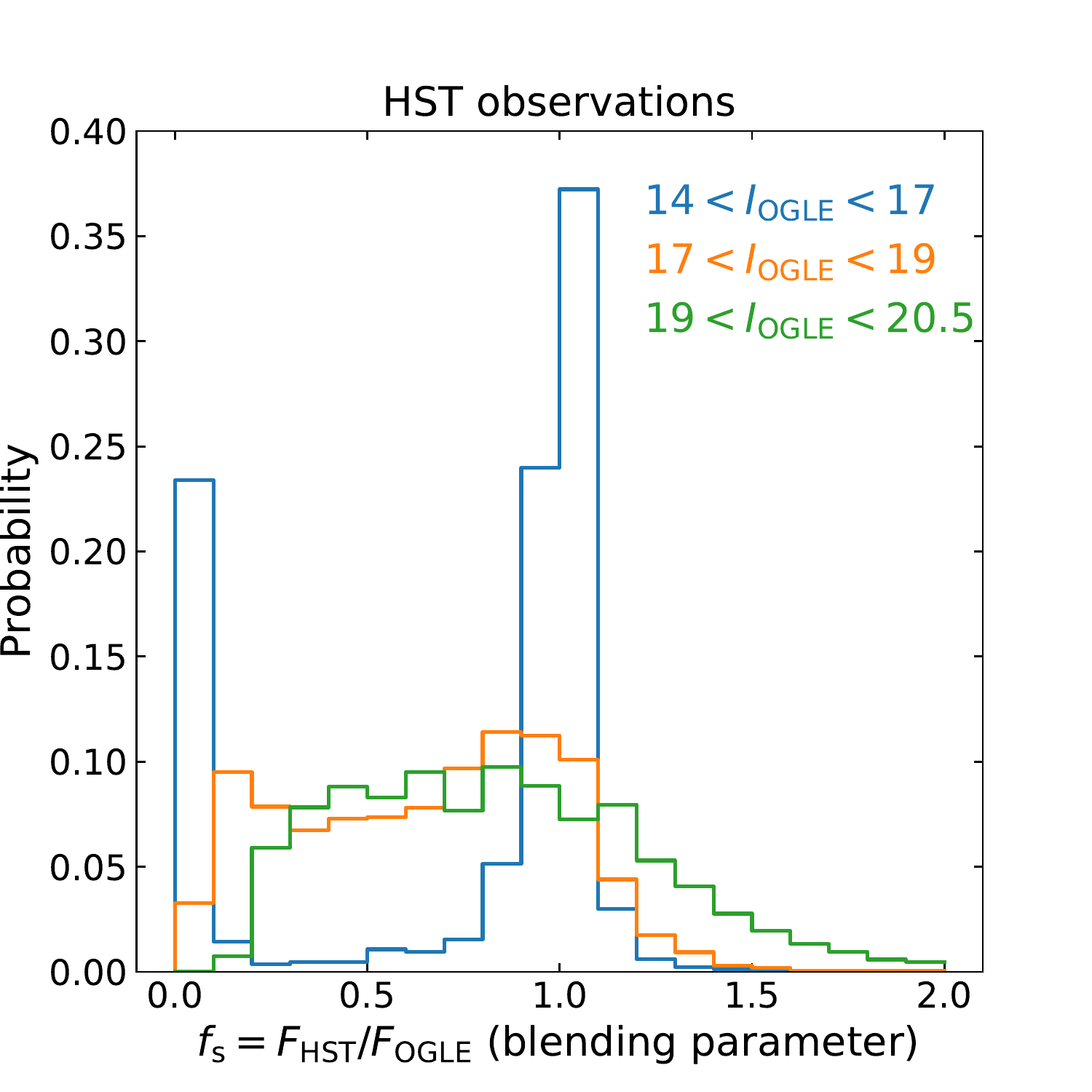} 
\caption{Comparison between distributions of blending parameter found using image-level simulations (left panels) and the \textit{HST} observations (right panels) of the same field. The \textit{HST} observations are taken from \citet{wyrzykowski2015}. Note that \citet{wyrzykowski2015} analyzed images obtained with the OGLE-III camera, which has the same pixel size as the OGLE-IV camera, but the reference images are different.}
\label{fig:blending_comparison}
\end{figure}

\begin{figure}
\centering
\includegraphics[height=0.4\textwidth]{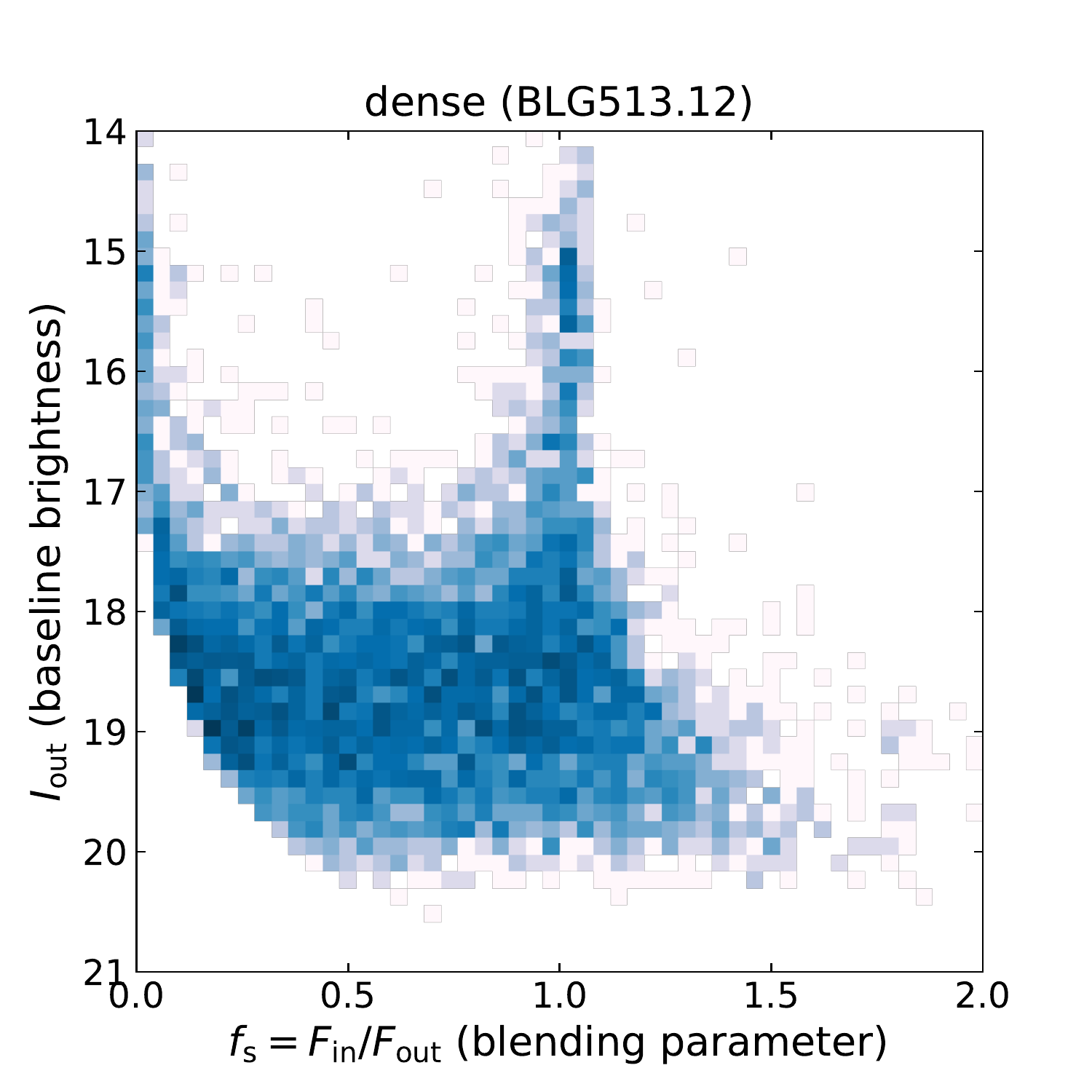}
\includegraphics[height=0.4\textwidth]{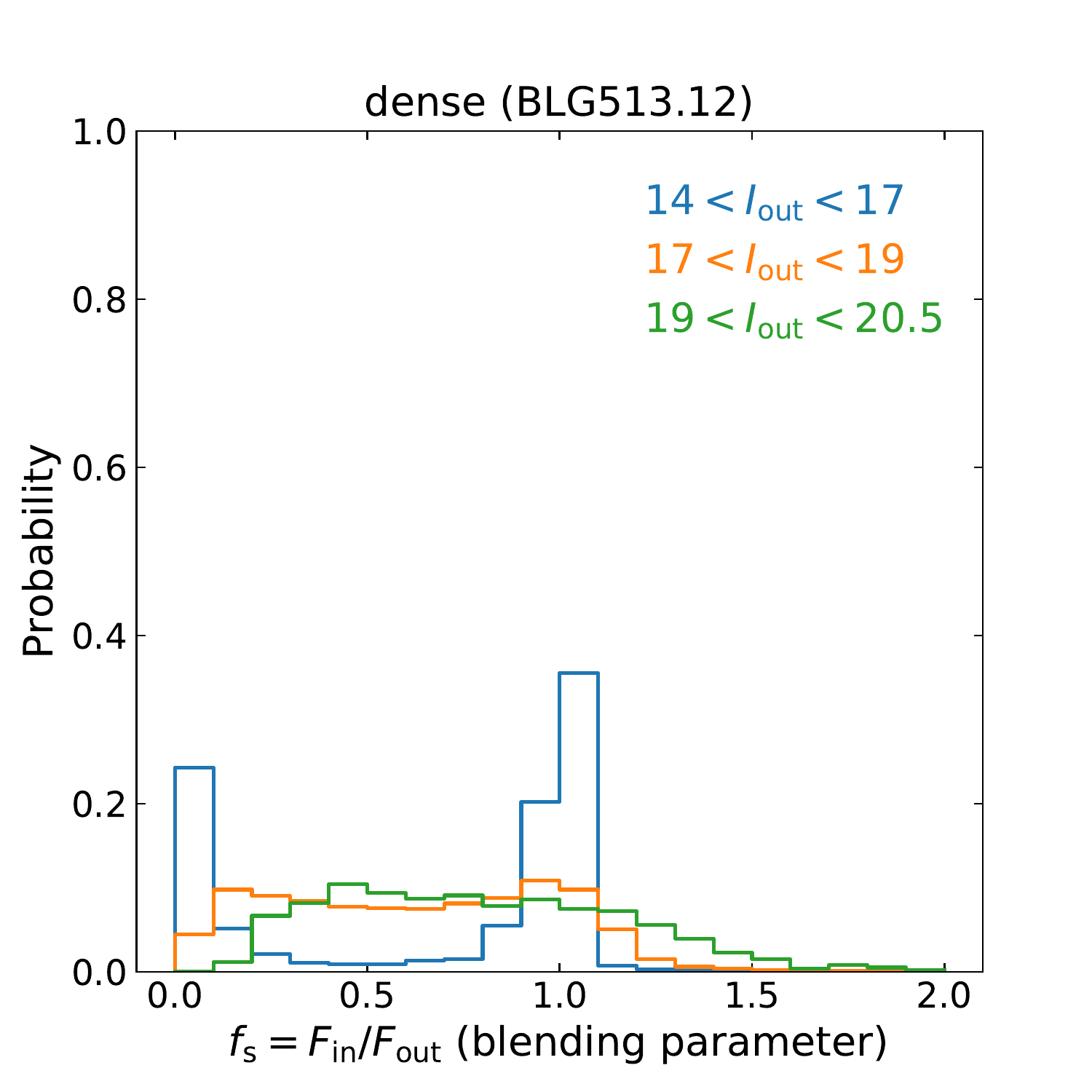}\\
\includegraphics[height=0.4\textwidth]{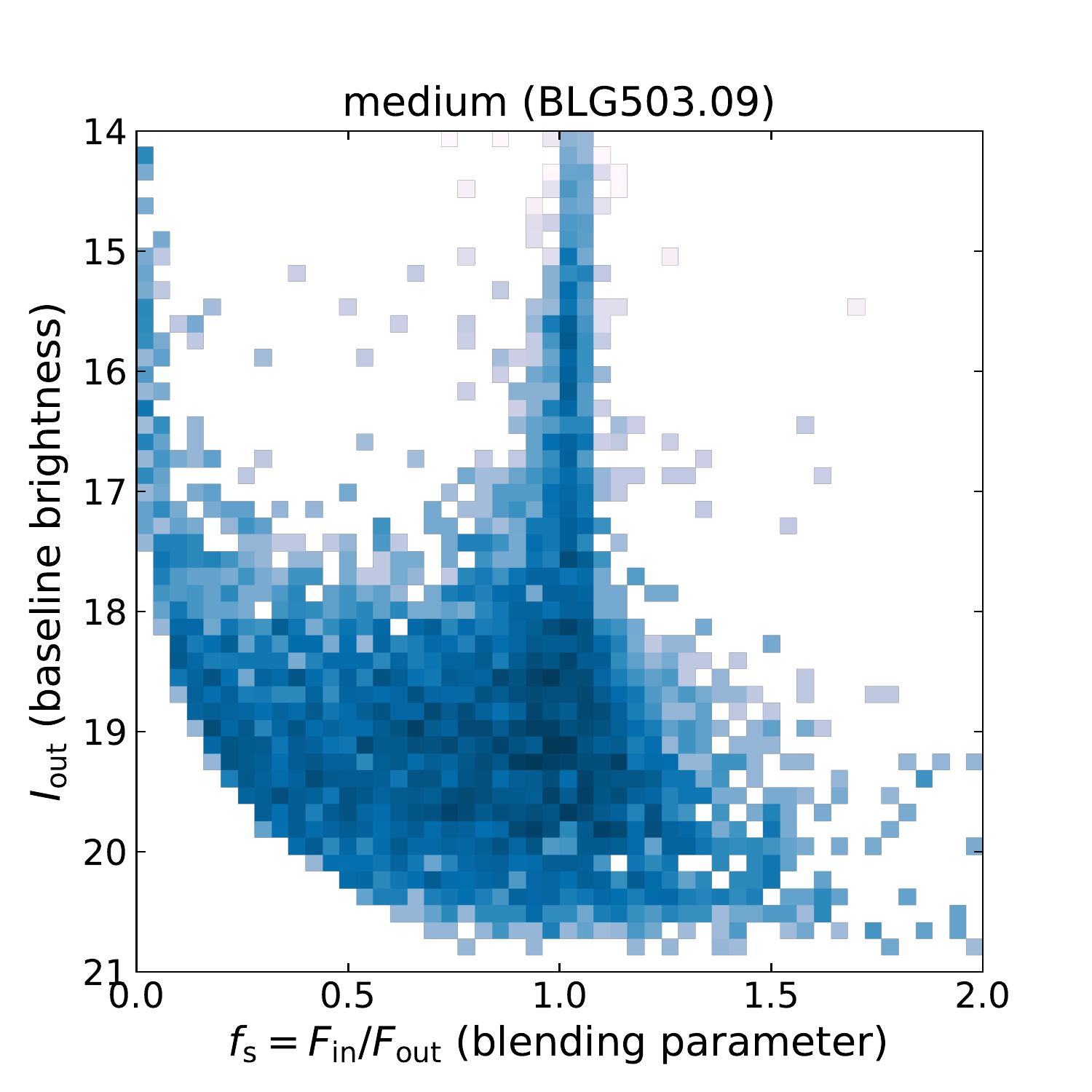}
\includegraphics[height=0.4\textwidth]{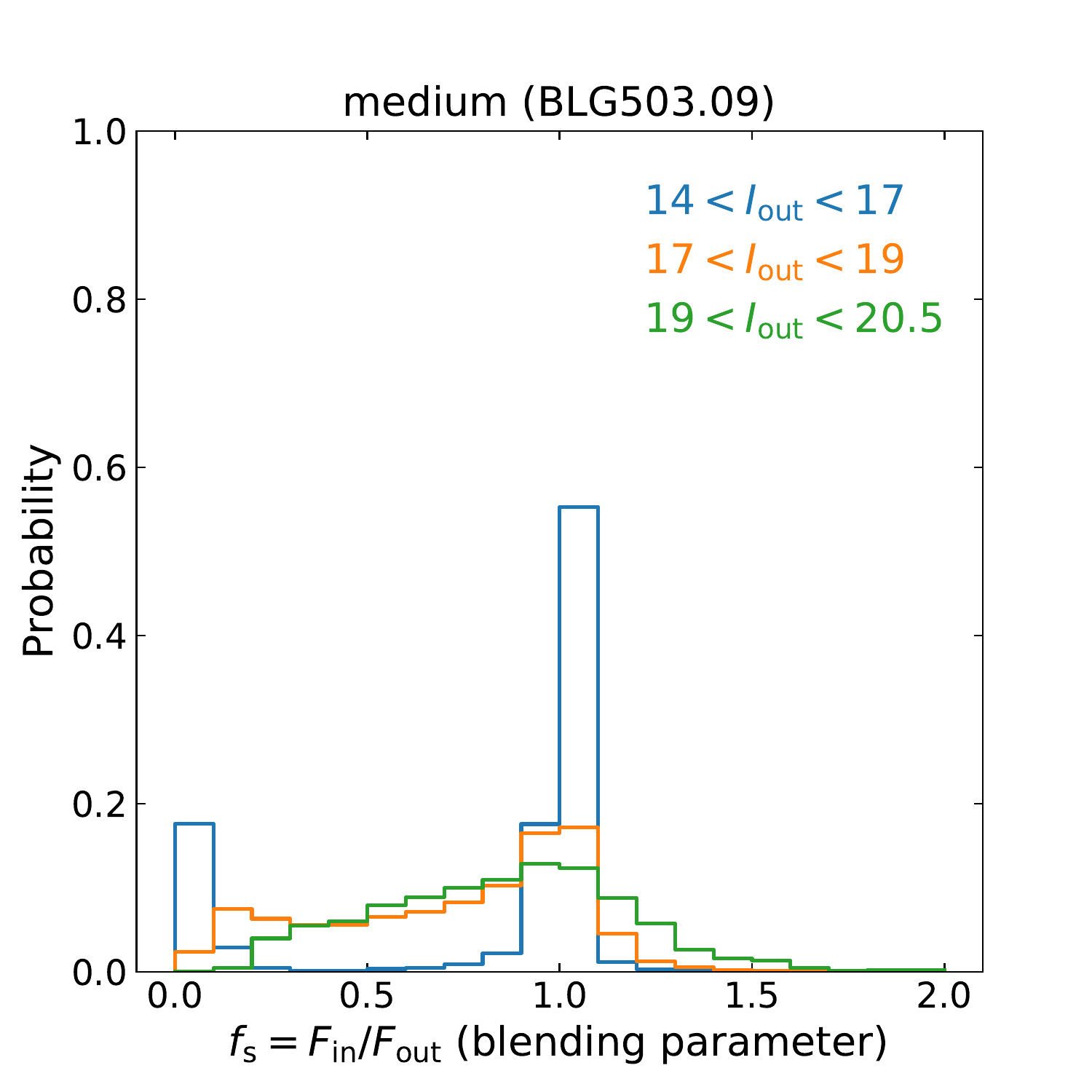}\\
\includegraphics[height=0.4\textwidth]{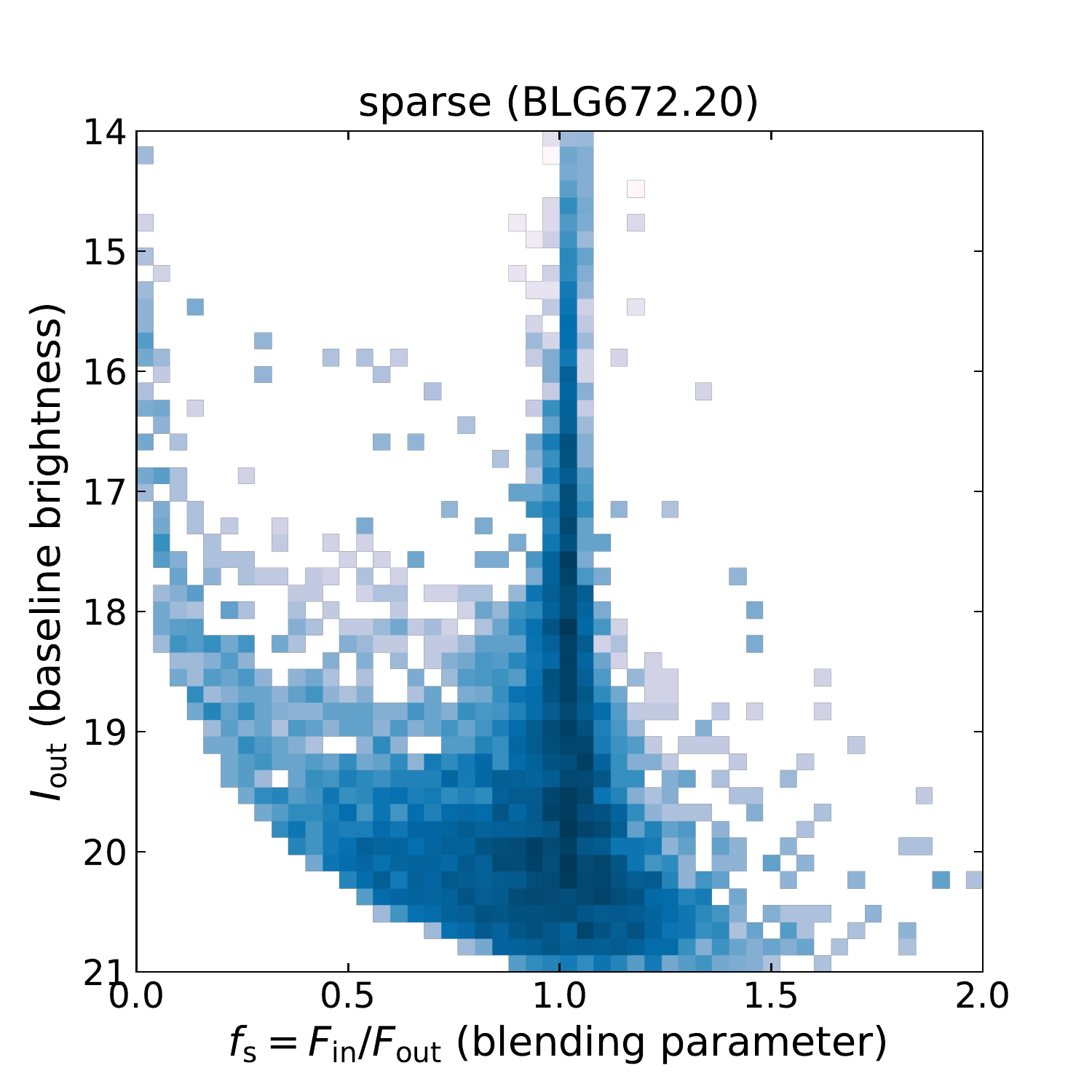}
\includegraphics[height=0.4\textwidth]{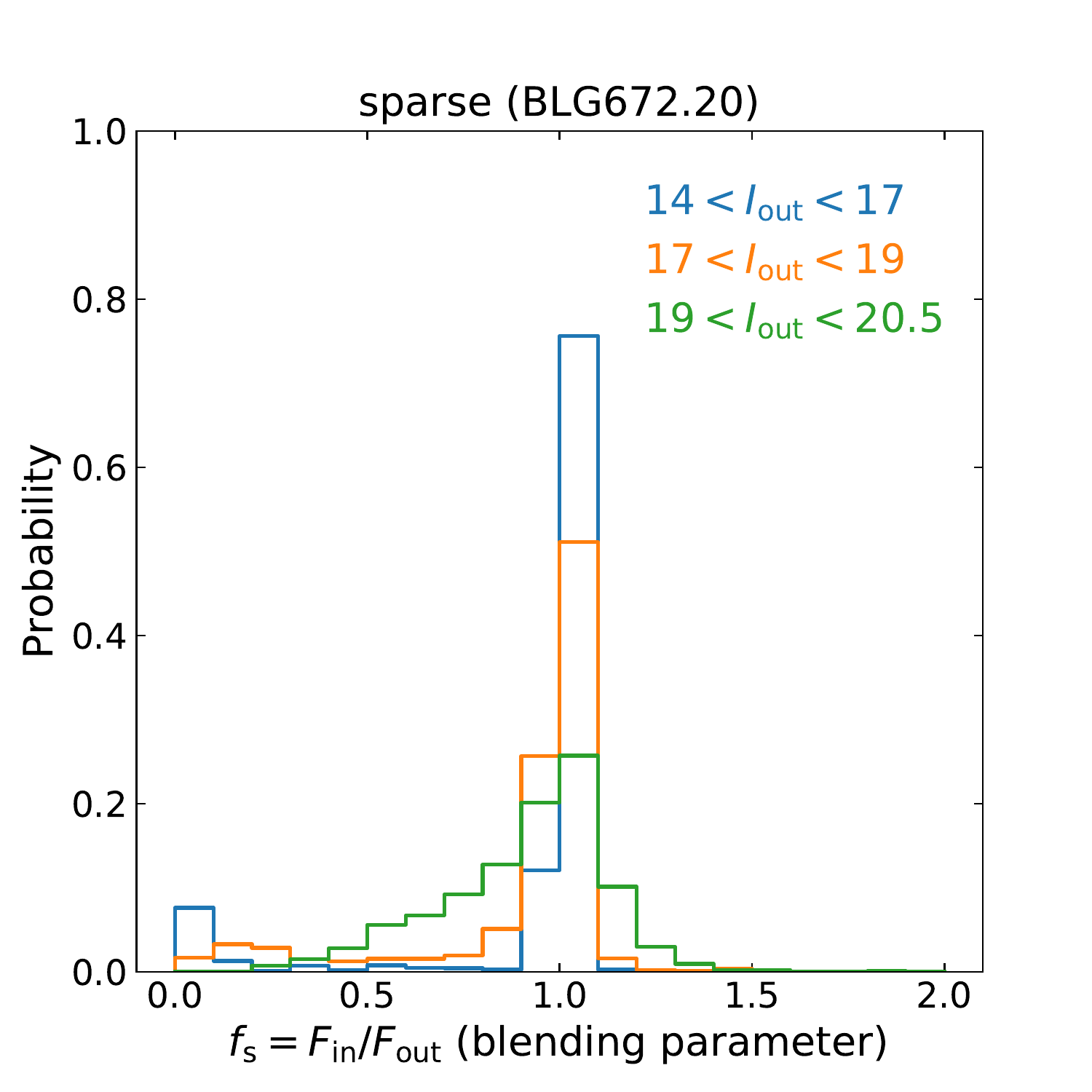}
\caption{Distribution of blending parameter as a function of the baseline brightness for three selected OGLE fields, which represent regions of high, medium, and low star density (see also Figure~\ref{fig:cts_completeness}).}
\label{fig:blending_sims}
\end{figure}

Figure~\ref{fig:blending_comparison} presents the comparison between the distribution of the blending parameter obtained from our image-level simulations and that from the empirical study of \citet{wyrzykowski2015} based on \textit{HST} images. Both distributions are very similar. The distribution of $f_{\rm s}$ of bright stars is bimodal \citep[e.g.,][]{smith2007}; typically, the entire flux comes from the source ($f_{\rm s}\approx 1$), or the source is much fainter than the blend ($f_{\rm s} \approx 0$). For fainter stars, the blending parameter is distributed more uniformly. There are small differences between the results of our simulations and the distributions of \citet{wyrzykowski2015}, which are likely caused by different template images (the OGLE-III reference image was slightly deeper and had better seeing than the OGLE-IV one). Figure~\ref{fig:blending_sims} shows the distributions of $f_{\rm s}$ in three fields with different star densities.

\section{Catalog-level Simulations}
\label{sec:cat_sim}

In the previous work \citep{mroz2017}, image-level simulations provided us with robust measurements of the detection efficiency of microlensing events. These calculations (i.e., injecting microlensing events into real images, performing image-subtraction photometry, and creating photometric databases) require a significant amount of computational resources. In fact, simulations of event detection efficiency in nine high-cadence fields \citep{mroz2017} lasted nearly 4~months on over 800 modern CPUs. As we aimed to measure detection efficiencies in the remaining 112 fields in a finite amount of time, we decided to carry out catalog-level simulations.

We injected microlensing events on top of the light curves of objects from the \mbox{OGLE-IV} photometric databases with the source flux drawn from the derived blending distribution. Each data point and its error bar were rescaled by the expected magnification, which depends on the microlensing model and blending. Our method conserves the variability and noise in the original light curves, as well as information on the quality of individual measurements. Let $\Fs$ be the flux of the source and $\Fb$ the unmagnified flux from possible blended stars and/or the lens itself. The flux of the baseline object ($F_0=\Fs+\Fb$) is magnified during a microlensing event by a factor
\begin{equation}
A'(t) = \frac{\Fs A(t)+\Fb}{\Fs+\Fb}=1+f_{\rm s}(A(t)-1),
\end{equation}
where $A(t)$ is the model magnification and $f_{\rm s}=\Fs/(\Fs+\Fb)$ is the blending parameter. If there is no blending ($f_{\rm s}=1$), then $A'(t)=A(t)$; if the blending is very strong ($f_{\rm s}\rightarrow 0$), the observed magnification $A'(t)\rightarrow 1$.

\begin{figure}[htb]
\centering
\includegraphics[width=\textwidth]{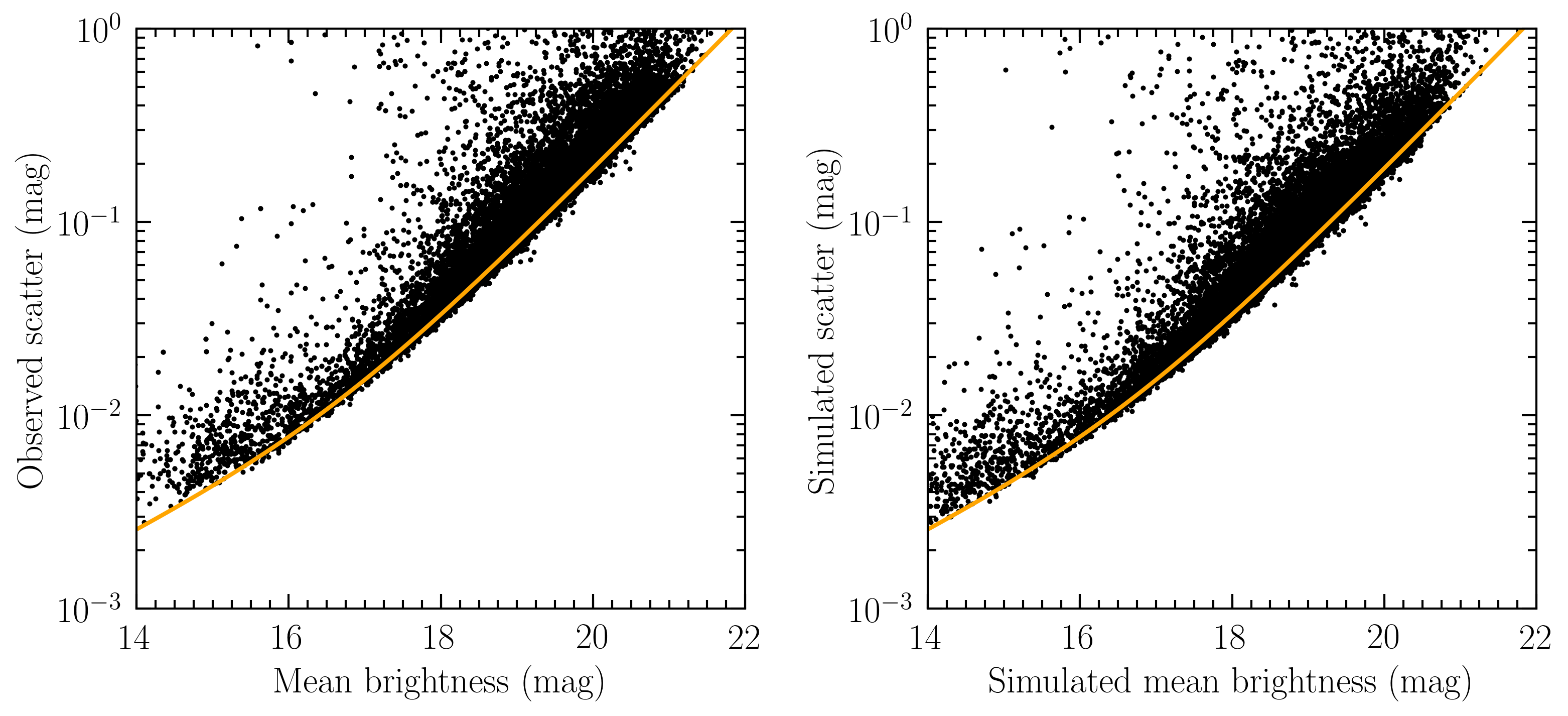}
\caption{The rms light-curve scatter as a function of the mean magnitude of a star in the $I$ band in the OGLE-IV field BLG521.12. Left panel: real data. Right panel: constant stars injected on top of real light curves ($m\rightarrow m+\Delta m$, where $0 \leq \Delta m \leq 0.75$ is drawn from a uniform distribution). The orange line is the best-fit model from \citet{skowron2016}.}
\label{fig:constant_star}
\end{figure}

To inject a microlensing event into the database light curve, we needed to transform the observed flux and its uncertainty $(F_i,\sigma_i)$ to $(F'_i,\sigma'_i)$. The naive transformation $F_i\rightarrow F'_i = F_i A'(t_i)$ is incorrect because it preserves the original photon noise during the magnified part of the event. To illustrate this, let us consider a constant $I=19.5$ star with a typical rms light-curve scatter of 0.1~mag (left panel of Figure~\ref{fig:constant_star}). The transformation $A'(t)=100$ would shift the mean magnitude to $I'=14.5$, but it would preserve the original scatter, whereas the observed scatter of constant stars of that magnitude is much smaller (Figure~\ref{fig:constant_star}).

Assuming that every observed fluctuation in brightness of the initial light curve from its mean brightness ($F_i-F_0$) is due to the observational noise, the following transformation ensures that the photon noise is properly scaled:
\begin{equation}
F_i \rightarrow F'_i = F_0 A'(t_i) + (F_i-F_0)\frac{\sigma_{\rm model}(F'_0)}{\sigma_{\rm model}(F_0)},
\end{equation}
where $F_0$ and $F'_0=F_0 A'(t_i)$ are the mean flux in the baseline and the magnified mean flux, respectively (the corresponding magnitudes are $m_0$ and $m'_0$). The ratio $\sigma_{\rm model}(F'_0)/\sigma_{\rm model}(F_0)$ can be calculated by assuming the photometric noise model of \citet{skowron2016}:
\begin{equation}
\frac{\sigma_{\rm model}(F'_0)}{\sigma_{\rm model}(F_0)}=\frac{1+10^{0.4(m_B-m'_0)}}{1+10^{0.4(m_B-m_0)}}.
\end{equation}
In this simple model, the observed scatter is the sum of the Poisson noise contributions from the object and background (parameterized by $m_B$). To verify the proposed model, we injected ``constant'' stars on top of real light curves from the database (which correspond to the transformation $A'(t)=\mathrm{const}$). The right panel of Figure~\ref{fig:constant_star} shows the rms scatter of simulated light curves, which is consistent with that of real data. Our simple model underpredicts the scatter of the brightest stars ($I\lesssim 15$), likely because the accuracy of their photometry is limited by the accuracy of modeling the PSF and thus the quality of image subtractions, not by the photon noise. This effect does not influence our detection efficiency simulations, because the vast majority of events are fainter than $I=15$.

Similarly, we used the noise model of \citet{skowron2016} to transform the uncertainties:
\begin{equation}
\sigma'_i=\frac{1+10^{0.4(m_B-m')}}{1+10^{0.4(m_B-m)}}\sigma_i.
\end{equation}
When the predicted uncertainties were equal to or below 0.003~mag, we assumed $\sigma'_i=0.003$~mag, as in the photometric databases. 

The proposed transformations preserve information on seeing and sky transparency (included in the reported error bars), as well as the variability in the original data. Catalog-level simulations, however, do not account for the fact that a blended event might have lower signal-to-noise if its centroid is displaced from the centroid of the reference image--identified star. Catalog-level simulations also do not reproduce events detected as ``new'' objects (i.e., undetected on reference images). We analyzed microlensing events detected through the ``new object'' channel by the EWS (about 7-8\% of all events) and found that the majority of them ($\sim 2/3$) occurred on sources fainter than $I=21$. Thus, by neglecting ``new object'' channel events in our catalog-level simulations, detection efficiencies are overestimated by at most $\sim 3\%$, which is consistent with the tests based on the image-level simulations (see Section \ref{sec:im_sim} for details). As this bias is much smaller than other sources of statistical and systematic errors, we did not explicitly correct it.

The catalog-level simulations were carried out with the following steps.
\begin{enumerate}
\item We drew the random parameters of a microlensing event from uniform distributions: $t_0 \sim U(2,455,377,2,458,118)$ (i.e., between 2010 June 29 and 2017 December 31), $u_0 \sim U(0.0,1.0)$, and $\log\tE\sim U(0.0,2.5)$.
\item We drew a random star from the database and calculated its mean magnitude.
\item We drew a random blending parameter $f_{\rm s}$ corresponding to the mean magnitude of the baseline object (Section~\ref{sec:blending}). This parameter describes what fraction of light comes from the source. 
\item We simulated a microlensing event on top of the light curve of the selected object using the procedure that was described above. Subsequently, we checked if the event passes our selection criteria (Table~\ref{tab:crit_2}).
\end{enumerate}
We properly weighted the simulated events so that the simulated distribution of $I_{\rm s}$ was consistent with the LF of the given field. We took into account sources brighter than $I_{\rm s}=21$. We simulated 25,000 events for each CCD detector, which yielded a total of 800,000 events per field. Examples of the detection efficiency curves are shown in the left panel of Figure~\ref{fig:tau_eff_image}.

\section{Image-level Simulations}
\label{sec:im_sim}

We carried out additional image-level simulations to check the accuracy of the catalog-level detection efficiencies for a subset of our low-cadence fields (BLG513, BLG518, BLG521, BLG535, BLG612, and BLG660). The image-level simulations were conducted using the pipeline described by \citet{mroz2017}, which was prepared for high-cadence fields. We injected artificial microlensing events into real images; sources were drawn from the LF of a given subfield and were placed in random locations within the field. We then constructed reference images and calculated image-subtraction photometry for all injected events. Finally, we measured the fraction of events that pass our selection criteria (Table~\ref{tab:crit_2}) as a function of an event timescale. Simulations span the period 2010--2015.

\begin{figure}
\centering
\includegraphics[width=0.49\textwidth]{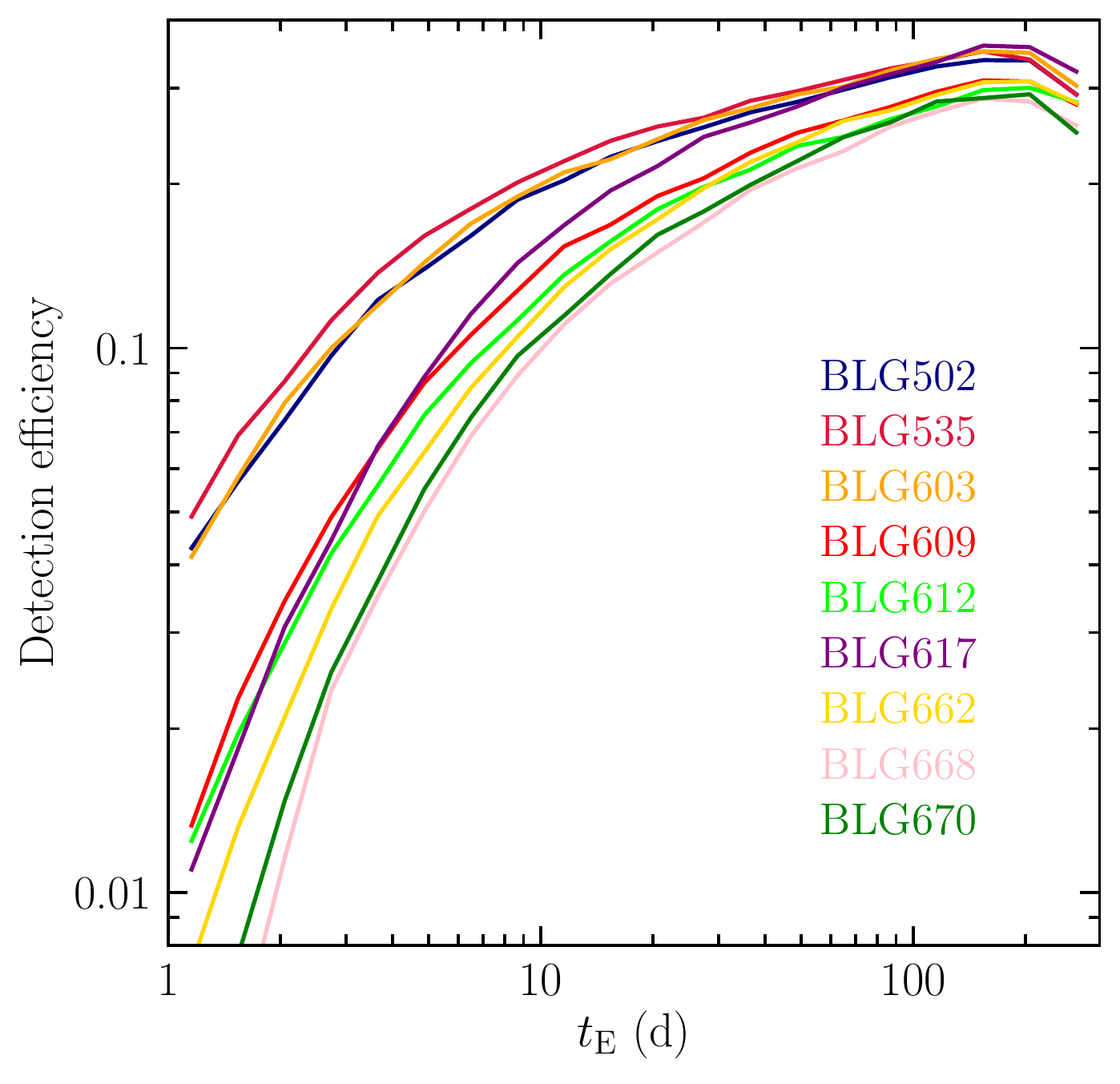}
\includegraphics[width=0.49\textwidth]{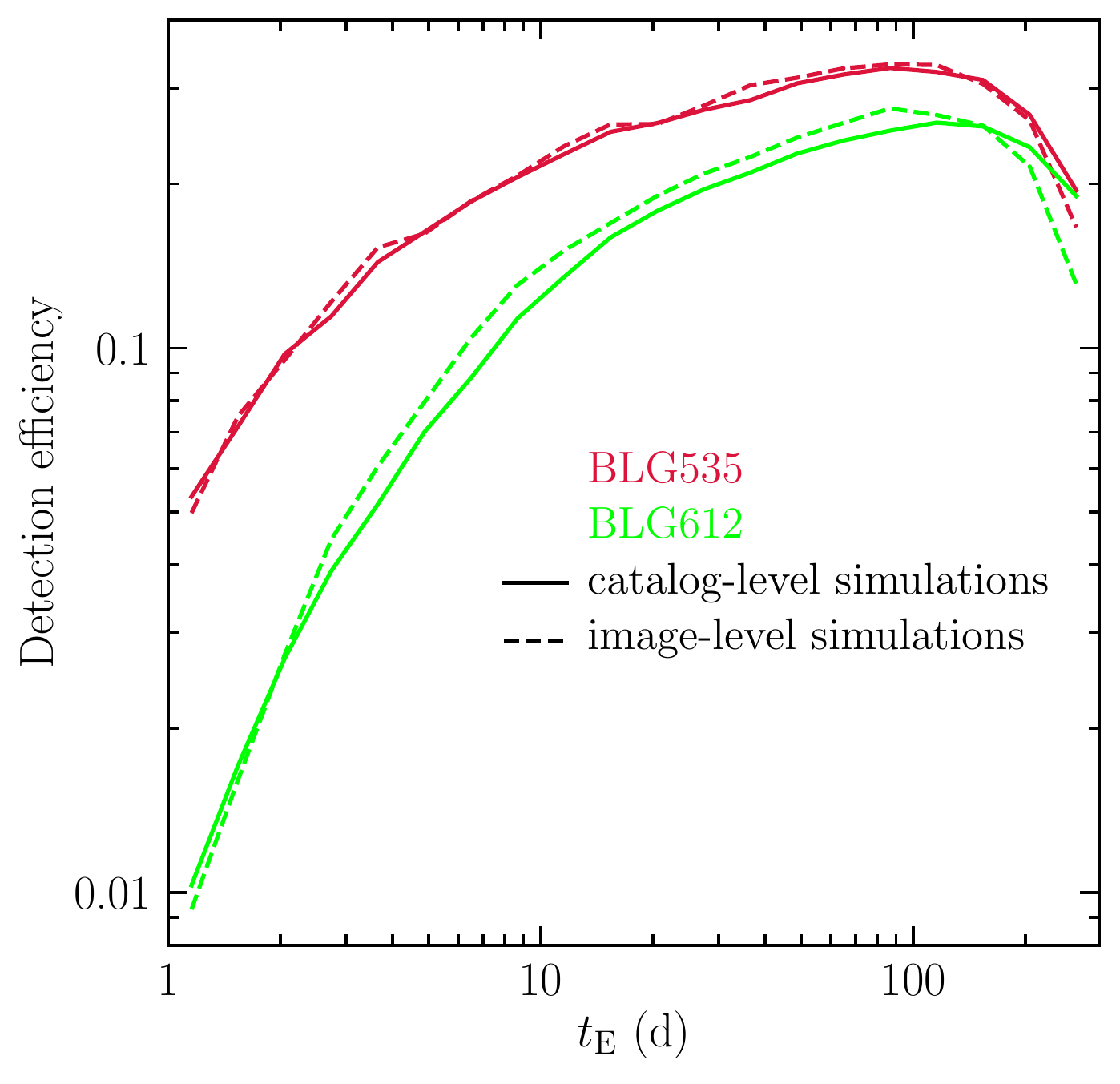}
\caption{Left: catalog-level detection efficiencies for selected OGLE-IV fields for sources brighter than $I=21$ for the period 2010--2017. Right: comparison between detection efficiencies calculated based on image-level and catalog-level simulations. These simulations span the period 2010--2015.}
\label{fig:tau_eff_image}
\end{figure}

The right panel of Figure~\ref{fig:tau_eff_image} presents the measured detection efficiencies in two representative fields, where, for comparison, we also present detection efficiencies calculated using catalog-level simulations for the same range of event parameters. Both curves agree surprisingly well, given that the effort put into image-level simulations is substantially higher. We estimate that microlensing optical depths and event rates measured using image- and catalog-level simulations agree within 3\%, a difference much smaller than that of other sources of statistical and systematic errors.

\section{Results and Discussion}
\label{sec:results}

We used catalog-level simulations to measure the detection efficiencies for 5790 events in low-cadence fields. For nine high-cadence fields (BLG500, BLG501, BLG504, BLG505, BLG506, BLG511, BLG512, BLG534, and BLG611), we used image-level simulations that were previously published by \citet{mroz2017}. The sample of events from high-cadence fields was restricted to sources brighter than $I_{\rm s}=21$ and events longer than $\tE=0.5$~days. Their detection efficiencies were accordingly recalculated to include these revised cuts. The restricted sample is comprised of 2212 events (i.e., 85\% of the original data set). All $5790+2212=8002$ events were thus used for the construction of event rate and optical depth maps.

\subsection{Timescale Distribution}

\begin{figure}
\centering
\includegraphics[width=0.8\textwidth]{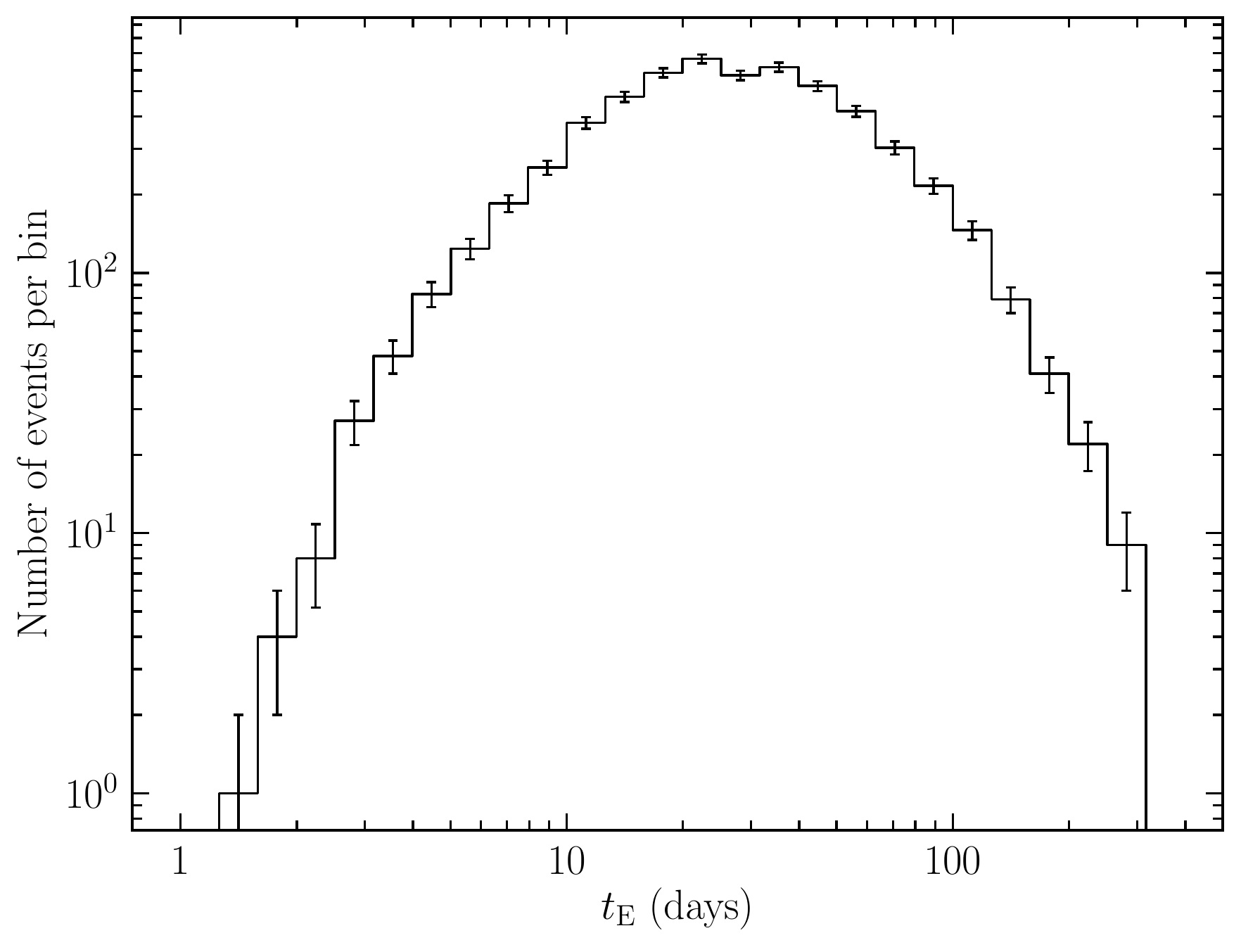}\\
\includegraphics[width=0.8\textwidth]{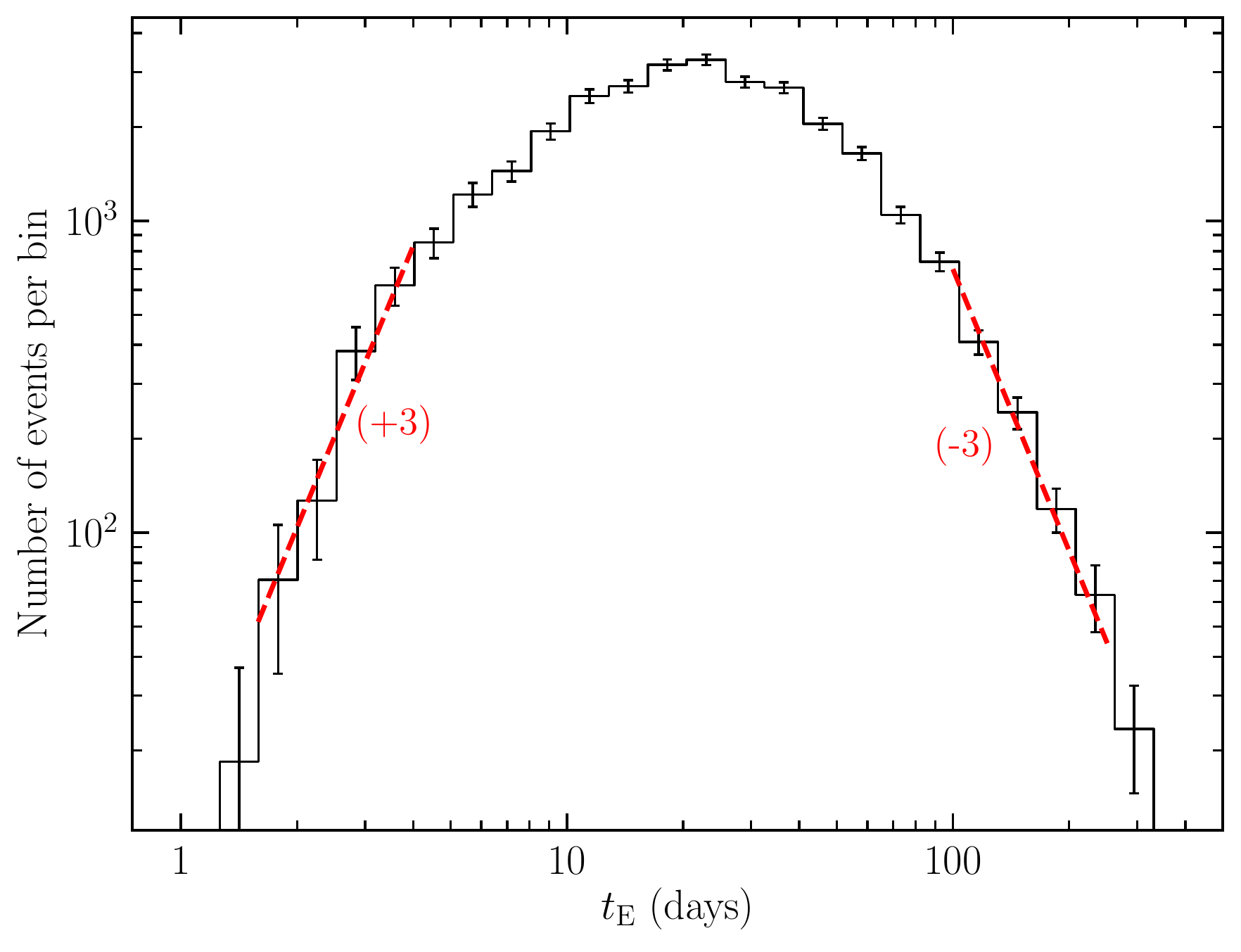}
\caption{Distribution of timescales of 5790 microlensing events detected in low-cadence OGLE-IV fields. Upper panel: observed timescales. Lower panel: timescales corrected for the detection efficiency. The short- and long-timescale distribution tails can be well described by power-law distributions with slopes of $+3$ and $-3$, as expected from theory \citep{mao1996}.}
\label{fig:timescales_distribution}
\end{figure}

The distribution of the timescales of the 5790 microlensing events detected in low-cadence fields is shown in the upper panel of Figure~\ref{fig:timescales_distribution}. The majority of events have timescales between 10~and~40~days, and the number of events falls smoothly at shorter and longer timescales. The short-$\tE$ end of the distribution appears to be steeper than the long-$\tE$ tail, which reflects the fact that the detection efficiency quickly declines as the timescale decreases (Figure~\ref{fig:tau_eff_image}). Indeed, the distribution of event timescales corrected for detection bias (which is constructed by assigning each event a weight $1/\varepsilon(t_{{\rm E},i})$, where $\varepsilon(t_{{\rm E},i})$ is the detection efficiency) is more symmetric (see the lower panel of Figure~\ref{fig:timescales_distribution}). The short- and long-timescale distribution tails can be well described by power-law distributions with slopes of $+3$ and $-3$, as expected from theory \citep{mao1996}. The shape of the timescale distribution does not perfectly match that presented in \citet[][see Figures~1 and~2 therein]{mroz2017} because the current sample contains events from a much larger region. As we will demonstrate, the mean timescales of microlensing events grow with increasing angular distance from the Galactic center. Moreover, the sensitivity to long-timescale events is larger than in the previous work because we were able to use longer light curves (and so we searched for microlensing events using longer 2 yr windows).

The short-$\tE$ end of the timescale distribution reveals no excess of short-duration ($\tE \leq 2$\,d) events. Because some analyzed fields are observed two to three times per night (these fields are marked in green in Figure~\ref{fig:fields}), there is still some sensitivity to short-timescale events, as shown in Figure~\ref{fig:tau_eff_image}. The nondetection of any excess of short-timescale events strengthens our conclusions from \citet{mroz2017} that there is no large population of free-floating or wide-orbit Jupiter-mass planets, in contrast to the \citet{sumi2011} results. 

The long-$\tE$ end of the timescale distribution is more uncertain, even though it can be well described by a power law. A number of long-timescale events exhibit a strong annual parallax effect due to the orbital motion of Earth. Therefore, they do not pass our strict selection criteria on fit quality. However, our detection efficiency simulations did not include the parallax effect (we simulated only point-source point-lens events without any second-order effects); thus the detection efficiencies of long-timescale events are systematically overestimated.

\begin{figure}
\centering
\includegraphics[width=\textwidth]{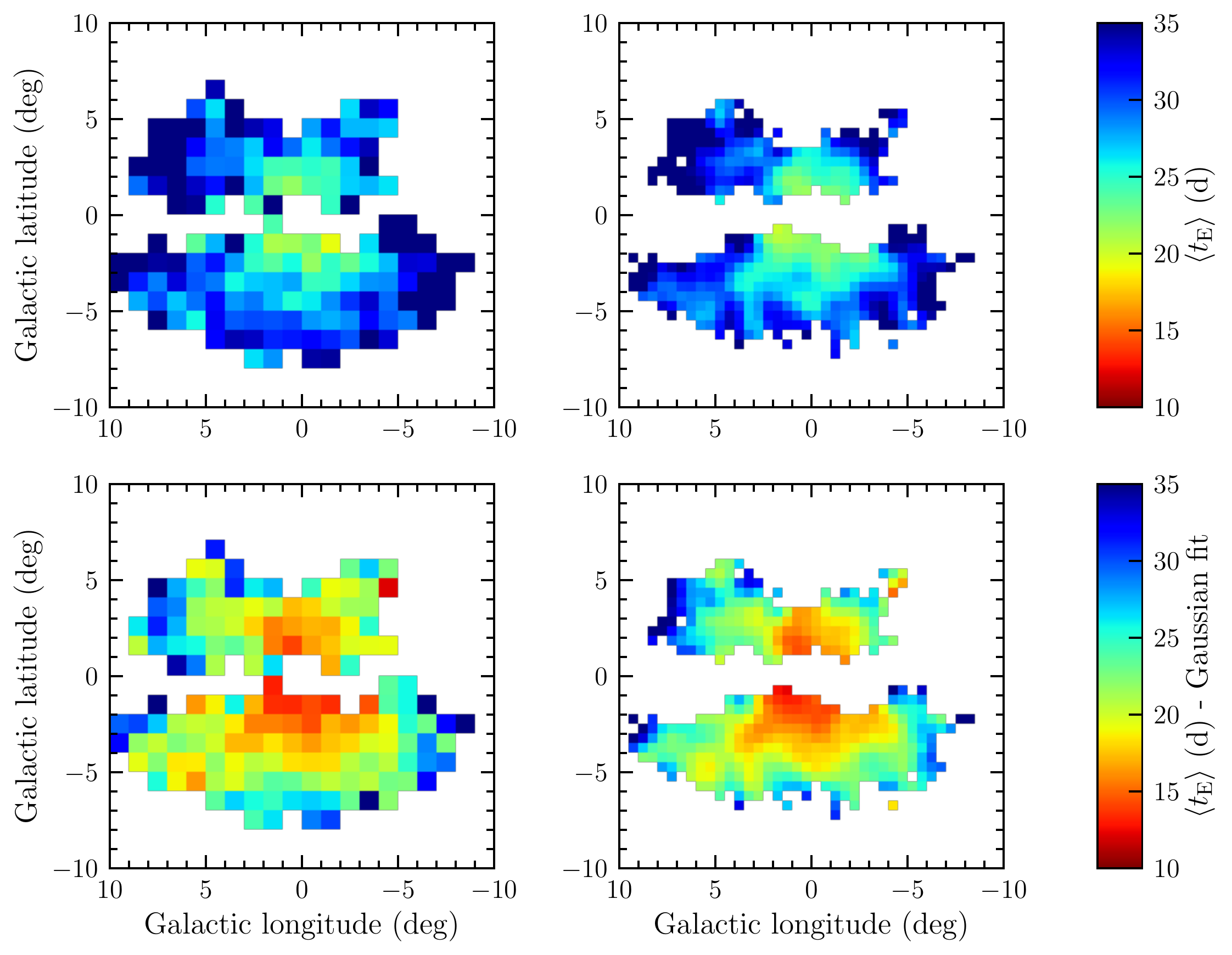}
\caption{Mean timescales of microlensing events in $60'\times 60'$ (left column) or $30' \times 30'$ (right) bins. The upper panels show the detection efficiency--corrected mean timescales, and the lower panels report the mean timescales from the Gaussian fit. All maps were smoothed with a Gaussian with $\sigma=0.5^{\circ}$.}
\label{fig:map_timescales}
\end{figure}

Galactic models predict that the mean timescales of microlensing events should depend on the location, since $\langle\tE\rangle = 2\tau / \pi\Gamma$ and $\tau$ and $\Gamma$ jointly depend on the population of lenses, sources, and their kinematics, which are all expected to change over the large area of the Galactic bulge. We divided the analyzed area into $30' \times 30'$ and $60' \times 60'$ bins and calculated the mean Einstein timescale provided tht at least five events were located in the bin:
\begin{equation}
\langle\tE\rangle = \frac{\sum\limits_{i}\frac{t_{{\rm E},i}}{\varepsilon(t_{{\rm E},i})}}{\sum\limits_{i}\frac{1}{\varepsilon(t_{{\rm E},i})}}.
\end{equation}
Following \citet{wyrzykowski2015}, we also calculated the timescale corresponding to the mean $\langle\log\tE\rangle$, which they call ``the mean timescale based on the Gaussian model'' (i.e., the maximum of the event timescale histograms corrected for the detection efficiency; see Figure~\ref{fig:timescales_distribution}). The latter value is smaller than $\langle\tE\rangle$ and less prone to large statistical fluctuations due to rare very long (or very short) timescale events. Figure~\ref{fig:map_timescales} shows how the mean timescales of the microlensing events vary with the location in the sky. Low-resolution bins ($60' \times 60'$) contain 5--324 events (median 33), and high-resolution bins ($30' \times 30'$) contain 5--106 (median 13.5). All maps were smoothed with a Gaussian with $\sigma=30'$.

The distribution of average timescales agrees well with that found by \citet{wyrzykowski2015} based on a smaller OGLE-III data set. Their sample of standard microlensing events comprised 3718 events from 2001 to 2009, most of which were located primarily in the southern Galactic hemisphere in the region \mbox{$-5^{\circ}<l<5^{\circ}$}, \mbox{$-5^{\circ}<b<-1^{\circ}$}. Similarly, \citet{sumi2013} analyzed the distribution of mean timescales of 474 microlensing events in the MOA-II sample, all of which are located below the Galactic plane. The mean timescale map of \citet{sumi2013} does not show any systematic trends with location, however, because their sample of events is too small.

\begin{figure}
\centering
\includegraphics[height=0.48\textwidth]{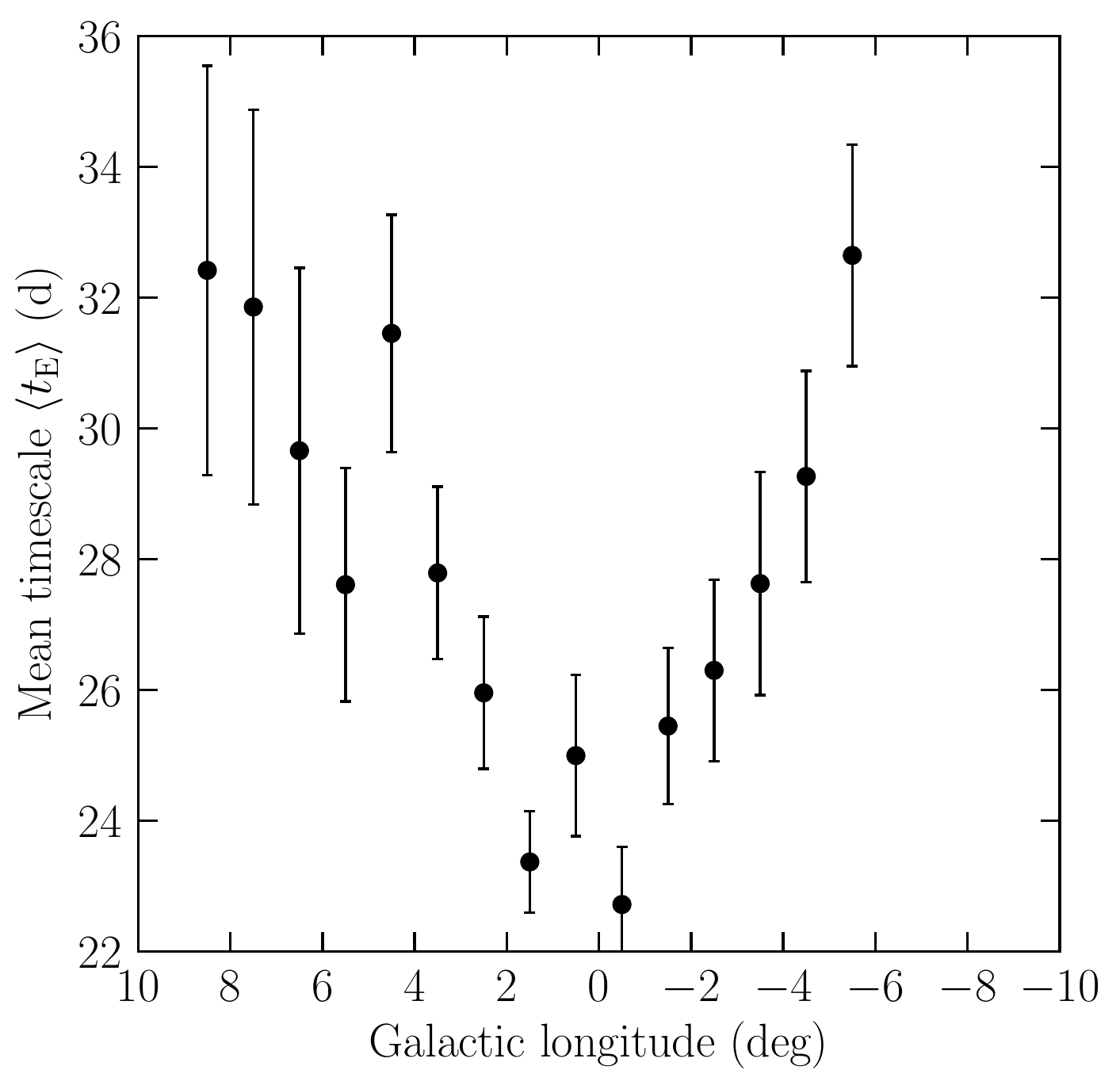}
\includegraphics[height=0.48\textwidth]{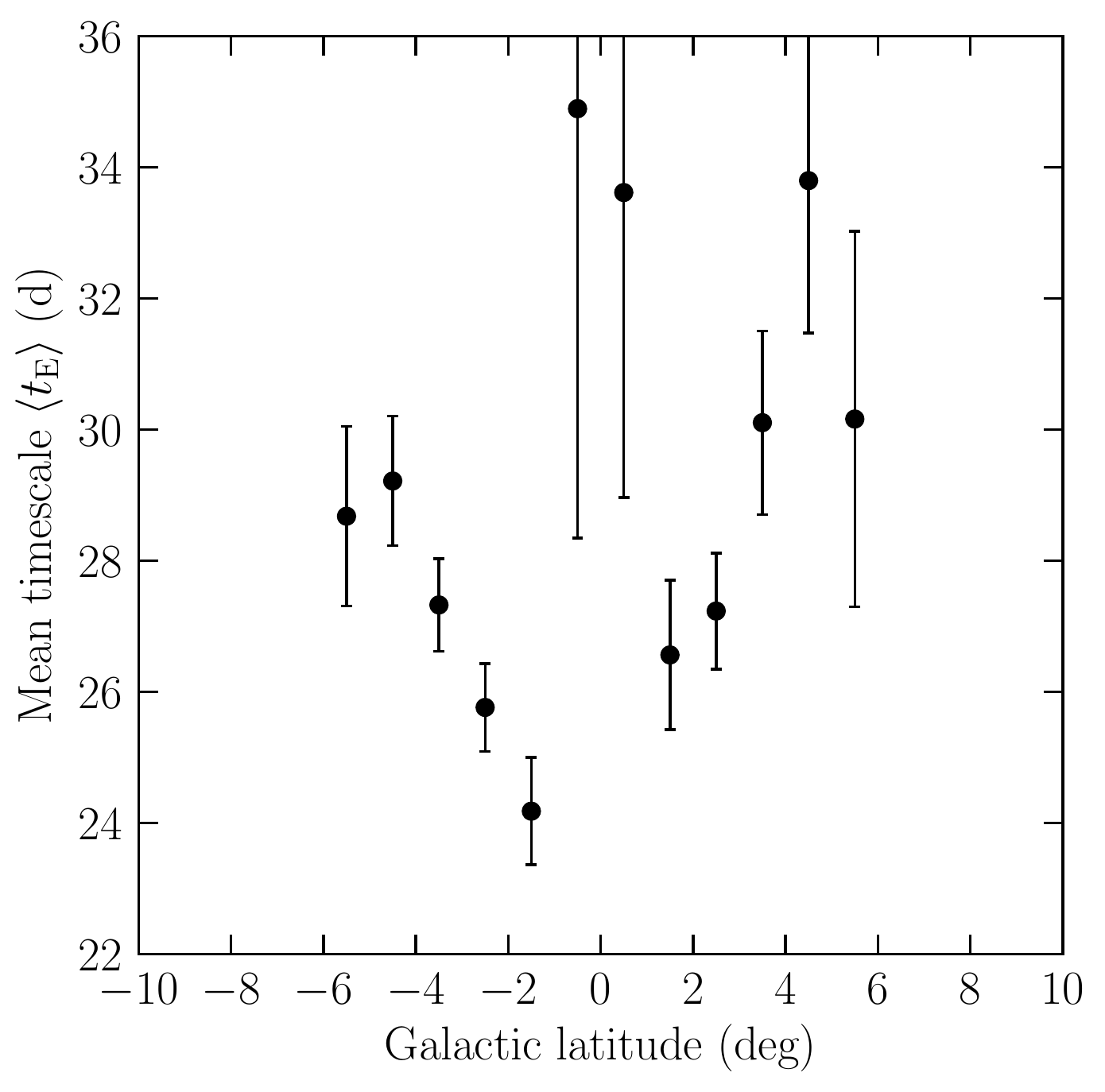} 
\caption{Mean timescales of microlensing events as a function of Galactic longitude (\mbox{$-6^{\circ} \leq b \leq -1^{\circ}$}; left) and Galactic latitude (\mbox{$-6^{\circ} \leq l \leq +10^{\circ}$}; right). Error bars represent the standard error of the mean.}
\label{fig:timescales1d}
\end{figure}

The mean timescales of microlensing events are the shortest in the central bins (located within $\sim3^{\circ}$ of the Galactic center), and they grow with the increasing angular distance from the Galactic center from 22 to 32~days (Figures~\ref{fig:map_timescales} and~\ref{fig:timescales1d}). A similar trend was noticed earlier by \citet{wyrzykowski2015} based on OGLE-III data, although they did not analyze events at $|l|\geq 5^{\circ}$ and $b>0^{\circ}$. One may argue that this is a systematic effect arising from the fact that we were unable to detect the shortest-timescale events in low-cadence fields, but this is not the case. The average event timescales in fields BLG580, BLG518, and BLG522 ($l\approx 5^{\circ}$, $b\approx -3^{\circ}$) are 28.5--30.0~days, and the shortest detected events have $\tE\approx 3$~days. Each of these three fields contains over 100 events, so the low-number statistics cannot also be blamed.

The average timescales increase with increasing Galactic longitude (left panel of Figure~\ref{fig:timescales1d}) with $\langle\tE\rangle\approx 32$~days at $l \approx 8^{\circ}$ and $l \approx -6^{\circ}$, near the edge of the analyzed fields. Currently, OGLE is observing a larger area around the Galactic bulge as part of the OGLE Galaxy Variability Survey \citep{udalski2015}. These observations will tell us whether the average event timescales outside the analyzed area reach a plateau or increase. The distribution of timescales is asymmetric in Galactic longitude (Figure~\ref{fig:map_timescales}) -- events located at positive longitudes appear to be, on average, slightly shorter than those at negative longitudes, which is qualitatively consistent with the theoretical mean timescale maps of \citet{wegg2016,wegg2017} and \citet{awiphan2016}. This asymmetry stems from the fact that the Galactic bar is inclined to the line of sight, resulting in typically larger Einstein radii at negative longitudes \citep{awiphan2016}.

The mean event timescales also vary with Galactic latitude (see Figure~\ref{fig:map_timescales} and the right panel of Figure~\ref{fig:timescales1d}), with shorter average values closer to the Galactic plane, which is in agreement with theoretical expectations. Events located very close to the Galactic plane ($|b| \leq 1.5^{\circ}$) are, on average, much longer ($\langle\tE\rangle\approx 34$~days) than those in neighboring fields ($\langle\tE\rangle\approx 25$~days). They are probably caused by lenses and sources located in the foreground disk (not the Galactic bulge, which is invisible due to large extinction). We expect that disk--disk events have, on average, longer timescales because both the lens and source are moving in a similar direction.

The previous MOA-II \citep{sumi_penny2016} and OGLE-III \citep{wyrzykowski2016} average timescale maps covered the area below the Galactic plane. Although the extent of the OGLE-IV fields in the northern Galactic hemisphere is smaller than that in the south, Figure~\ref{fig:map_timescales} suggests that the average timescale distribution may be slightly asymmetric about the Galactic plane. Events located above the Galactic equator appear to be slightly longer than those below it. We compared the timescale distributions of events below and above the Galactic plane using the Kolmogorov--Smirnov test (the test implementation for weighted data is discussed by \citealt{monahan}) and found $p$-values of 0.01 and 0.27 for $0^{\circ}\leq l \leq 5^{\circ}$ and $-5^{\circ}\leq l \leq 0^{\circ}$, respectively. This confirms a small asymmetry for positive longitudes. The difference may be partly caused by the nonuniform interstellar extinction.

\subsection{Microlensing Optical Depth and Event Rate}

The microlensing optical depths and event rates were calculated using Equations~(\ref{eq:tau_obs}) and~(\ref{eq:gamma_obs}) for each OGLE field (see Table~\ref{tab:results} in Appendix~\ref{ch:app2}). The uncertainties of these quantities can be calculated as follows. \citet{han1995_stat} derived the formula for the statistical error in estimating the optical depth, and they demonstrated that it is substantially higher than the naive Poisson estimate. We may derive a similar expression for the statistical error of the event rate. Recall that the event rate can be written as
\begin{equation}
\Gamma = \frac{1}{N_{\rm s}\Delta T}\sum_{j}\frac{n_j}{\varepsilon_j},
\end{equation}
where $N_{\rm s}$ is the number of monitored sources, $\Delta T$ is the duration of the survey, $n_j$ is the number of events in a $j$th timescale bin, and $\varepsilon_j$ is the event detection efficiency in that bin. The summation is performed over all timescale bins. Since the $\varepsilon_j$s are constants and the number of events obeys Poisson statistics ($\sigma_{n_j}=\sqrt{n_j}$), the uncertainty of $\Gamma$ can be evaluated using the standard error propagation,
\begin{equation}
(\sigma_{\Gamma})^2 = \sum_j\left(\frac{\partial\Gamma}{\partial n_j}\sqrt{n_j}\right)^2 = \frac{1}{N^2_{\rm s}\Delta T^2}\sum_{j}\frac{n_j}{\varepsilon^2_j},
\end{equation}
and hence
\begin{equation}
\frac{\sigma_{\Gamma}}{\Gamma}=\frac{\sqrt{\sum\limits_{j}\frac{n_j}{\varepsilon^2_j}}}{\sum\limits_{j}\frac{n_j}{\varepsilon_j}}.
\end{equation}

\noindent For the construction of microlensing maps, we used the source counts estimated from our image-level simulations (Table~\ref{tab:stars}). As discussed in Section~\ref{sec:counts}, these counts may suffer from systematic errors at the 10\% level. We assumed $\Delta T=2011$~days or $\Delta T = 2741$~days when using 2010--2015 or 2010--2017 light curves, respectively.

Our microlensing maps were constructed using the sample of point-source point-lens events with timescales shorter than $\tE = 300$~days, whereas the longest known microlensing events have timescales $\tE \geq 500$\,days \citep{mao2002,poindexter2005}. The contribution of the longest events to the optical depth may be nonnegligible, so we added a subscript ``300'' to the measured values of $\tau$ to make it explicit that such a limit is in place. About 10\% of all events are anomalous (binary lenses, events with parallax); although they were initially detected by our search algorithm, they were rejected by cuts imposing conditions on the point-source point-lens model fit quality. The measured optical depths and event rates may thus be slightly ($\sim 10\%$) underestimated. To account for binary lens events, we rescaled optical depths and event rates (and their uncertainties) by a factor of 1.09 \citep{sumi2013}.

\begin{figure}
\centering
\includegraphics[width=0.8\textwidth]{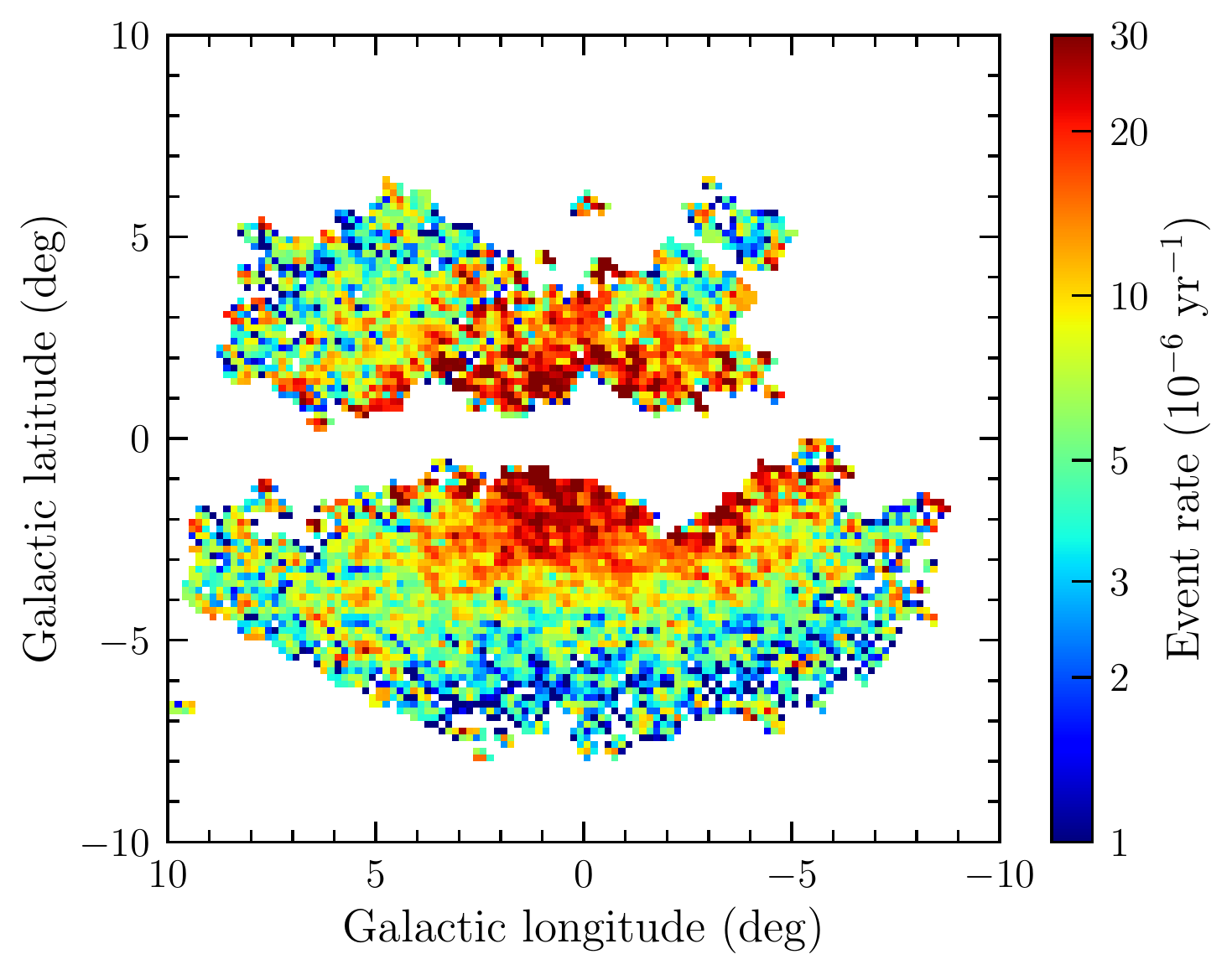}\\
\includegraphics[width=0.8\textwidth]{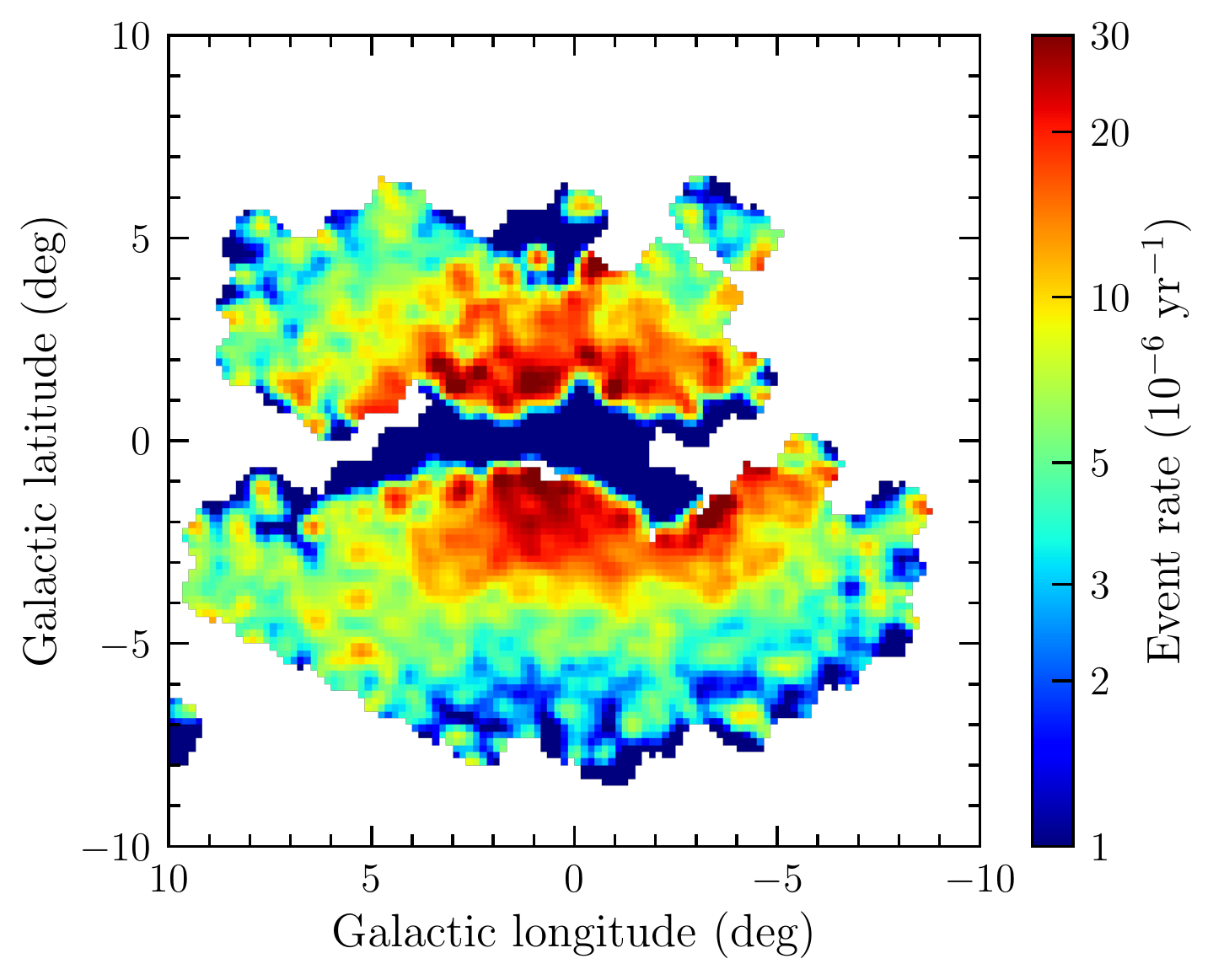}
\caption{Microlensing event rate per star in $10' \times 10'$ bins. The lower map was smoothed with a Gaussian with $\sigma=10'$.}
\label{fig:gamma}
\end{figure}

\begin{figure}
\centering
\includegraphics[width=0.8\textwidth]{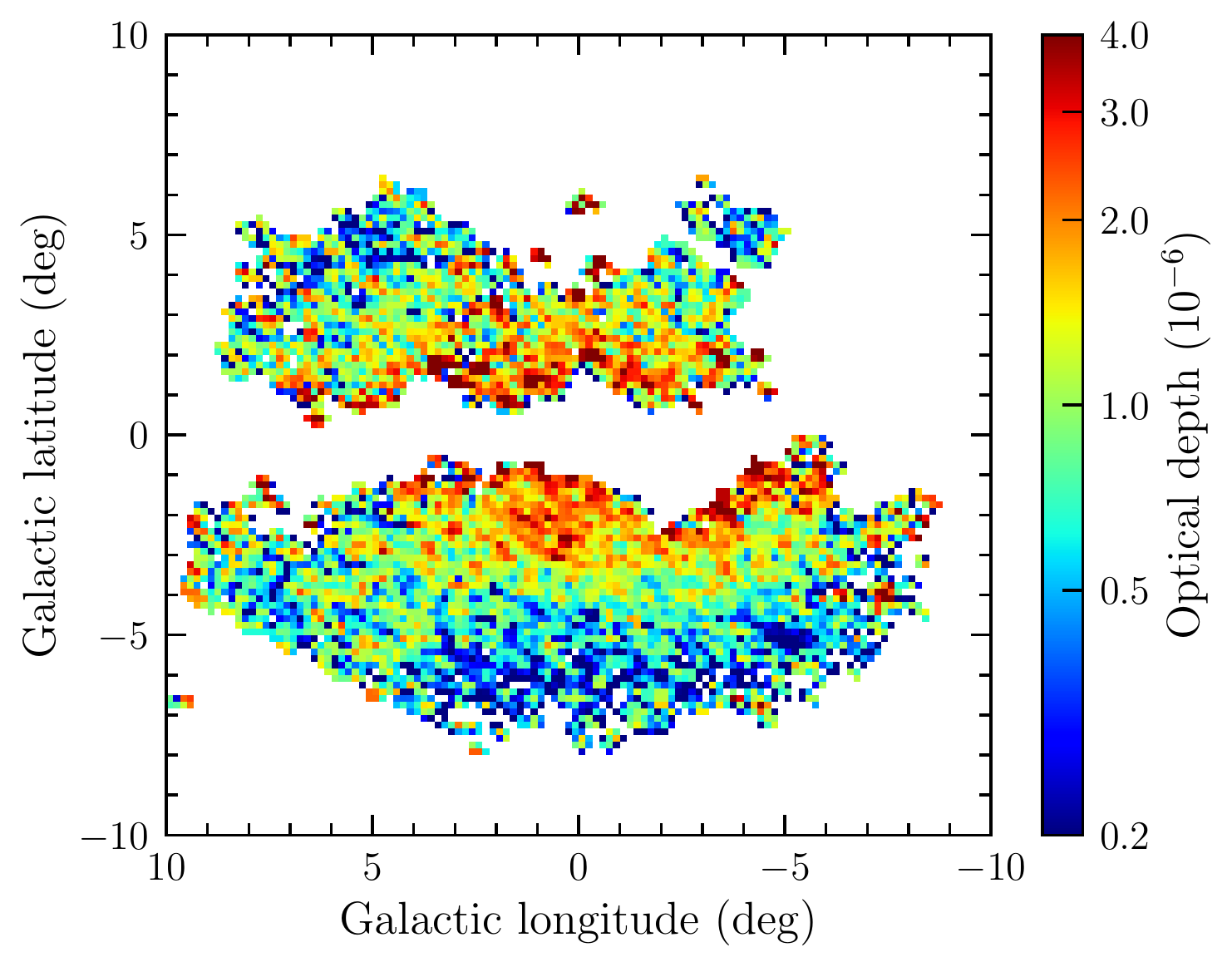}\\
\includegraphics[width=0.8\textwidth]{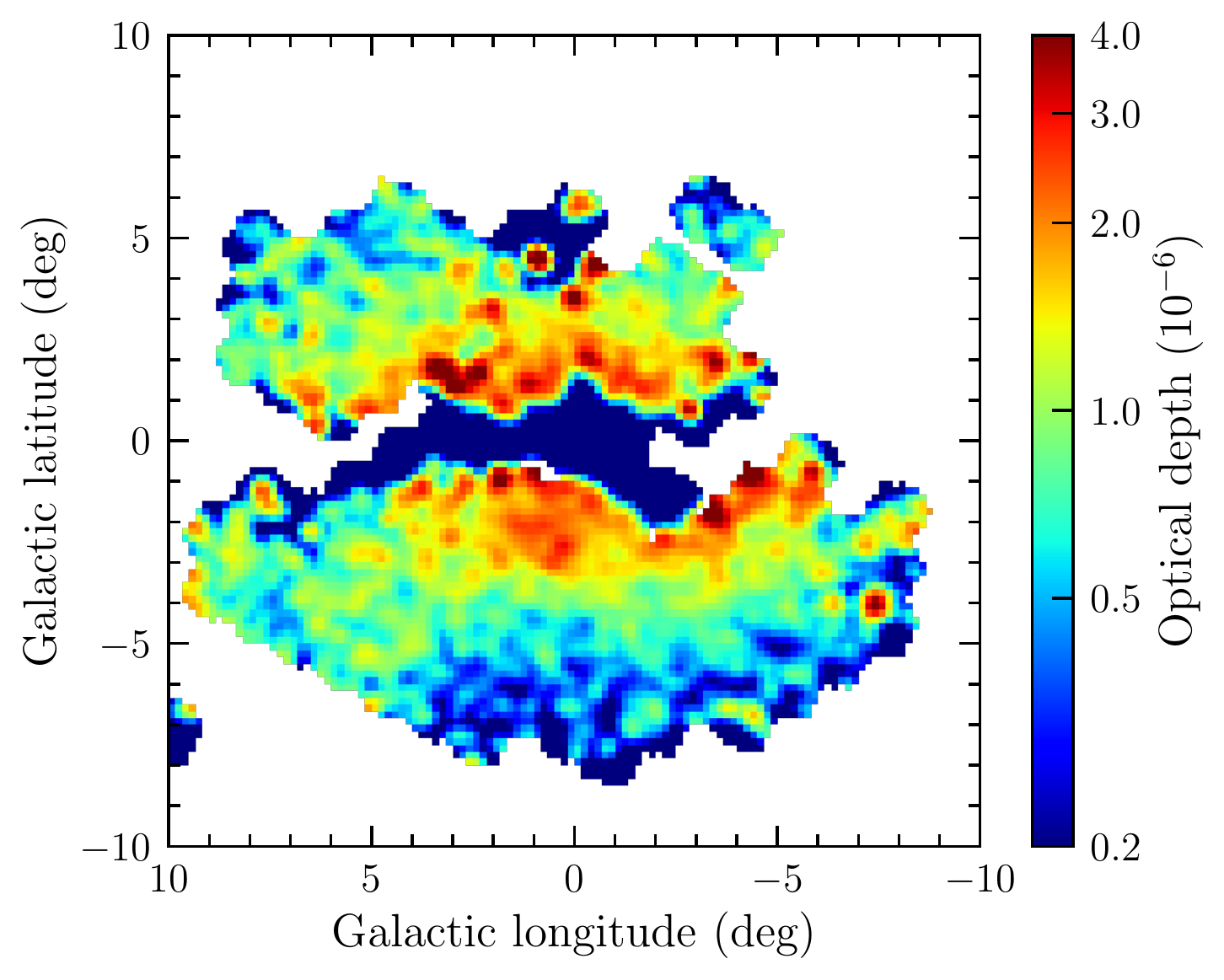}
\caption{Microlensing optical depth per star in $10' \times 10'$ bins. The lower map was smoothed with a Gaussian with $\sigma=10'$.}
\label{fig:tau}
\end{figure}

\begin{figure}
\centering
\includegraphics[width=0.8\textwidth]{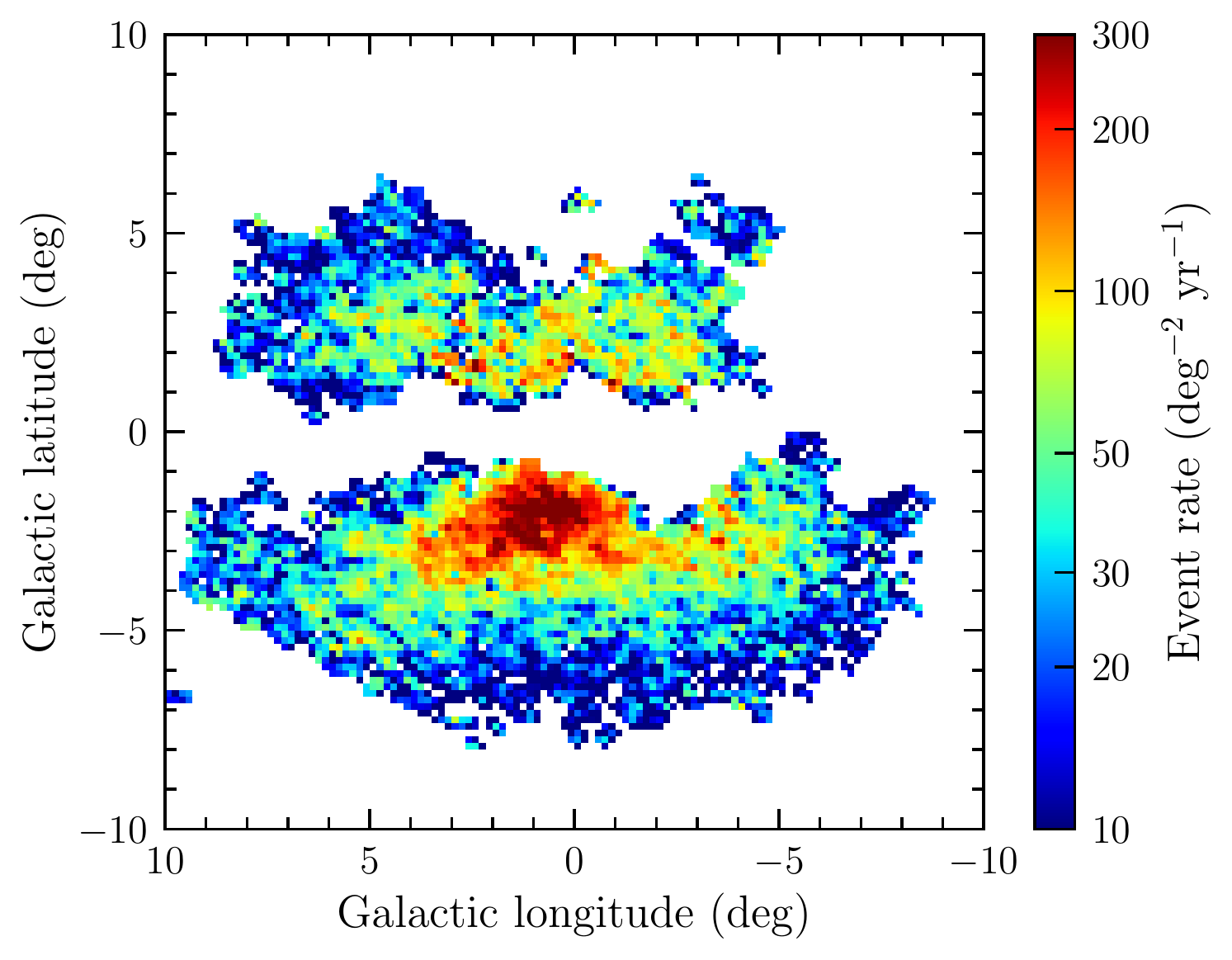}
\caption{Microlensing event rate per unit area in $10' \times 10'$ bins. }
\label{fig:gamma_deg}
\end{figure}

The optical depths and event rates were calculated in individual OGLE fields, and the most robust comparisons with the Galactic models can be performed on a field-to-field basis. For illustration purposes, we constructed high-resolution maps ($10'\times 10'$) showing the distribution of $\Gamma$ and $\tau$ in the sky in the Galactic coordinates (Figures~\ref{fig:gamma} and~\ref{fig:tau}). There are two versions of each map: unbinned and smoothed with a Gaussian with $\sigma=10'$. We also calculated microlensing event rates per unit area $\Gamma_{\mathrm{deg}^2}$ (Figure~\ref{fig:gamma_deg}).

The first important conclusion that can be drawn from these images is that both maps are continuous. Recall that maps were constructed from two independent samples of microlensing events that were selected using different criteria. Moreover, for events from high-cadence fields, we used image-level simulations to assess the detection efficiencies, while for low-cadence fields, catalog-level simulations were used. Figures~\ref{fig:gamma} and~\ref{fig:tau} reveal no discontinuities, which could indicate systematic errors in the analysis. The optical depth map is more granular (especially the Gaussian smoothed version) because optical depths are prone to large statistical fluctuations due to rare very long timescale events. Event rates do not directly depend on timescales, so Figure~\ref{fig:gamma} is smoother.

\begin{figure}
\centering
\includegraphics[width=0.8\textwidth]{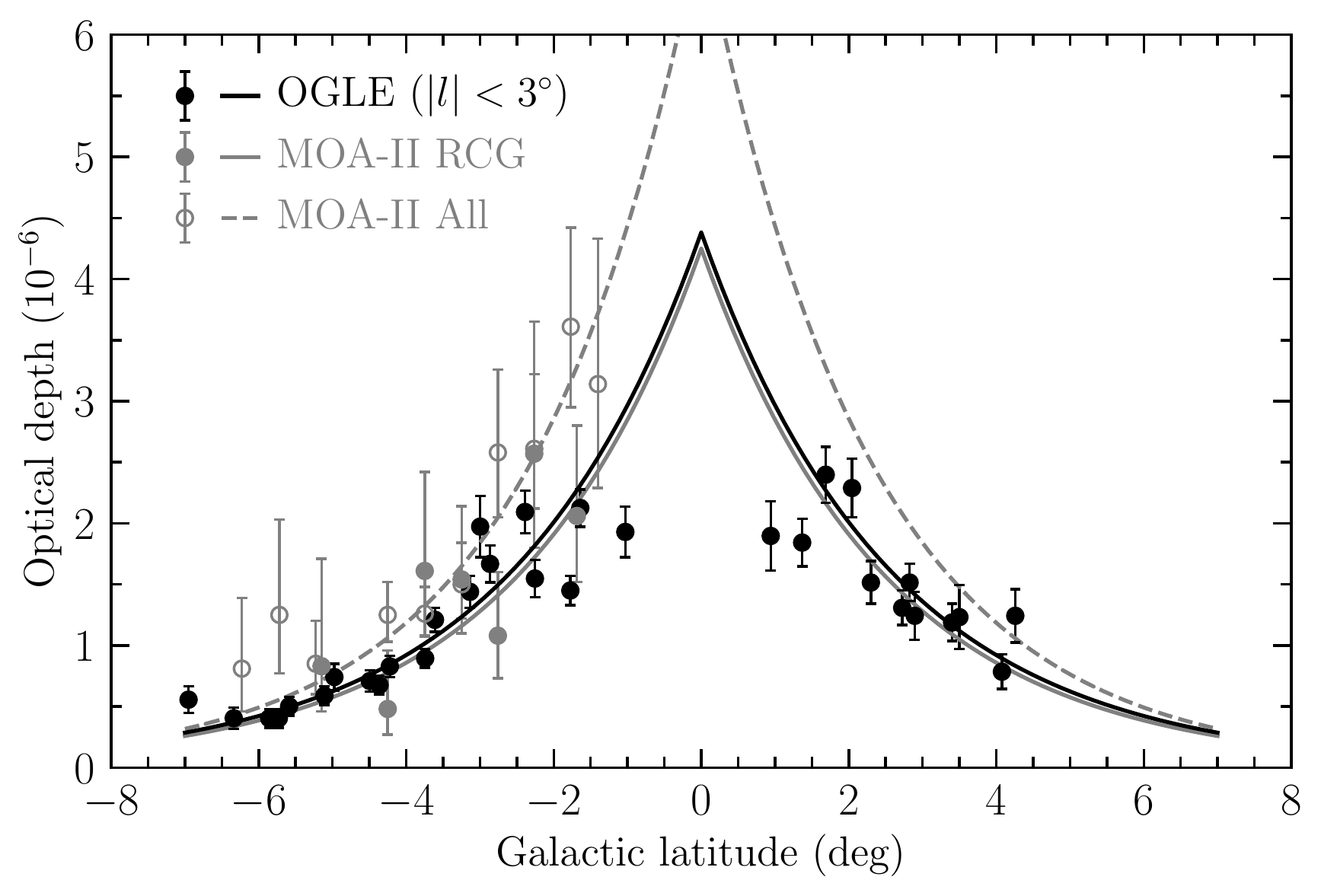}
\includegraphics[width=0.8\textwidth]{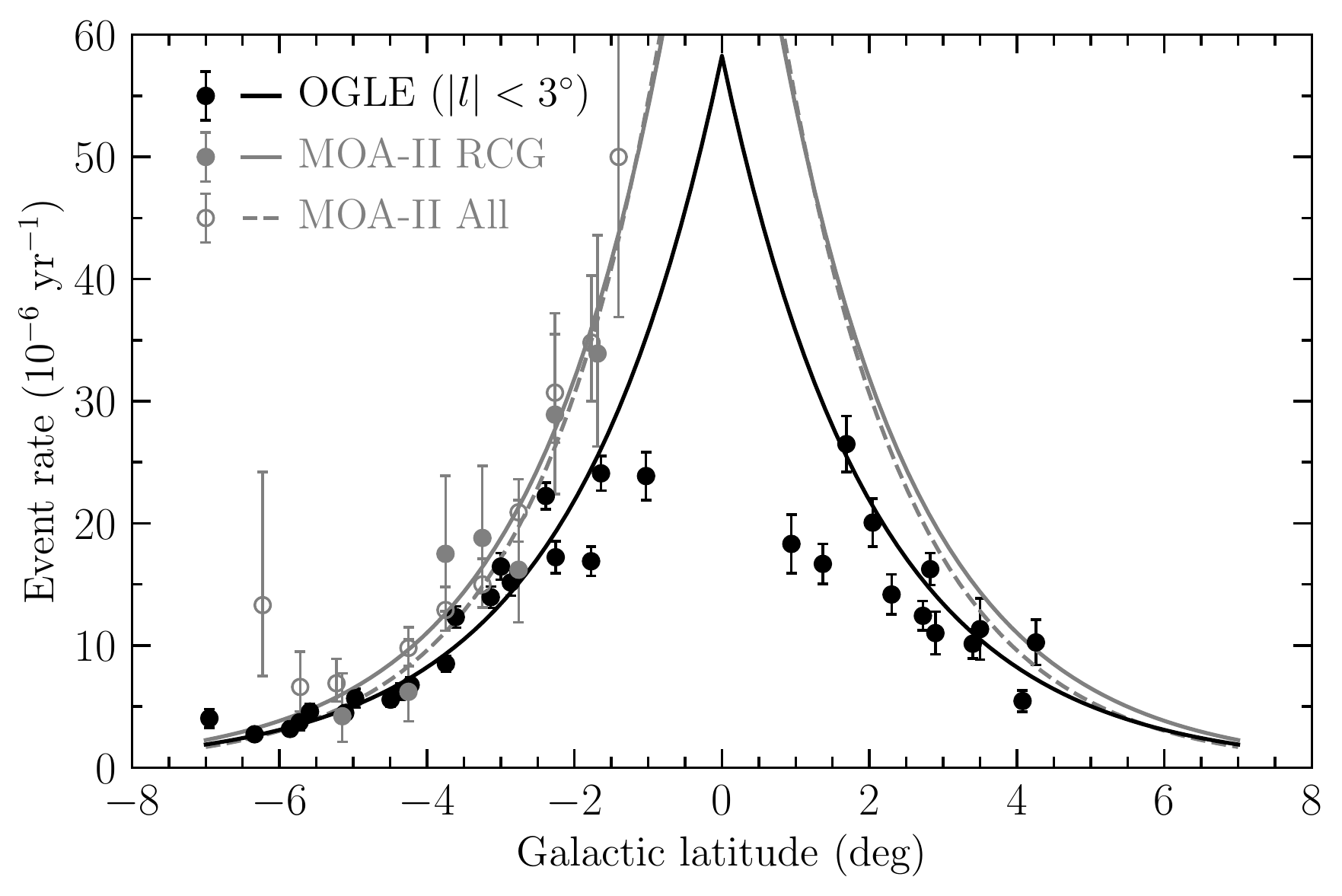} 
\caption{Comparison between microlensing optical depth and event rates measured using OGLE data (black data points) and previous measurements based on MOA-II observations \citep{sumi_penny2016}. Filled gray circles and a gray solid line are MOA measurements based on red clump giant (RCG) stars; open points and a gray dashed line are based on an all-source sample of events. Black lines are the best-fit exponential models to the OGLE data: $\tau_{300}=(1.36\pm 0.04)\times 10^{-6} \exp((0.39\pm0.03)\times(3^{\circ}-|b|))$ and $\Gamma=(13.4\pm 0.3)\times 10^{-6}\ \mathrm{yr}^{-1} \exp((0.49 \pm 0.02)\times(3^{\circ}-|b|))$. }
\label{fig:sumi_comparison}
\end{figure}

Figures~\ref{fig:gamma}, \ref{fig:tau}, and~\ref{fig:sumi_comparison} show that both the optical depth and event rate decrease with increasing angular distance from the Galactic center. We fitted the simple exponential models $\tau_{300} = \tau_0 \exp(c_{\tau}(3^{\circ}-|b|)$ and $\Gamma = \Gamma_0 \exp(c_{\Gamma}(3^{\circ}-|b|)$ for fields located within $|l|<3^{\circ}$ and $|b|>2^{\circ}$, where variations of $\tau$ and $\Gamma$ with Galactic longitude are small. We found $\tau_0=(1.36 \pm 0.04)\times 10^{-6}$ and $c_{\tau}=0.39 \pm 0.03$. The optical depth is symmetric (within the error bars) with respect to the Galactic plane, as illustrated by these separate measurements: $\tau_0=(1.32 \pm 0.06)\times 10^{-6}$, $c_{\tau}=0.35 \pm 0.08$ for $b>0^{\circ}$ and $\tau_0=(1.39 \pm 0.05)\times 10^{-6}$, $c_{\tau}=0.41 \pm 0.03$ for $b<0^{\circ}$. The event rate can also be described using the exponential model with $\Gamma_0=(13.4 \pm 0.3)\times 10^{-6}$\,yr$^{-1}$ and $c_{\Gamma}=0.49 \pm 0.02$. The best-fitting models for the northern ($\Gamma_0=(11.8 \pm 0.5)\times 10^{-6}$\,yr$^{-1}$, $c_{\Gamma}=0.49 \pm 0.07$) and southern ($\Gamma_0=(14.3 \pm 0.4)\times 10^{-6}$\,yr$^{-1}$ and $c_{\Gamma}=0.52 \pm 0.02$) hemispheres are marginally consistent.

We found that optical depths and event rates weakly depend on the limiting magnitude. We chose a sample of 5463 events with sources brighter than $I=20$ and recalculated detection efficiencies. We found that the optical depths calculated using sources brighter than $I=20$ are, on average, equal to those calculated using all events, $\tau_{I \leq 20}/\tau_{I \leq 21}=0.980 \pm 0.017$. Similarly, the event rates are, on average, equal, $\Gamma_{I \leq 20}/\Gamma_{I \leq 21}=0.996 \pm 0.013$. However, both optical depths and event rates in individual fields may vary by up to $\sim 30\%$.

To aid the comparisons with the previous works \citep{sumi2013,sumi_penny2016}, we also calculated optical depths and event rates using a sample of 7970 events with timescales shorter than $\tE=200$\,days, but we found that measured quantities are weakly affected by this choice. We found $\tau_0=(1.36 \pm 0.04)\times 10^{-6}$, $c_{\tau}=0.40 \pm 0.03$, $\Gamma_0=(13.4 \pm 0.3)\times 10^{-6}$\,yr$^{-1}$, and $c_{\Gamma}=0.49 \pm 0.02$, which are virtually identical to the values based on the entire sample. Our sample of events is an order of magnitude larger than that used in the earlier analyses, so statistical fluctuations due to very long timescale events are mitigated. However, some long-timescale events may have exhibited second-order effects (such as annual parallax) and were excluded from our sample. A comprehensive analysis of microlensing events exhibiting annual parallax will be published elsewhere.

\clearpage

The number of observed events sharply decreases at low Galactic latitudes ($|b|\leq 1^{\circ}$) owing to extremely large interstellar extinction. In these regions, we detected only a few microlensing events with nearby sources. Figure~\ref{fig:sumi_comparison} shows that both $\tau$ and $\Gamma$ turn over from the simple exponential models at $|b|\approx 1.5^{\circ}$ because extinction limits the number of observable sources (in the optical band). Source stars of events detected in this region are located closer than those at larger Galactic latitudes, and hence the number of potential lenses (and so the optical depth) is smaller. The situation should be different in the infrared bands. We expect that $\tau$ and $\Gamma$ should follow the rising trend that is observed at larger latitudes. These regions can be probed by infrared surveys, such as VVV \citep{navarro2017,navarro2018} or UKIRT \citep{ukirt2017}.

\begin{figure}
\centering
\includegraphics[height=0.48\textwidth]{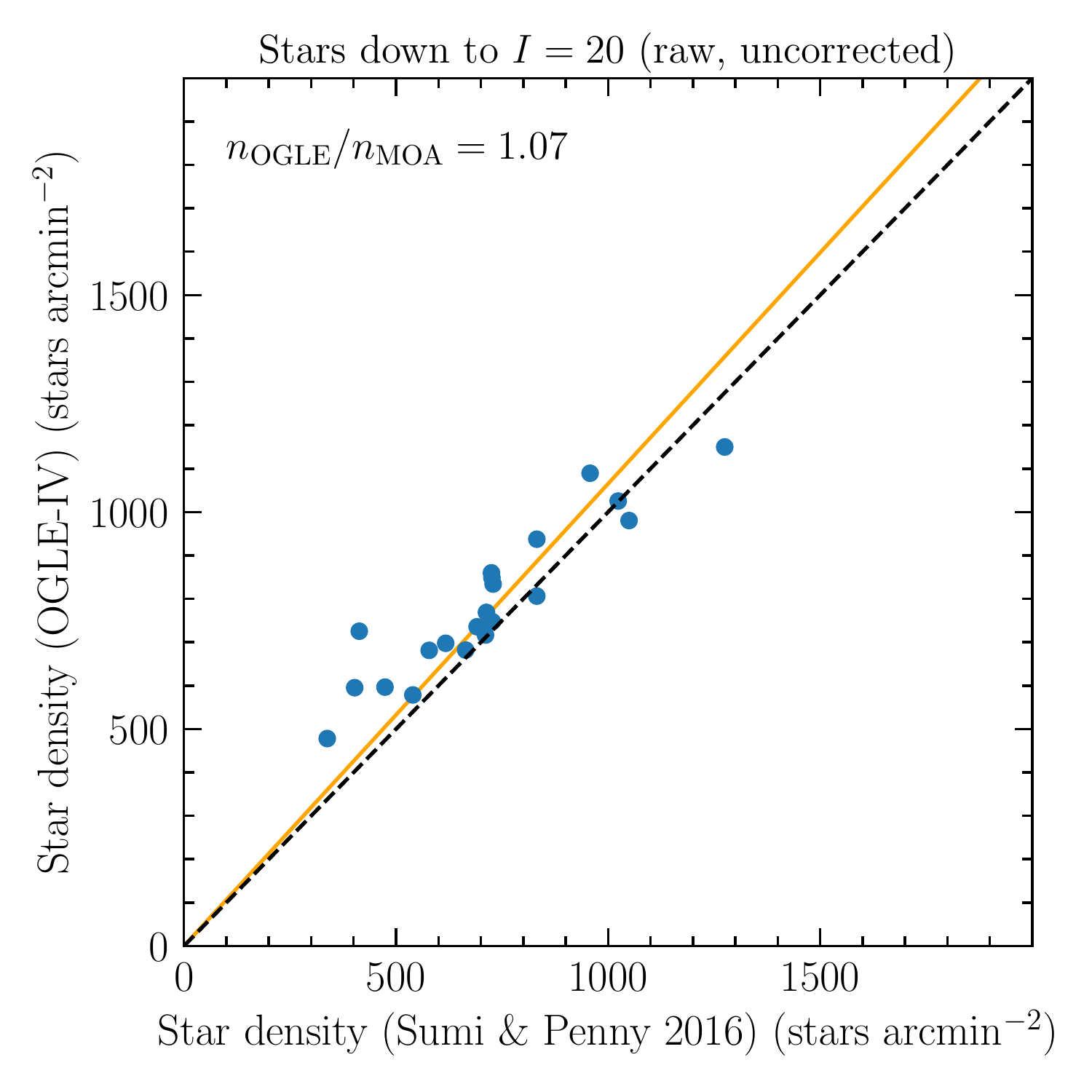}
\includegraphics[height=0.48\textwidth]{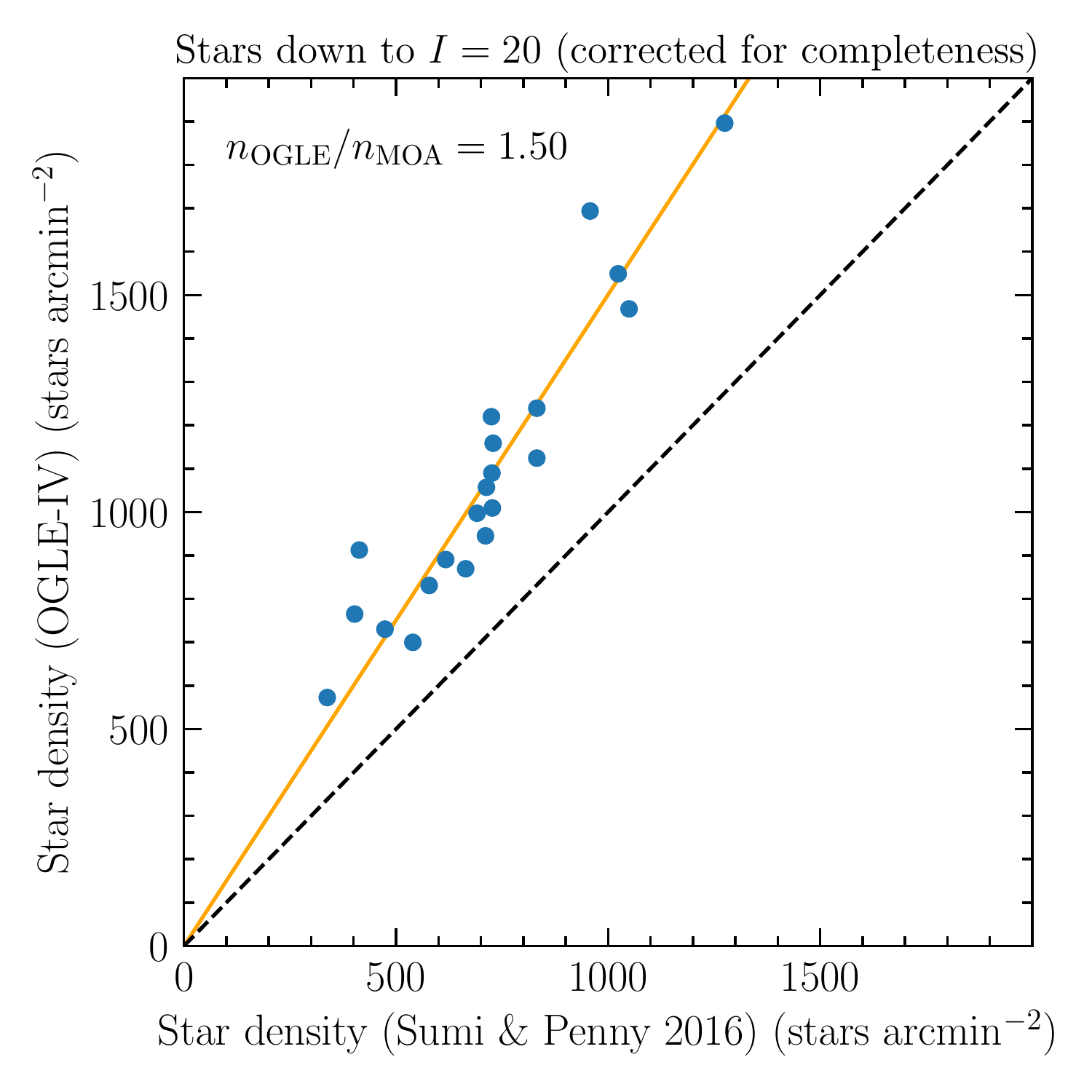} 
\caption{Comparison between the source star surface density in MOA-II fields from \citet{sumi_penny2016} and that calculated using OGLE star catalogs uncorrected (left) or corrected (right) for incompleteness.}
\label{fig:compare_moa_stars}
\end{figure}

Our optical depths and event rates are smaller than previous determinations based on all-star samples of events (Table~\ref{tab:tau1}), but they are consistent (within $1.5\sigma$) with EROS-2 measurements based on bright (red clump) stars \citep{afonso2003,hamadache2006}. Figure~\ref{fig:sumi_comparison} shows the comparison between the measured $\tau$ and $\Gamma$ in the central Galactic bulge fields ($|l|<3^{\circ}$) and the recent measurements of \citet{sumi_penny2016}, which are based on a sample of 474 events from the MOA-II survey. \citet{sumi_penny2016} carried out two types of measurements: one based on red clump giant stars and another using all stars brighter than $I=20$. Their microlensing optical depths based on red clump stars were systematically lower than those based on the all-source sample (see Figure~\ref{fig:sumi_comparison}). Our values are consistent with the MOA-II red clump sample and are a factor of $\sim 1.4$ lower than those based on all MOA-II events. Similarly, our event rates are systematically lower (also by a factor of $\sim 1.4$) than those measured by \citet{sumi_penny2016}. 

We tried to determine the cause of this difference. We suspected the cause was the number of sources used for optical depth and event rate calculations. The MOA-II fields toward the Galactic bulge (with the exception of gb21) overlap with the currently analyzed OGLE-IV fields. First, we used the data reported in Table~1 of \citet{sumi_penny2016} -- number of stars down to $I=20$ and number of subfields used -- to estimate the surface density of stars in their fields. Each MOA-II subfield has an area of 98.1\,arcmin$^{2}$, so the calculated surface density varies from 338 (gb21) to 1275 (gb9) arcmin$^{-2}$. Then, we measured the number of OGLE objects as detected on the reference images that are brighter than $I=20$ and fall into the MOA-II fields. As shown in the left panel of Figure~\ref{fig:compare_moa_stars}, the surface density of OGLE sources is 7\% larger than that used by \citet{sumi_penny2016}, while we still did not correct the OGLE star counts for incompleteness. Thus, the discrepancy between the numbers of sources should be even larger. 

\begin{figure}
\centering
\includegraphics[width=0.49\textwidth]{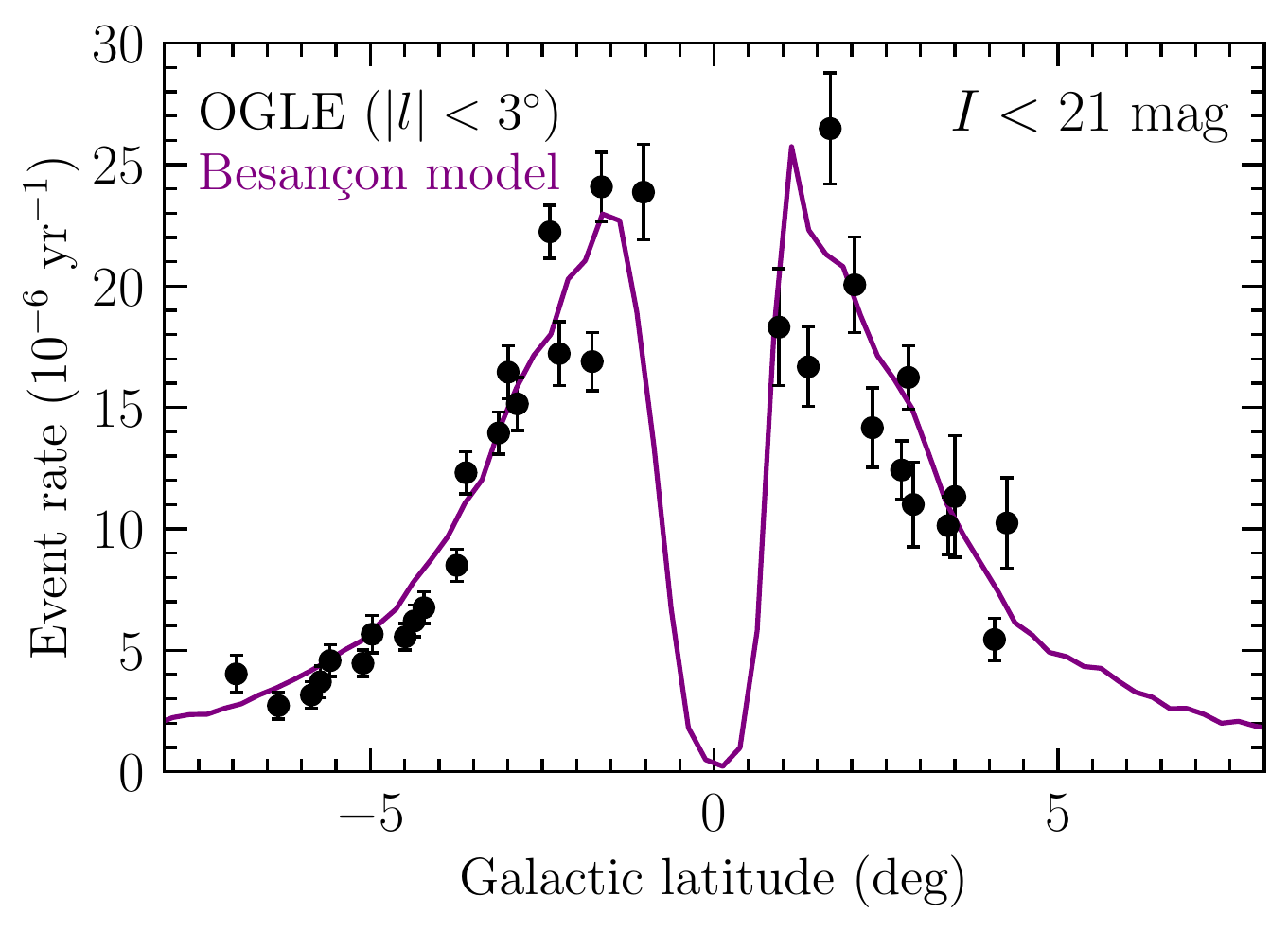}
\includegraphics[width=0.49\textwidth]{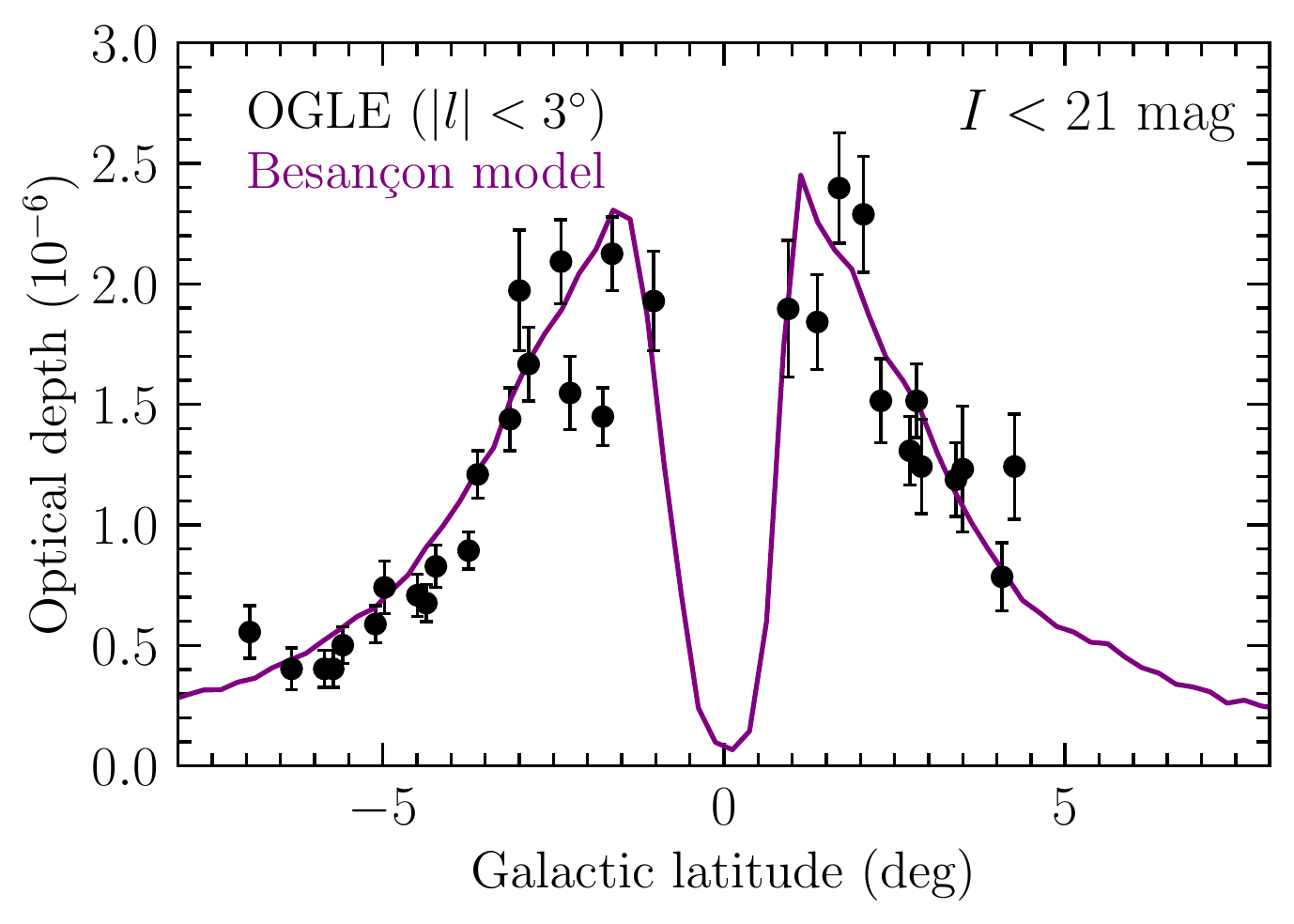} \\
\includegraphics[width=0.49\textwidth]{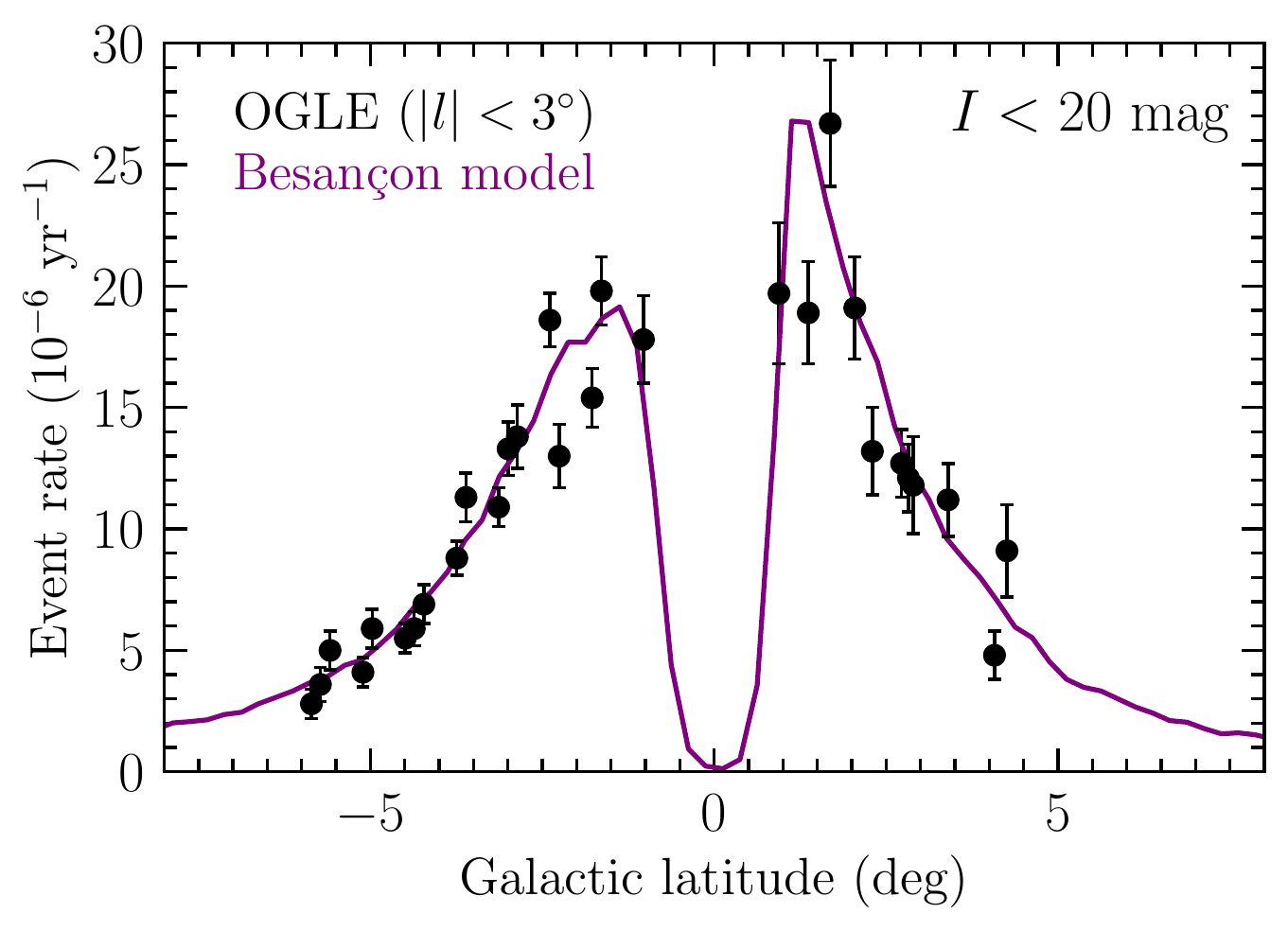}
\includegraphics[width=0.49\textwidth]{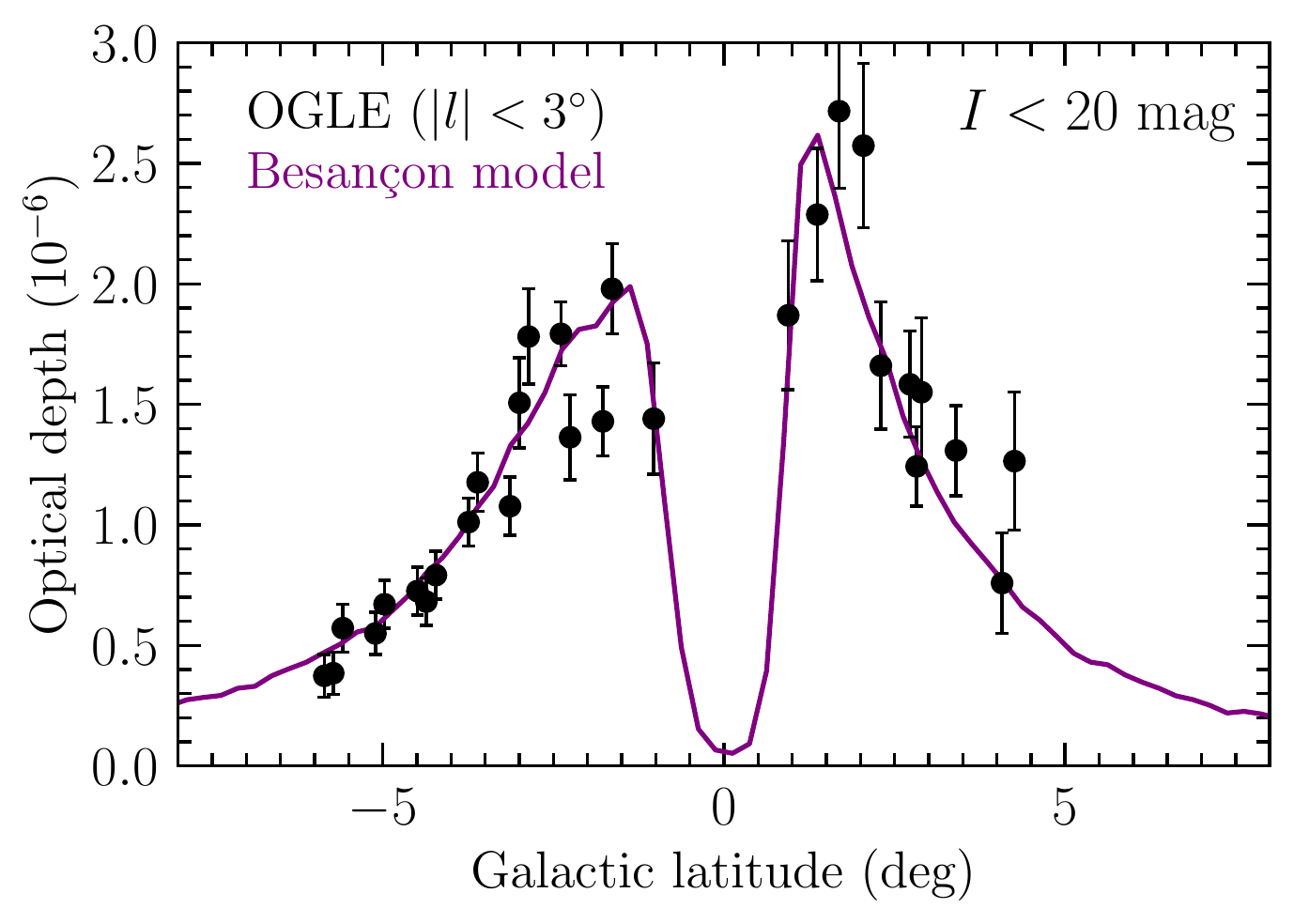} \\
\caption{Comparison between the observed microlensing event rate and optical depth and predictions based on the Manchester--Besan\c{c}on Microlensing Simulator \citep{awiphan2016} (for events shorter than $\tE=300$~days and located in the region $|l|<3^{\circ}$). Upper panels: sources brighter than $I=21$. Lower panels: $I=20$.}
\label{fig:comparison_besancon}
\end{figure}

Indeed, we found that star counts (down to $I=20$) estimated using our image-level simulations were a factor of 1.5 larger than those reported by \citet{sumi_penny2016} (see the right panel of Figure~\ref{fig:compare_moa_stars}), which explains the constant systematic difference between the optical depths and event rates. For example, for the MOA field gb9 (with 1275\,arcmin$^{-2}$ according to \citealt{sumi_penny2016}), we measured the star density of 1896\,arcmin$^{-2}$ using our image-level simulations and 1747\,arcmin$^{-2}$ by matching the LF template. As shown in Table~\ref{tab:hst_density}, star counts calculated using our two independent approaches are consistent within 10\% with the ``ground truth'' based on very deep images taken with \textit{HST}. We are therefore confident that the larger source star counts (and so smaller optical depths and event rates) are correct.

Another line of evidence comes from the comparison of event rates per unit area that do not explicitly depend on the source counts. We restricted our sample to events with sources brighter than $I=20$ and calculated $\Gamma_{\mathrm{deg}^2}$ in the MOA fields. We found that event rates per unit area are, on average, identical, $\Gamma^{\mathrm{OGLE}}_{\mathrm{deg}^2}/\Gamma^{\mathrm{MOA}}_{\mathrm{deg}^2}=0.97 \pm 0.06$, meaning that the difference between our results and those of \citet{sumi_penny2016} is mostly caused by the different number of sources. 

The previous measurements of the microlensing optical depth toward the Galactic bulge (Table~\ref{tab:tau1}) were based on modest samples of events, and their uncertainties were dominated by the Poissonian errors. The accuracy of the optical depths and event rates presented in this paper is mostly limited by the accuracy of the determination of the number of sources. As we demonstrated in Section~\ref{sec:counts}, our star counts may suffer from systematic errors at a 10\% level. However, the event rates per unit area (Figure~\ref{fig:gamma_deg}) do not explicitly depend on the number of sources, so they are free from this possible bias. Other plausible sources of systematic errors include neglecting binary and anomalous events, especially long-timescale events exhibiting a strong annual parallax effect (to account for binary lens events, we rescaled the optical depths and event rates by a constant factor of 1.09). The remaining sources of systematic errors -- such as neglecting ``new object'' channel events in catalog-level simulations (Section~\ref{sec:cat_sim}) and using catalog-level simulations instead of more time-consuming image-level simulations (Section~\ref{sec:im_sim}) -- are negligible ($<3\%$).

Theoretical models of the Galactic bulge were not able to explain the large optical depths calculated using MOA-II observations. For example, the revised optical depths \citep[Figure~14 in][]{sumi_penny2016} were a factor of 1.5 larger than the predictions of the Besan\c{c}on model by \citet{awiphan2016}. The Galactic model of \citet{penny2019} underpredicted the microlensing event rate \citep{sumi_penny2016} by a factor of 2.11. Our new measurements of the microlensing optical depth and event rate based on a large sample of 8002 events from OGLE-IV will allow strict tests of the current models.

As an example, we used the Manchester--Besan\c{c}on Microlensing Simulator\footnote{http://www.mabuls.net/} \citep{awiphan2016} to confront the predictions based on the recent version of the Besan\c{c}on Galactic model \citep{robin2014} with our observations. We simulated events with sources brighter than $I=21$ and timescales shorter than $\tE=300$~days using version 1307 of the Besan\c{c}on model. The model is described in detail by \citet{robin2014} and \citet{awiphan2016}. It consists of a thin disk, thick disk, boxy bulge (bar), and stellar halo. The model also includes a 3D extinction map \citep{marshall2006} based on star counts from the Two Micron All Sky Survey (2MASS).

\begin{figure}
\centering
\includegraphics[width=0.8\textwidth]{gamma_10_smooth.pdf}
\includegraphics[width=0.8\textwidth]{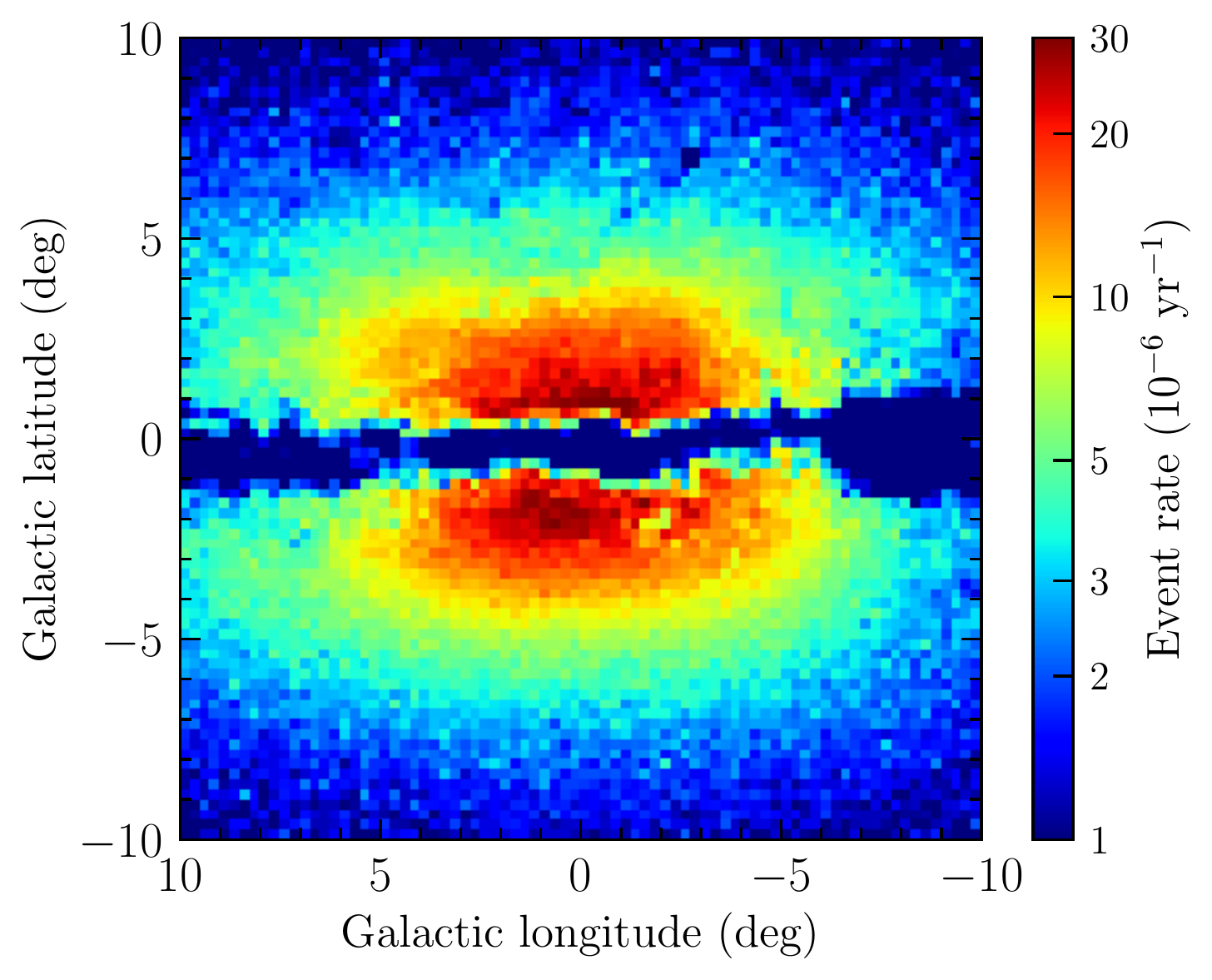}
\caption{Comparison between the observed microlensing event rate (upper panel) and predictions based on the Manchester--Besan\c{c}on Microlensing Simulator \citep{awiphan2016} for sources brighter than $I=21$ and events shorter than $\tE=300$~days (lower panel). Both maps have the same color scale.}
\label{fig:gamma_besa}
\end{figure}

\begin{figure}
\centering
\includegraphics[width=0.8\textwidth]{tau_10_smooth.pdf}
\includegraphics[width=0.8\textwidth]{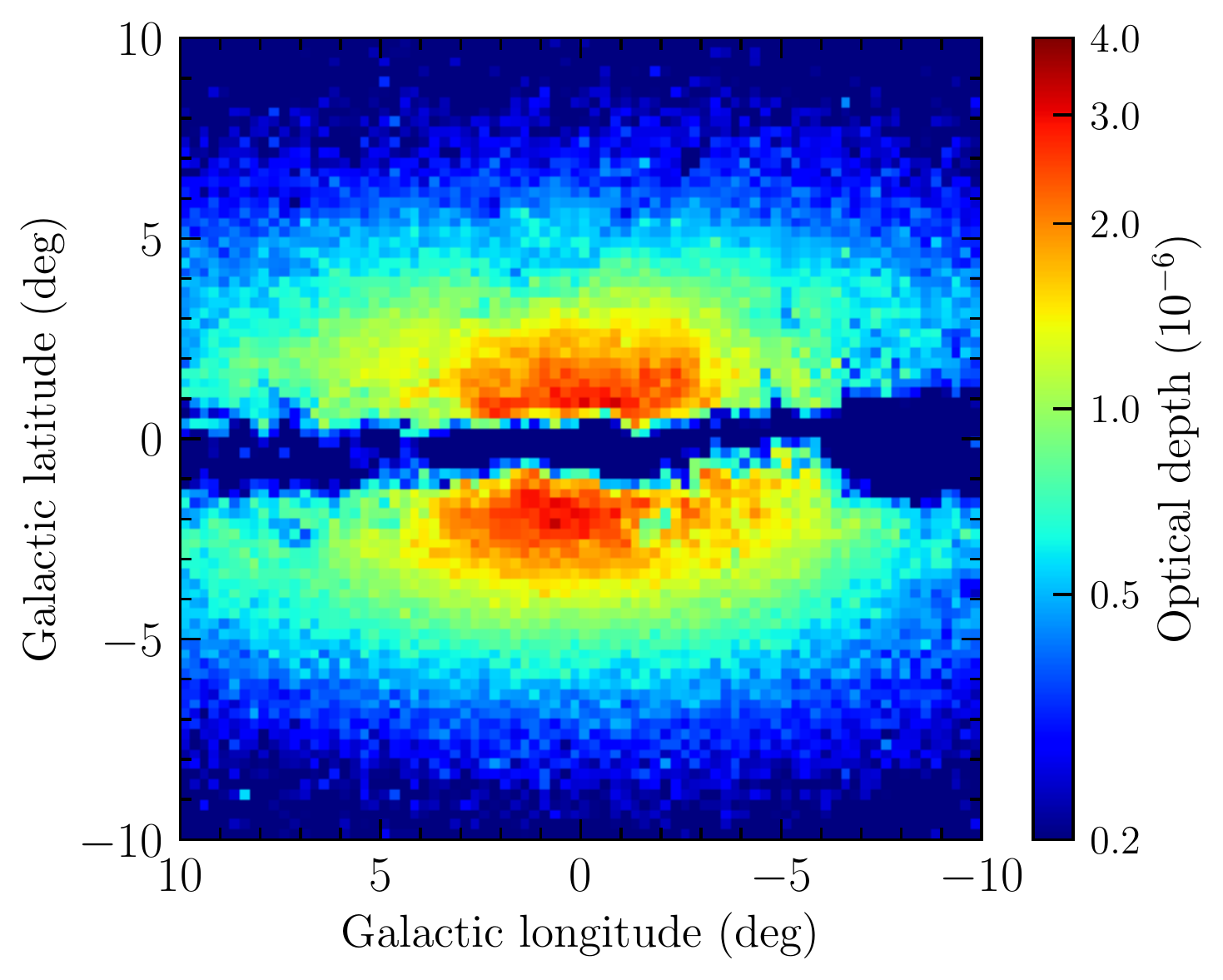}
\caption{Comparison between the observed microlensing optical depth (upper panel) and predictions based on the Manchester--Besan\c{c}on Microlensing Simulator \citep{awiphan2016} for sources brighter than $I=21$ and events shorter than $\tE=300$~days (lower panel). Both maps have the same color scale.}
\label{fig:tau_besa}
\end{figure}

Our observations agree remarkably well with the predictions based on the Besan\c{c}on model (see Figures~\ref{fig:comparison_besancon}, \ref{fig:gamma_besa} and~\ref{fig:tau_besa}), especially at positive Galactic latitudes. We did not adjust the model's parameters; instead, we overplotted its predictions on our data. It is also noteworthy that the model predicts some detailed features of the maps, such as the increased event rate at $(l,b)\approx(1^{\circ},-2^{\circ})$ (see Figure~\ref{fig:gamma_besa}). According to model predictions, $\Gamma$ and $\tau$ should turn over at $|b| < 1.5 ^{\circ}$ owing to the increasing impact of extinction, which agrees well with our data. However, the minuscule details of maps at low Galactic latitudes are different and can be fixed by incorporating recent extinction maps \citep[e.g.,][]{gonzalez2012,nataf2013} into the model. Extensive tests of the Besan\c{c}on model also have to take into account LFs, star density, stellar kinematics, etc. Such tests are beyond the scope of this paper.

\subsection{Microlensing Events in the Direction of the Sgr dSph Galaxy}

The main focus of this paper is the study of microlensing events toward the Galactic bulge. In this section, however, we will discuss microlensing events in the direction of the Sagittarius dwarf spheroidal (Sgr~dSph) galaxy \citep{ibata1994}, for reasons that will become clear later. One of the closest neighbors of the Milky Way, Sgr dSph it is located at a distance of $26.7 \pm 1.3$~kpc \citep{hamanowicz2016} on the opposite site of the Galactic center from Earth, near the Galactic bulge in the sky. The galaxy is extended, but its core (corresponding to the globular cluster M54) is located at the Galactic coordinates of $(l,b)=(+5.6^{\circ},-14.1^{\circ})$. 

During 2011--2014, OGLE carried out a dedicated survey of the central regions of Sgr~dSph with the aim of detecting variable stars and constructing a 3D picture of the galaxy and its stream. Seven fields (BLG705--BLG711) covering an area of about 10~deg$^2$ were observed. The survey's results were published by \citet{hamanowicz2016}. 

\begin{figure}[t]
\centering
\includegraphics[width=0.8\textwidth]{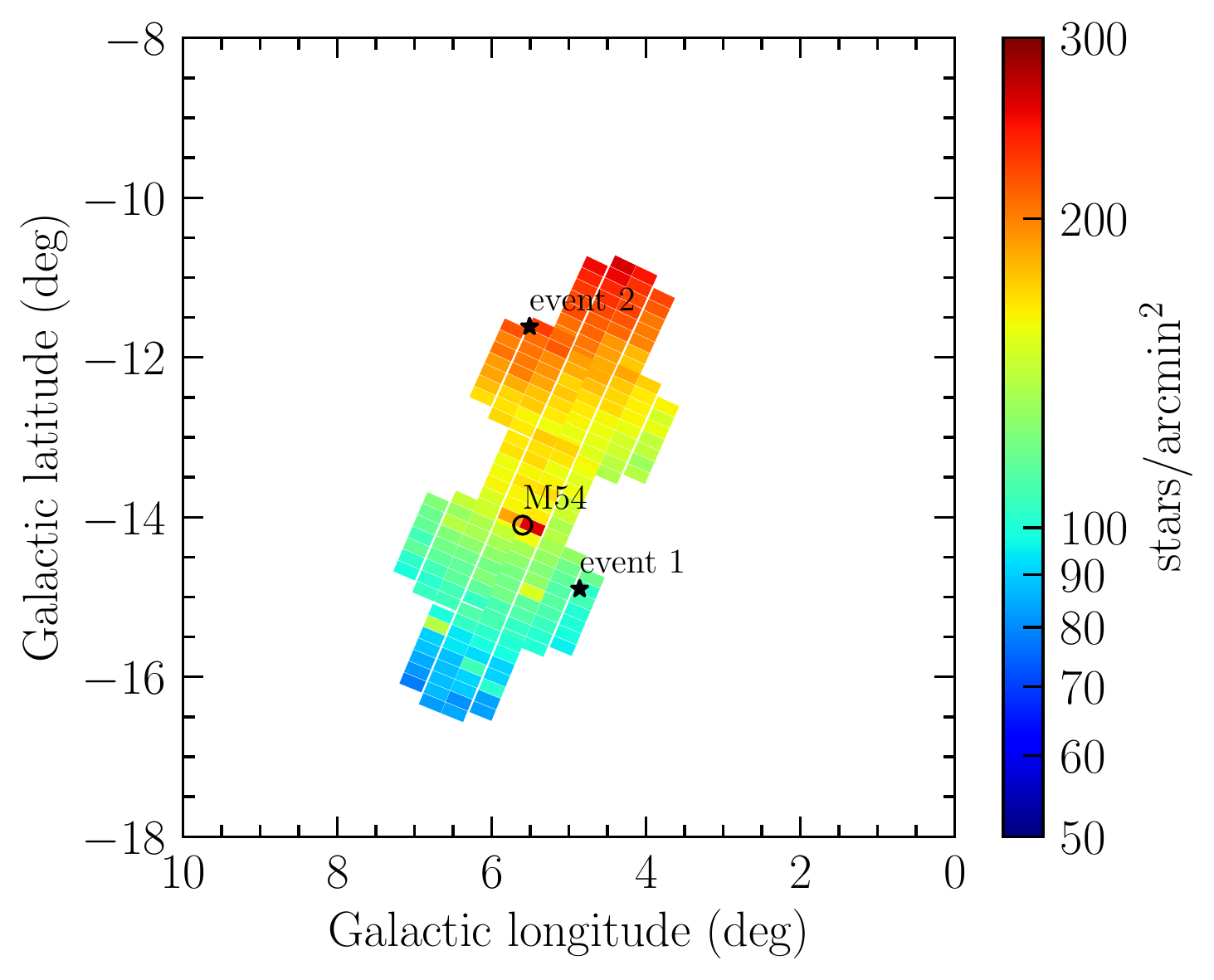}
\caption{Star surface density (down to $I=21$) in OGLE fields toward the Sgr~dSph galaxy. Stars mark the locations of detected microlensing events. The core of the galaxy (globular cluster M54) is marked with a black circle.}
\label{fig:sgr}
\end{figure}

We used these observations to search for microlensing events. Figure~\ref{fig:sgr} shows the surface density of stars in the Sgr~dSph fields, which decreases from 250 to 80~stars arcmin$^{-2}$ with increasing Galactic latitude. This suggests that the majority of observed sources are in fact located in the Milky Way, which will allow us to estimate the optical depth and event rate in the ``field,'' far from the Galactic plane. It has to be stressed, however, that the center of Sgr~dSph (globular cluster M54) is clearly visible in the star counts map. 

We detected two microlensing events. Event~1 (Galactic coordinates $l=4.86^{\circ}, b=-14.90^{\circ}$) is located close to the core of Sgr~dSph, so we checked whether the source can belong to the dwarf galaxy. According to the microlensing model, the majority of light came from the source, which is included in the \textit{Gaia}~DR2. Its proper motion ($\mu_{\alpha}=-1.00 \pm 1.51$, $\mu_{\delta}=-9.77 \pm 1.71\,\mathrm{mas\,yr}^{-1}$) is inconsistent with that of Sgr~dSph ($\mu_{\alpha}\approx-2.69$, $\mu_{\delta}\approx-1.36\,\mathrm{mas\,yr}^{-1}$ with a dispersion of 0.16\,mas\,yr$^{-1}$ in both directions) \citep{helmi2018}. Moreover, because the event was simultaneously observed in the $V$ and $I$ bands, we were able to measure its color $(V-I)_{\rm s}=0.92^{+0.23}_{-0.27}$ (from the model-independent regression) and brightness $I_{\rm s}=20.18^{+0.64}_{-0.76}$. The source position on the color--magnitude diagram is also consistent with that of Milky Way stars.

Event~2 ($l=5.51^{\circ}, b=-11.61^{\circ}$) is located $13.7''$ from a bright $V=11.1$ star that is saturated in OGLE images. We nonetheless checked the individual CCD images of the field and verified this as a genuine transient event, not a diffraction spike from the neighboring star. The event has a timescale of $\tE=10.0^{+2.4}_{-1.6}$~days and source brightness $I_{\rm s}=20.73 \pm 0.29$. There are no magnified $V$-band observations, and the source is not included in the \textit{Gaia}~DR2 catalog. We cannot rule out that the source is located in Sgr~dSph, but the event is located far from its center in the region of high density of Galactic stars. We will assume it also belongs to the Milky Way.

Both detected events likely occurred on sources located in the thick disk of the Milky Way. Taking into account the number of observed sources down to $I=21$ and the duration of the survey, we estimate that the microlensing optical depth in this direction ($l\approx 5^{\circ}$, $b\approx -14^{\circ}$) is \mbox{$\tau=(0.09 \pm 0.07)\times 10^{-6}$}, while the event rate \mbox{$\Gamma=(0.8 \pm 0.6)\times 10^{-6}$\,yr$^{-1}$}. These estimates are consistent with the predictions based on the Manchester--Besan\c{c}on Microlensing Simulator \citep{awiphan2016}. \citet{awiphan2016} calculated their model in the latitude range $|b|<10^{\circ}$, but from extrapolation to ($l\approx 5^{\circ}$, $b\approx -14^{\circ}$), we found $\tau_{\rm model}=(0.044 \pm 0.003) \times 10^{-6}$ and $\Gamma_{\rm model}=(0.29 \pm 0.03) \times 10^{-6}$\,yr$^{-1}$.

\citet{alard2001}, who studied microlensing toward Sgr~dSph, argued that at high Galactic latitudes ($b\approx -9^{\circ}$), microlensing events with Sgr~dSph sources may outnumber Milky Way events by a factor of 5 or larger. Our observations suggest that the optical depth in these regions is smaller than \citeauthor{alard2001}'s (\citeyear{alard2001}) predictions and is consistent with Milky Way events. We cannot rule out, however, that one of the detected events occurred on an Sgr~dSph source. Moreover, the majority of Sgr~dSph sources should be fainter than $I\sim 20$ ($V\sim 21$) \citep{alard2001}, while the limiting magnitude of our experiment is only $I=21$. OGLE is currently monitoring the Sgr stream at Galactic latitudes $-12^{\circ}<b<-6^{\circ}$ \citep{udalski2015}, which will allow us to constrain its contribution to the observed microlensing event rate.

\section{Summary}

The searches for gravitational microlensing of stars in the Galactic bulge were proposed almost 30~yr ago by \citet{paczynski1991} and \citet{griest1991}. However, only the last decade has brought about developments that have allowed us to fully appreciate the power of microlensing. Thanks to the installation of new, large field-of-view detectors, microlensing surveys have been able to monitor hundreds of millions of stars toward the Galactic bulge with a cadence as short as several minutes. Thousands of detected microlensing events allow us to make robust statistical inferences and to detect rare phenomena.

Here we have presented the largest homogeneous sample of 8000 microlensing events that were detected toward the Galactic bulge by the OGLE-IV survey during 2010--2017. Our sample comprises 2212 events from high-cadence fields that were previously published by \citet{mroz2017} and an additional 5790 events from low-cadence fields. We conducted extensive image- and catalog-level simulations that allowed us to measure the detection efficiency of microlensing events as a function of their timescales. Consequently, we were able to precisely measure the microlensing optical depth and event rate toward more than 100 sight lines toward the Galactic bulge. 

Previous measurements, based on samples of up to a few hundred events, were larger than the expectations from the Galactic models and were difficult to reconcile with other constraints on the Galactic structure. The new optical depth and event rate maps ease the tension between the previous measurements and Galactic models. They are consistent with some earlier calculations based on bright stars \citep{afonso2003,hamadache2006} and are systematically $\sim 30\%$ smaller than the other estimates based on ``all-source'' samples of microlensing events \citep{sumi_penny2016}. The difference is probably caused by the wrong number of source stars used for calculations.

Our new maps will allow strict tests of the current models of the inner regions of the Milky Way. For example, we found that the new maps agree well with predictions based on the Besan\c{c}on model of the Galaxy \citep{robin2014,awiphan2016}. Our results may have numerous other applications, such as the measurement of the initial mass function \citep{calchi_novati2008,wegg2017} or constraining the dark matter content in the Milky Way center \citep{wegg2016}. The new maps will also inform the planning of the future space-based microlensing experiments by revising the expected number of events \citep{penny2019}.

The data presented in this paper are publicly available to the astronomical community,

\centerline{\textit{http://www.astrouw.edu.pl/ogle/ogle4/microlensing\_maps}}

\noindent and in an interactive online interface on the OGLE website,

\centerline{\textit{http://ogle.astrouw.edu.pl/cgi-ogle/get\_o4\_tau.py}}

\section*{Acknowledgements}

We thank the referee, Matthew Penny, for constructive comments that improved the presentation of our results. We acknowledge comments from K.~Z. Stanek, M. R\'o\.zyczka, and D.~P. Bennett.
We would like to thank M. Kubiak, G. Pietrzy{\'n}ski, and M. Pawlak for their
contribution to the collection of the OGLE photometric data analyzed in
this paper. P.M. acknowledges support from the Foundation for Polish Science (program START) and the National Science Center, Poland (grant ETIUDA 2018/28/T/ST9/00096). The OGLE project has received funding from the National Science Centre, Poland, grant MAESTRO 2014/14/A/ST9/00121 to A.U.

\clearpage
\appendix

\section{OGLE-IV Fields}
\label{ch:app1}

\startlongtable
\begin{deluxetable}{lrrrrrr}
\tablecaption{Basic Information about Analyzed Fields \label{tab:allfields}}
\tablehead{
\colhead{Field} & \colhead{R.A.} & \colhead{Decl.} & \colhead{$l$} &
\colhead{$b$} & \colhead{$N_{\mathrm{stars}}$} & \colhead{$N_{\mathrm{epochs}}$} \\
\colhead{} & \colhead{(J2000)} & \colhead{(J2000)} & \colhead{(deg)} &
\colhead{(deg)} & \colhead{($10^6$)} & \colhead{}
}
\startdata
BLG500 & 17$^{\mathrm{h}}$51$^{\mathrm{m}}$60$^{\mathrm{s}}$ & $-28^{\circ}36'35''$ & $ 0.9999$ & $-1.0293$ &  4.1 &  4701 \\
BLG501 & 17$^{\mathrm{h}}$51$^{\mathrm{m}}$56$^{\mathrm{s}}$ & $-29^{\circ}50'00''$ & $-0.0608$ & $-1.6400$ &  5.3 & 12099 \\
BLG502 & 17$^{\mathrm{h}}$51$^{\mathrm{m}}$39$^{\mathrm{s}}$ & $-33^{\circ}32'15''$ & $-3.2832$ & $-3.4735$ &  5.0 &  1700 \\
BLG503 & 17$^{\mathrm{h}}$51$^{\mathrm{m}}$34$^{\mathrm{s}}$ & $-34^{\circ}46'05''$ & $-4.3547$ & $-4.0831$ &  4.9 &   745 \\
BLG504 & 17$^{\mathrm{h}}$57$^{\mathrm{m}}$33$^{\mathrm{s}}$ & $-27^{\circ}59'40''$ & $ 2.1491$ & $-1.7747$ &  5.9 &  6429 \\
BLG505 & 17$^{\mathrm{h}}$57$^{\mathrm{m}}$34$^{\mathrm{s}}$ & $-29^{\circ}13'15''$ & $ 1.0870$ & $-2.3890$ &  7.0 & 12066 \\
BLG506 & 17$^{\mathrm{h}}$57$^{\mathrm{m}}$31$^{\mathrm{s}}$ & $-30^{\circ}27'23''$ & $ 0.0103$ & $-2.9974$ &  5.4 &  4706 \\
BLG507 & 17$^{\mathrm{h}}$57$^{\mathrm{m}}$30$^{\mathrm{s}}$ & $-31^{\circ}41'30''$ & $-1.0641$ & $-3.6101$ &  5.5 &  1882 \\
BLG508 & 17$^{\mathrm{h}}$57$^{\mathrm{m}}$30$^{\mathrm{s}}$ & $-32^{\circ}55'20''$ & $-2.1341$ & $-4.2222$ &  4.3 &  1490 \\
BLG509 & 17$^{\mathrm{h}}$57$^{\mathrm{m}}$30$^{\mathrm{s}}$ & $-34^{\circ}09'10''$ & $-3.2058$ & $-4.8329$ &  4.6 &   807 \\
BLG510 & 17$^{\mathrm{h}}$57$^{\mathrm{m}}$30$^{\mathrm{s}}$ & $-35^{\circ}23'00''$ & $-4.2794$ & $-5.4419$ &  3.9 &   559 \\
BLG511 & 18$^{\mathrm{h}}$03$^{\mathrm{m}}$02$^{\mathrm{s}}$ & $-27^{\circ}22'49''$ & $ 3.2835$ & $-2.5219$ &  5.6 &  4588 \\
BLG512 & 18$^{\mathrm{h}}$03$^{\mathrm{m}}$04$^{\mathrm{s}}$ & $-28^{\circ}36'39''$ & $ 2.2154$ & $-3.1355$ &  7.1 & 10253 \\
BLG513 & 18$^{\mathrm{h}}$03$^{\mathrm{m}}$06$^{\mathrm{s}}$ & $-29^{\circ}50'40''$ & $ 1.1399$ & $-3.7432$ &  4.6 &  1909 \\
BLG514 & 18$^{\mathrm{h}}$03$^{\mathrm{m}}$11$^{\mathrm{s}}$ & $-31^{\circ}04'27''$ & $ 0.0747$ & $-4.3626$ &  4.6 &  1532 \\
BLG515 & 18$^{\mathrm{h}}$03$^{\mathrm{m}}$15$^{\mathrm{s}}$ & $-32^{\circ}18'25''$ & $-0.9993$ & $-4.9741$ &  4.4 &  1350 \\
BLG516 & 18$^{\mathrm{h}}$03$^{\mathrm{m}}$20$^{\mathrm{s}}$ & $-33^{\circ}32'15''$ & $-2.0711$ & $-5.5870$ &  4.2 &   634 \\
BLG517 & 18$^{\mathrm{h}}$03$^{\mathrm{m}}$25$^{\mathrm{s}}$ & $-34^{\circ}46'05''$ & $-3.1453$ & $-6.1976$ &  3.6 &   197 \\
BLG518 & 18$^{\mathrm{h}}$08$^{\mathrm{m}}$26$^{\mathrm{s}}$ & $-26^{\circ}46'10''$ & $ 4.4046$ & $-3.2761$ &  4.9 &  1937 \\
BLG519 & 18$^{\mathrm{h}}$08$^{\mathrm{m}}$30$^{\mathrm{s}}$ & $-28^{\circ}00'00''$ & $ 3.3316$ & $-3.8823$ &  5.8 &  1988 \\
BLG520 & 18$^{\mathrm{h}}$08$^{\mathrm{m}}$36$^{\mathrm{s}}$ & $-29^{\circ}13'50''$ & $ 2.2603$ & $-4.4933$ &  5.6 &  1625 \\
BLG521 & 18$^{\mathrm{h}}$08$^{\mathrm{m}}$44$^{\mathrm{s}}$ & $-30^{\circ}27'40''$ & $ 1.1905$ & $-5.1086$ &  5.1 &   738 \\
BLG522 & 18$^{\mathrm{h}}$13$^{\mathrm{m}}$47$^{\mathrm{s}}$ & $-26^{\circ}09'15''$ & $ 5.5202$ & $-4.0329$ &  4.7 &   999 \\
BLG523 & 18$^{\mathrm{h}}$13$^{\mathrm{m}}$55$^{\mathrm{s}}$ & $-27^{\circ}23'05''$ & $ 4.4477$ & $-4.6423$ &  6.3 &   779 \\
BLG524 & 18$^{\mathrm{h}}$14$^{\mathrm{m}}$02$^{\mathrm{s}}$ & $-28^{\circ}36'55''$ & $ 3.3712$ & $-5.2462$ &  6.1 &   644 \\
BLG525 & 18$^{\mathrm{h}}$14$^{\mathrm{m}}$12$^{\mathrm{s}}$ & $-29^{\circ}50'45''$ & $ 2.2975$ & $-5.8574$ &  4.6 &   558 \\
BLG526 & 18$^{\mathrm{h}}$14$^{\mathrm{m}}$25$^{\mathrm{s}}$ & $-31^{\circ}04'35''$ & $ 1.2260$ & $-6.4752$ &  3.1 &   516 \\
BLG527 & 18$^{\mathrm{h}}$19$^{\mathrm{m}}$05$^{\mathrm{s}}$ & $-23^{\circ}04'40''$ & $ 8.8082$ & $-3.6426$ &  3.2 &   385 \\
BLG528 & 18$^{\mathrm{h}}$19$^{\mathrm{m}}$08$^{\mathrm{s}}$ & $-24^{\circ}18'30''$ & $ 7.7241$ & $-4.2297$ &  4.4 &   439 \\
BLG529 & 18$^{\mathrm{h}}$19$^{\mathrm{m}}$11$^{\mathrm{s}}$ & $-25^{\circ}32'20''$ & $ 6.6383$ & $-4.8152$ &  5.6 &   449 \\
BLG530 & 18$^{\mathrm{h}}$19$^{\mathrm{m}}$14$^{\mathrm{s}}$ & $-26^{\circ}46'10''$ & $ 5.5505$ & $-5.3987$ &  5.6 &   441 \\
BLG531 & 18$^{\mathrm{h}}$19$^{\mathrm{m}}$26$^{\mathrm{s}}$ & $-28^{\circ}00'00''$ & $ 4.4760$ & $-6.0094$ &  5.3 &   396 \\
BLG532 & 18$^{\mathrm{h}}$19$^{\mathrm{m}}$36$^{\mathrm{s}}$ & $-29^{\circ}13'50''$ & $ 3.3951$ & $-6.6107$ &  3.8 &   358 \\
BLG533 & 17$^{\mathrm{h}}$52$^{\mathrm{m}}$00$^{\mathrm{s}}$ & $-27^{\circ}23'05''$ & $ 2.0542$ & $-0.4054$ &  0.7 &    46 \\
BLG534 & 17$^{\mathrm{h}}$51$^{\mathrm{m}}$51$^{\mathrm{s}}$ & $-31^{\circ}04'15''$ & $-1.1356$ & $-2.2547$ &  4.3 &  4646 \\
BLG535 & 17$^{\mathrm{h}}$51$^{\mathrm{m}}$44$^{\mathrm{s}}$ & $-32^{\circ}18'25''$ & $-2.2129$ & $-2.8632$ &  3.7 &  1947 \\
BLG536 & 17$^{\mathrm{h}}$57$^{\mathrm{m}}$30$^{\mathrm{s}}$ & $-36^{\circ}36'50''$ & $-5.3552$ & $-6.0490$ &  3.2 &   197 \\
BLG539 & 18$^{\mathrm{h}}$03$^{\mathrm{m}}$30$^{\mathrm{s}}$ & $-35^{\circ}59'55''$ & $-4.2223$ & $-6.8055$ &  2.8 &   191 \\
BLG543 & 18$^{\mathrm{h}}$13$^{\mathrm{m}}$47$^{\mathrm{s}}$ & $-22^{\circ}27'45''$ & $ 8.7716$ & $-2.2752$ &  3.4 &   395 \\
BLG544 & 18$^{\mathrm{h}}$13$^{\mathrm{m}}$47$^{\mathrm{s}}$ & $-23^{\circ}41'35''$ & $ 7.6890$ & $-2.8622$ &  3.0 &   384 \\
BLG545 & 18$^{\mathrm{h}}$13$^{\mathrm{m}}$47$^{\mathrm{s}}$ & $-24^{\circ}55'25''$ & $ 6.6053$ & $-3.4482$ &  3.9 &   731 \\
BLG546 & 18$^{\mathrm{h}}$14$^{\mathrm{m}}$39$^{\mathrm{s}}$ & $-32^{\circ}18'25''$ & $ 0.1530$ & $-7.0928$ &  2.8 &   244 \\
BLG547 & 18$^{\mathrm{h}}$14$^{\mathrm{m}}$52$^{\mathrm{s}}$ & $-33^{\circ}32'15''$ & $-0.9252$ & $-7.7037$ &  3.3 &   195 \\
BLG566 & 18$^{\mathrm{h}}$34$^{\mathrm{m}}$55$^{\mathrm{s}}$ & $-23^{\circ}41'35''$ & $ 9.9310$ & $-7.1538$ &  2.4 &   198 \\
BLG573 & 18$^{\mathrm{h}}$08$^{\mathrm{m}}$54$^{\mathrm{s}}$ & $-31^{\circ}41'30''$ & $ 0.1216$ & $-5.7276$ &  4.5 &   584 \\
BLG580 & 18$^{\mathrm{h}}$08$^{\mathrm{m}}$22$^{\mathrm{s}}$ & $-25^{\circ}32'20''$ & $ 5.4762$ & $-2.6684$ &  4.7 &  1803 \\
BLG588 & 18$^{\mathrm{h}}$09$^{\mathrm{m}}$02$^{\mathrm{s}}$ & $-32^{\circ}55'20''$ & $-0.9534$ & $-6.3377$ &  3.7 &   492 \\
BLG597 & 18$^{\mathrm{h}}$09$^{\mathrm{m}}$12$^{\mathrm{s}}$ & $-34^{\circ}09'10''$ & $-2.0280$ & $-6.9510$ &  2.6 &   359 \\
BLG599 & 17$^{\mathrm{h}}$51$^{\mathrm{m}}$29$^{\mathrm{s}}$ & $-35^{\circ}59'55''$ & $-5.4275$ & $-4.6916$ &  4.1 &   374 \\
BLG600 & 17$^{\mathrm{h}}$51$^{\mathrm{m}}$23$^{\mathrm{s}}$ & $-37^{\circ}13'45''$ & $-6.5036$ & $-5.2961$ &  3.6 &   370 \\
BLG603 & 17$^{\mathrm{h}}$45$^{\mathrm{m}}$46$^{\mathrm{s}}$ & $-34^{\circ}09'10''$ & $-4.4401$ & $-2.7412$ &  4.7 &  1631 \\
BLG604 & 17$^{\mathrm{h}}$45$^{\mathrm{m}}$36$^{\mathrm{s}}$ & $-35^{\circ}23'00''$ & $-5.5113$ & $-3.3505$ &  4.4 &   817 \\
BLG605 & 17$^{\mathrm{h}}$45$^{\mathrm{m}}$25$^{\mathrm{s}}$ & $-36^{\circ}36'50''$ & $-6.5850$ & $-3.9567$ &  3.8 &   371 \\
BLG606 & 17$^{\mathrm{h}}$45$^{\mathrm{m}}$13$^{\mathrm{s}}$ & $-37^{\circ}50'40''$ & $-7.6615$ & $-4.5597$ &  2.9 &   195 \\
BLG609 & 17$^{\mathrm{h}}$39$^{\mathrm{m}}$50$^{\mathrm{s}}$ & $-34^{\circ}46'05''$ & $-5.6069$ & $-2.0233$ &  3.6 &   725 \\
BLG610 & 17$^{\mathrm{h}}$39$^{\mathrm{m}}$35$^{\mathrm{s}}$ & $-35^{\circ}59'55''$ & $-6.6777$ & $-2.6334$ &  3.4 &   524 \\
BLG611 & 17$^{\mathrm{h}}$35$^{\mathrm{m}}$33$^{\mathrm{s}}$ & $-27^{\circ}09'41''$ & $ 0.3282$ & $ 2.8242$ &  5.1 &  4526 \\
BLG612 & 17$^{\mathrm{h}}$30$^{\mathrm{m}}$00$^{\mathrm{s}}$ & $-28^{\circ}00'00''$ & $-1.0464$ & $ 3.4008$ &  3.1 &   768 \\
BLG613 & 17$^{\mathrm{h}}$30$^{\mathrm{m}}$00$^{\mathrm{s}}$ & $-29^{\circ}13'50''$ & $-2.0762$ & $ 2.7248$ &  4.1 &   811 \\
BLG614 & 17$^{\mathrm{h}}$24$^{\mathrm{m}}$27$^{\mathrm{s}}$ & $-28^{\circ}36'55''$ & $-2.2382$ & $ 4.0765$ &  3.6 &   620 \\
BLG615 & 17$^{\mathrm{h}}$24$^{\mathrm{m}}$23$^{\mathrm{s}}$ & $-29^{\circ}50'45''$ & $-3.2679$ & $ 3.3992$ &  4.0 &   800 \\
BLG616 & 17$^{\mathrm{h}}$16$^{\mathrm{m}}$00$^{\mathrm{s}}$ & $-28^{\circ}30'00''$ & $-3.1997$ & $ 5.6697$ &  3.6 &   225 \\
BLG617 & 17$^{\mathrm{h}}$16$^{\mathrm{m}}$00$^{\mathrm{s}}$ & $-29^{\circ}43'50''$ & $-4.2100$ & $ 4.9609$ &  3.8 &   726 \\
BLG619 & 17$^{\mathrm{h}}$25$^{\mathrm{m}}$00$^{\mathrm{s}}$ & $-26^{\circ}00'00''$ & $ 0.0082$ & $ 5.4341$ &  2.8 &    59 \\
BLG621 & 17$^{\mathrm{h}}$35$^{\mathrm{m}}$00$^{\mathrm{s}}$ & $-22^{\circ}13'50''$ & $ 4.4323$ & $ 5.5773$ &  3.6 &   477 \\
BLG622 & 17$^{\mathrm{h}}$35$^{\mathrm{m}}$00$^{\mathrm{s}}$ & $-23^{\circ}27'40''$ & $ 3.3887$ & $ 4.9182$ &  3.3 &   449 \\
BLG624 & 17$^{\mathrm{h}}$40$^{\mathrm{m}}$14$^{\mathrm{s}}$ & $-21^{\circ}36'55''$ & $ 5.6025$ & $ 4.8747$ &  3.8 &   502 \\
BLG625 & 17$^{\mathrm{h}}$40$^{\mathrm{m}}$17$^{\mathrm{s}}$ & $-22^{\circ}50'45''$ & $ 4.5592$ & $ 4.2168$ &  3.9 &   537 \\
BLG626 & 17$^{\mathrm{h}}$40$^{\mathrm{m}}$20$^{\mathrm{s}}$ & $-24^{\circ}04'35''$ & $ 3.5176$ & $ 3.5577$ &  4.6 &   755 \\
BLG629 & 17$^{\mathrm{h}}$45$^{\mathrm{m}}$23$^{\mathrm{s}}$ & $-19^{\circ}46'10''$ & $ 7.8134$ & $ 4.8089$ &  3.2 &   185 \\
BLG630 & 17$^{\mathrm{h}}$45$^{\mathrm{m}}$25$^{\mathrm{s}}$ & $-21^{\circ}00'00''$ & $ 6.7610$ & $ 4.1659$ &  3.5 &   564 \\
BLG631 & 17$^{\mathrm{h}}$45$^{\mathrm{m}}$30$^{\mathrm{s}}$ & $-22^{\circ}13'50''$ & $ 5.7162$ & $ 3.5117$ &  4.1 &   599 \\
BLG632 & 17$^{\mathrm{h}}$45$^{\mathrm{m}}$36$^{\mathrm{s}}$ & $-23^{\circ}27'40''$ & $ 4.6747$ & $ 2.8534$ &  5.2 &   877 \\
BLG633 & 17$^{\mathrm{h}}$45$^{\mathrm{m}}$42$^{\mathrm{s}}$ & $-24^{\circ}41'30''$ & $ 3.6341$ & $ 2.1945$ &  3.7 &   672 \\
BLG636 & 17$^{\mathrm{h}}$50$^{\mathrm{m}}$35$^{\mathrm{s}}$ & $-20^{\circ}23'05''$ & $ 7.9121$ & $ 3.4452$ &  2.8 &   200 \\
BLG637 & 17$^{\mathrm{h}}$50$^{\mathrm{m}}$39$^{\mathrm{s}}$ & $-21^{\circ}36'55''$ & $ 6.8600$ & $ 2.8041$ &  3.9 &   522 \\
BLG638 & 17$^{\mathrm{h}}$50$^{\mathrm{m}}$47$^{\mathrm{s}}$ & $-22^{\circ}50'45''$ & $ 5.8168$ & $ 2.1491$ &  3.7 &   575 \\
BLG639 & 17$^{\mathrm{h}}$50$^{\mathrm{m}}$56$^{\mathrm{s}}$ & $-24^{\circ}04'35''$ & $ 4.7762$ & $ 1.4906$ &  2.9 &   543 \\
BLG641 & 17$^{\mathrm{h}}$55$^{\mathrm{m}}$48$^{\mathrm{s}}$ & $-21^{\circ}00'00''$ & $ 7.9992$ & $ 2.0816$ &  3.0 &   358 \\
BLG642 & 17$^{\mathrm{h}}$55$^{\mathrm{m}}$55$^{\mathrm{s}}$ & $-22^{\circ}13'50''$ & $ 6.9487$ & $ 1.4394$ &  1.9 &   362 \\
BLG643 & 17$^{\mathrm{h}}$56$^{\mathrm{m}}$06$^{\mathrm{s}}$ & $-23^{\circ}27'40''$ & $ 5.9062$ & $ 0.7838$ &  1.7 &   358 \\
BLG644 & 17$^{\mathrm{h}}$57$^{\mathrm{m}}$30$^{\mathrm{s}}$ & $-25^{\circ}32'20''$ & $ 4.2691$ & $-0.5348$ &  0.8 &    36 \\
BLG645 & 17$^{\mathrm{h}}$57$^{\mathrm{m}}$30$^{\mathrm{s}}$ & $-26^{\circ}46'10''$ & $ 3.2039$ & $-1.1511$ &  2.0 &   550 \\
BLG646 & 18$^{\mathrm{h}}$02$^{\mathrm{m}}$55$^{\mathrm{s}}$ & $-26^{\circ}09'15''$ & $ 4.3401$ & $-1.8981$ &  2.6 &   690 \\
BLG647 & 17$^{\mathrm{h}}$52$^{\mathrm{m}}$04$^{\mathrm{s}}$ & $-26^{\circ}09'15''$ & $ 3.1204$ & $ 0.2092$ &  0.5 &   198 \\
BLG648 & 17$^{\mathrm{h}}$46$^{\mathrm{m}}$36$^{\mathrm{s}}$ & $-26^{\circ}46'10''$ & $ 1.9634$ & $ 0.9419$ &  1.6 &   644 \\
BLG649 & 17$^{\mathrm{h}}$46$^{\mathrm{m}}$28$^{\mathrm{s}}$ & $-28^{\circ}00'00''$ & $ 0.8963$ & $ 0.3281$ &  0.8 &    37 \\
BLG650 & 17$^{\mathrm{h}}$46$^{\mathrm{m}}$22$^{\mathrm{s}}$ & $-29^{\circ}13'50''$ & $-0.1665$ & $-0.2925$ &  0.7 &    45 \\
BLG651 & 17$^{\mathrm{h}}$46$^{\mathrm{m}}$14$^{\mathrm{s}}$ & $-30^{\circ}27'40''$ & $-1.2329$ & $-0.9073$ &  1.1 &    41 \\
BLG652 & 17$^{\mathrm{h}}$41$^{\mathrm{m}}$10$^{\mathrm{s}}$ & $-26^{\circ}09'15''$ & $ 1.8508$ & $ 2.3000$ &  3.0 &   646 \\
BLG653 & 17$^{\mathrm{h}}$35$^{\mathrm{m}}$32$^{\mathrm{s}}$ & $-28^{\circ}36'55''$ & $-0.8998$ & $ 2.0442$ &  3.2 &   778 \\
BLG654 & 17$^{\mathrm{h}}$35$^{\mathrm{m}}$36$^{\mathrm{s}}$ & $-29^{\circ}50'45''$ & $-1.9286$ & $ 1.3682$ &  2.7 &   920 \\
BLG655 & 17$^{\mathrm{h}}$35$^{\mathrm{m}}$41$^{\mathrm{s}}$ & $-31^{\circ}04'35''$ & $-2.9551$ & $ 0.6890$ &  1.3 &    56 \\
BLG657 & 18$^{\mathrm{h}}$19$^{\mathrm{m}}$50$^{\mathrm{s}}$ & $-30^{\circ}27'40''$ & $ 2.3179$ & $-7.2216$ &  4.3 &   208 \\
BLG659 & 17$^{\mathrm{h}}$46$^{\mathrm{m}}$04$^{\mathrm{s}}$ & $-31^{\circ}41'30''$ & $-2.3030$ & $-1.5164$ &  1.0 &    37 \\
BLG660 & 17$^{\mathrm{h}}$45$^{\mathrm{m}}$56$^{\mathrm{s}}$ & $-32^{\circ}55'20''$ & $-3.3695$ & $-2.1318$ &  3.2 &   687 \\
BLG661 & 17$^{\mathrm{h}}$40$^{\mathrm{m}}$05$^{\mathrm{s}}$ & $-33^{\circ}32'15''$ & $-4.5361$ & $-1.4138$ &  2.2 &   531 \\
BLG662 & 17$^{\mathrm{h}}$30$^{\mathrm{m}}$00$^{\mathrm{s}}$ & $-30^{\circ}27'40''$ & $-3.1047$ & $ 2.0479$ &  3.1 &   573 \\
BLG663 & 18$^{\mathrm{h}}$08$^{\mathrm{m}}$18$^{\mathrm{s}}$ & $-24^{\circ}18'30''$ & $ 6.5466$ & $-2.0595$ &  1.6 &   230 \\
BLG664 & 18$^{\mathrm{h}}$02$^{\mathrm{m}}$55$^{\mathrm{s}}$ & $-24^{\circ}55'25''$ & $ 5.4117$ & $-1.2924$ &  1.4 &   230 \\
BLG665 & 17$^{\mathrm{h}}$30$^{\mathrm{m}}$04$^{\mathrm{s}}$ & $-25^{\circ}19'15''$ & $ 1.2091$ & $ 4.8559$ &  1.9 &   114 \\
BLG666 & 17$^{\mathrm{h}}$30$^{\mathrm{m}}$03$^{\mathrm{s}}$ & $-26^{\circ}33'05''$ & $ 0.1737$ & $ 4.1858$ &  1.6 &    89 \\
BLG667 & 17$^{\mathrm{h}}$35$^{\mathrm{m}}$28$^{\mathrm{s}}$ & $-25^{\circ}56'10''$ & $ 1.3523$ & $ 3.4996$ &  1.8 &   456 \\
BLG668 & 17$^{\mathrm{h}}$29$^{\mathrm{m}}$57$^{\mathrm{s}}$ & $-31^{\circ}41'30''$ & $-4.1383$ & $ 1.3792$ &  1.4 &   323 \\
BLG670 & 17$^{\mathrm{h}}$33$^{\mathrm{m}}$57$^{\mathrm{s}}$ & $-34^{\circ}09'10''$ & $-5.7392$ & $-0.6699$ &  1.3 &   338 \\
BLG672 & 17$^{\mathrm{h}}$33$^{\mathrm{m}}$02$^{\mathrm{s}}$ & $-36^{\circ}37'00''$ & $-7.9091$ & $-1.8529$ &  1.8 &   343 \\
BLG675 & 17$^{\mathrm{h}}$40$^{\mathrm{m}}$58$^{\mathrm{s}}$ & $-27^{\circ}23'05''$ & $ 0.7819$ & $ 1.6875$ &  3.3 &   900 \\
BLG676 & 17$^{\mathrm{h}}$40$^{\mathrm{m}}$49$^{\mathrm{s}}$ & $-28^{\circ}36'55''$ & $-0.2799$ & $ 1.0642$ &  2.4 &    69 \\
BLG677 & 17$^{\mathrm{h}}$40$^{\mathrm{m}}$37$^{\mathrm{s}}$ & $-29^{\circ}50'45''$ & $-1.3467$ & $ 0.4490$ &  0.6 &    66 \\
BLG680 & 17$^{\mathrm{h}}$39$^{\mathrm{m}}$06$^{\mathrm{s}}$ & $-37^{\circ}13'45''$ & $-7.7734$ & $-3.2045$ &  2.0 &   175 \\
BLG683 & 17$^{\mathrm{h}}$46$^{\mathrm{m}}$39$^{\mathrm{s}}$ & $-25^{\circ}32'20''$ & $ 3.0214$ & $ 1.5708$ &  2.4 &   105 \\
BLG705 & 18$^{\mathrm{h}}$41$^{\mathrm{m}}$50$^{\mathrm{s}}$ & $-30^{\circ}27'40''$ & $ 4.4295$ & $-11.4798$ &  1.2 &   167 \\
BLG706 & 18$^{\mathrm{h}}$47$^{\mathrm{m}}$06$^{\mathrm{s}}$ & $-29^{\circ}50'45''$ & $ 5.4859$ & $-12.2582$ &  1.0 &   164 \\
BLG707 & 18$^{\mathrm{h}}$47$^{\mathrm{m}}$31$^{\mathrm{s}}$ & $-31^{\circ}04'35''$ & $ 4.3744$ & $-12.8458$ &  1.0 &   160 \\
BLG708 & 18$^{\mathrm{h}}$52$^{\mathrm{m}}$50$^{\mathrm{s}}$ & $-30^{\circ}27'40''$ & $ 5.4276$ & $-13.6408$ &  1.0 &   157 \\
BLG709 & 18$^{\mathrm{h}}$58$^{\mathrm{m}}$02$^{\mathrm{s}}$ & $-29^{\circ}50'45''$ & $ 6.4709$ & $-14.4267$ &  0.7 &   160 \\
BLG710 & 18$^{\mathrm{h}}$58$^{\mathrm{m}}$31$^{\mathrm{s}}$ & $-31^{\circ}04'35''$ & $ 5.3429$ & $-15.0054$ &  0.7 &   153 \\
BLG711 & 19$^{\mathrm{h}}$03$^{\mathrm{m}}$50$^{\mathrm{s}}$ & $-30^{\circ}27'40''$ & $ 6.3893$ & $-15.8210$ &  0.6 &   156 \\
BLG714 & 17$^{\mathrm{h}}$40$^{\mathrm{m}}$24$^{\mathrm{s}}$ & $-25^{\circ}18'25''$ & $ 2.4794$ & $ 2.8944$ &  2.8 &   345 \\
BLG715 & 17$^{\mathrm{h}}$35$^{\mathrm{m}}$00$^{\mathrm{s}}$ & $-24^{\circ}41'30''$ & $ 2.3472$ & $ 4.2575$ &  2.4 &   444 \\
BLG717 & 18$^{\mathrm{h}}$08$^{\mathrm{m}}$30$^{\mathrm{s}}$ & $-23^{\circ}04'40''$ & $ 7.6458$ & $-1.5033$ &  1.8 &   156 \\
\enddata
\tablecomments{Equatorial coordinates are given for the epoch J2000. Here $N_{\rm stars}$ is the number of stars in the database in millions, and $N_{\rm epochs}$ is the number of collected frames used in the analysis; $l$ and $b$ are Galactic longitude and latitude, respectively. (This table is available in machine-readable form.)}
\end{deluxetable}

\section{Microlensing Optical Depths and Event Rates in the OGLE-IV Fields}
\label{ch:app2}

\startlongtable
\begin{deluxetable}{lrrrrrrrr}
\tablecaption{Microlensing Optical Depths and Event Rates in the OGLE-IV Fields (Averaged over Sources Brighter than $I_{\rm s} = 21$) \label{tab:results}}
\tablehead{
\colhead{Field} & \colhead{$l$} & \colhead{$b$}  &
\colhead{$\tau_{300}$} & \colhead{$\Gamma$} & \colhead{$\Gamma_{\mathrm{deg}^2}$} &\colhead{$\langle\tE\rangle$} & 
\colhead{$N_{\rm ev}$} & \colhead{$N_{\rm s}$} \\
\colhead{} & \colhead{(deg)} & \colhead{(deg)}  &
\colhead{$(10^{-6})$} & \colhead{$(10^{-6}\,\mathrm{yr}^{-1})$} & \colhead{($\mathrm{deg}^{-2}\,\mathrm{yr}^{-1})$} & \colhead{(days)} & 
\colhead{} & \colhead{$(10^6)$}
}
\startdata
BLG500 & $ 0.9999$ & $-1.0293$ & $1.93 \pm 0.21$ & $23.9 \pm  2.0$ & $168.8 \pm  13.7$ & $18.8 \pm 1.6 $ & 164 &  6.78 \\
BLG501 & $-0.0608$ & $-1.6400$ & $2.13 \pm 0.15$ & $24.1 \pm  1.4$ & $222.9 \pm  12.9$ & $20.5 \pm 1.1 $ & 317 & 13.31 \\
BLG502 & $-3.2832$ & $-3.4735$ & $1.22 \pm 0.12$ & $11.0 \pm  0.9$ & $ 69.7 \pm   5.5$ & $25.7 \pm 1.8 $ & 171 & 10.02 \\
BLG503 & $-4.3547$ & $-4.0831$ & $0.74 \pm 0.10$ & $ 5.5 \pm  0.7$ & $ 37.4 \pm   4.2$ & $31.4 \pm 3.3 $ &  91 & 10.70 \\
BLG504 & $ 2.1491$ & $-1.7747$ & $1.45 \pm 0.12$ & $16.9 \pm  1.2$ & $134.3 \pm   9.1$ & $20.0 \pm 1.2 $ & 225 & 11.86 \\
BLG505 & $ 1.0870$ & $-2.3890$ & $2.09 \pm 0.17$ & $22.2 \pm  1.1$ & $265.3 \pm  12.8$ & $21.8 \pm 1.5 $ & 441 & 19.02 \\
BLG506 & $ 0.0103$ & $-2.9974$ & $1.97 \pm 0.25$ & $16.5 \pm  1.1$ & $137.4 \pm   8.9$ & $28.0 \pm 2.9 $ & 247 & 12.85 \\
BLG507 & $-1.0641$ & $-3.6101$ & $1.21 \pm 0.10$ & $12.3 \pm  0.9$ & $ 87.3 \pm   6.1$ & $22.9 \pm 1.4 $ & 216 & 11.22 \\
BLG508 & $-2.1341$ & $-4.2222$ & $0.83 \pm 0.09$ & $ 6.8 \pm  0.7$ & $ 42.3 \pm   4.1$ & $28.5 \pm 2.2 $ & 119 &  9.53 \\
BLG509 & $-3.2058$ & $-4.8329$ & $0.73 \pm 0.10$ & $ 4.6 \pm  0.5$ & $ 28.5 \pm   3.5$ & $36.6 \pm 4.2 $ &  79 &  9.82 \\
BLG510 & $-4.2794$ & $-5.4419$ & $0.48 \pm 0.09$ & $ 3.6 \pm  0.7$ & $ 17.5 \pm   3.1$ & $31.2 \pm 4.6 $ &  38 &  7.67 \\
BLG511 & $ 3.2835$ & $-2.5219$ & $1.43 \pm 0.15$ & $13.5 \pm  1.0$ & $113.9 \pm   8.1$ & $24.5 \pm 2.0 $ & 204 & 13.45 \\
BLG512 & $ 2.2154$ & $-3.1355$ & $1.44 \pm 0.13$ & $14.0 \pm  0.9$ & $148.5 \pm   9.0$ & $24.0 \pm 1.5 $ & 280 & 17.48 \\
BLG513 & $ 1.1399$ & $-3.7432$ & $0.89 \pm 0.08$ & $ 8.5 \pm  0.7$ & $ 76.0 \pm   5.4$ & $24.4 \pm 1.6 $ & 213 & 13.71 \\
BLG514 & $ 0.0747$ & $-4.3626$ & $0.68 \pm 0.08$ & $ 6.2 \pm  0.7$ & $ 39.8 \pm   4.0$ & $25.2 \pm 2.0 $ & 108 & 10.27 \\
BLG515 & $-0.9993$ & $-4.9741$ & $0.74 \pm 0.11$ & $ 5.7 \pm  0.8$ & $ 29.6 \pm   3.8$ & $30.5 \pm 4.3 $ &  80 &  8.10 \\
BLG516 & $-2.0711$ & $-5.5870$ & $0.50 \pm 0.08$ & $ 4.6 \pm  0.7$ & $ 21.5 \pm   3.2$ & $25.3 \pm 2.6 $ &  48 &  7.61 \\
BLG517 & $-3.1453$ & $-6.1976$ & $0.36 \pm 0.11$ & $ 2.6 \pm  0.9$ & $  9.7 \pm   3.2$ & $32.6 \pm 8.6 $ &  12 &  5.96 \\
BLG518 & $ 4.4046$ & $-3.2761$ & $0.94 \pm 0.10$ & $ 7.3 \pm  0.5$ & $ 56.7 \pm   4.6$ & $30.0 \pm 2.3 $ & 160 & 12.34 \\
BLG519 & $ 3.3316$ & $-3.8823$ & $1.04 \pm 0.10$ & $ 9.3 \pm  0.7$ & $ 77.8 \pm   5.4$ & $25.9 \pm 1.7 $ & 223 & 12.93 \\
BLG520 & $ 2.2603$ & $-4.4933$ & $0.71 \pm 0.09$ & $ 5.6 \pm  0.5$ & $ 40.5 \pm   3.7$ & $29.5 \pm 2.6 $ & 124 & 11.63 \\
BLG521 & $ 1.1905$ & $-5.1086$ & $0.59 \pm 0.08$ & $ 4.5 \pm  0.5$ & $ 24.7 \pm   3.1$ & $30.9 \pm 2.6 $ &  69 &  9.18 \\
BLG522 & $ 5.5202$ & $-4.0329$ & $0.90 \pm 0.11$ & $ 7.4 \pm  0.8$ & $ 41.5 \pm   4.3$ & $28.5 \pm 2.8 $ & 108 &  8.62 \\
BLG523 & $ 4.4477$ & $-4.6423$ & $0.87 \pm 0.11$ & $ 6.0 \pm  0.7$ & $ 39.2 \pm   4.2$ & $34.1 \pm 3.4 $ &  99 & 10.40 \\
BLG524 & $ 3.3712$ & $-5.2462$ & $0.72 \pm 0.11$ & $ 6.0 \pm  0.8$ & $ 36.7 \pm   4.5$ & $28.2 \pm 3.4 $ &  75 &  9.66 \\
BLG525 & $ 2.2975$ & $-5.8574$ & $0.40 \pm 0.08$ & $ 3.2 \pm  0.5$ & $ 15.1 \pm   2.6$ & $29.9 \pm 3.4 $ &  36 &  7.84 \\
BLG526 & $ 1.2260$ & $-6.4752$ & $0.31 \pm 0.08$ & $ 2.3 \pm  0.5$ & $  7.8 \pm   1.8$ & $31.3 \pm 4.9 $ &  20 &  5.25 \\
BLG527 & $ 8.8082$ & $-3.6426$ & $0.92 \pm 0.15$ & $ 5.5 \pm  0.9$ & $ 21.9 \pm   3.5$ & $39.5 \pm 5.0 $ &  48 &  6.36 \\
BLG528 & $ 7.7241$ & $-4.2297$ & $0.63 \pm 0.11$ & $ 5.5 \pm  0.9$ & $ 24.7 \pm   4.0$ & $26.9 \pm 3.7 $ &  50 &  7.39 \\
BLG529 & $ 6.6383$ & $-4.8152$ & $0.65 \pm 0.12$ & $ 5.6 \pm  1.0$ & $ 26.8 \pm   4.7$ & $27.2 \pm 3.9 $ &  39 &  7.79 \\
BLG530 & $ 5.5505$ & $-5.3987$ & $1.00 \pm 0.16$ & $ 8.5 \pm  1.3$ & $ 39.2 \pm   6.3$ & $27.6 \pm 3.9 $ &  54 &  7.37 \\
BLG531 & $ 4.4760$ & $-6.0094$ & $0.68 \pm 0.13$ & $ 5.1 \pm  1.0$ & $ 22.2 \pm   4.1$ & $30.7 \pm 4.1 $ &  34 &  6.85 \\
BLG532 & $ 3.3951$ & $-6.6107$ & $0.41 \pm 0.10$ & $ 2.7 \pm  0.7$ & $ 10.3 \pm   2.2$ & $34.8 \pm 6.0 $ &  23 &  6.02 \\
BLG533 & $ 2.0542$ & $-0.4054$ & $1.23 \pm 1.23$ & $ 7.1 \pm  7.1$ & $  4.7 \pm   4.7$ & $40.5 \pm 40.5 $ &   1 &  1.02 \\
BLG534 & $-1.1356$ & $-2.2547$ & $1.55 \pm 0.15$ & $17.2 \pm  1.3$ & $104.8 \pm   8.1$ & $20.8 \pm 1.6 $ & 176 &  9.06 \\
BLG535 & $-2.2129$ & $-2.8632$ & $1.67 \pm 0.15$ & $15.2 \pm  1.1$ & $ 85.2 \pm   6.4$ & $25.7 \pm 1.7 $ & 189 &  7.57 \\
BLG536 & $-5.3552$ & $-6.0490$ & $0.51 \pm 0.14$ & $ 3.8 \pm  1.1$ & $ 12.2 \pm   3.4$ & $31.3 \pm 6.0 $ &  15 &  5.21 \\
BLG539 & $-4.2223$ & $-6.8055$ & $0.61 \pm 0.21$ & $ 4.0 \pm  1.4$ & $ 11.6 \pm   4.0$ & $34.9 \pm 11.9 $ &  13 &  4.52 \\
BLG543 & $ 8.7716$ & $-2.2752$ & $1.07 \pm 0.20$ & $ 5.9 \pm  1.1$ & $ 17.7 \pm   3.1$ & $41.8 \pm 5.5 $ &  36 &  4.76 \\
BLG544 & $ 7.6890$ & $-2.8622$ & $0.77 \pm 0.13$ & $ 6.1 \pm  1.0$ & $ 22.1 \pm   3.7$ & $29.2 \pm 3.5 $ &  40 &  5.66 \\
BLG545 & $ 6.6053$ & $-3.4482$ & $0.85 \pm 0.13$ & $ 6.1 \pm  0.9$ & $ 25.0 \pm   3.7$ & $32.3 \pm 4.2 $ &  57 &  6.47 \\
BLG546 & $ 0.1530$ & $-7.0928$ & $0.45 \pm 0.13$ & $ 3.4 \pm  1.0$ & $  8.4 \pm   2.5$ & $30.7 \pm 4.5 $ &  12 &  4.25 \\
BLG547 & $-0.9252$ & $-7.7037$ & $0.19 \pm 0.11$ & $ 1.3 \pm  0.8$ & $  3.1 \pm   1.8$ & $32.6 \pm 12.6 $ &   3 &  3.67 \\
BLG566 & $ 9.9310$ & $-7.1538$ & $0.14 \pm 0.10$ & $ 0.9 \pm  0.7$ & $  1.6 \pm   1.1$ & $38.5 \pm 13.6 $ &   2 &  3.03 \\
BLG573 & $ 0.1216$ & $-5.7276$ & $0.40 \pm 0.08$ & $ 3.7 \pm  0.7$ & $ 16.9 \pm   3.0$ & $25.0 \pm 2.8 $ &  33 &  7.16 \\
BLG580 & $ 5.4762$ & $-2.6684$ & $1.00 \pm 0.11$ & $ 8.0 \pm  0.8$ & $ 43.1 \pm   4.1$ & $29.1 \pm 2.2 $ & 116 &  8.47 \\
BLG588 & $-0.9534$ & $-6.3377$ & $0.40 \pm 0.09$ & $ 2.7 \pm  0.5$ & $  9.1 \pm   1.9$ & $33.9 \pm 4.0 $ &  23 &  5.39 \\
BLG597 & $-2.0280$ & $-6.9510$ & $0.56 \pm 0.11$ & $ 4.0 \pm  0.8$ & $ 11.1 \pm   2.2$ & $31.6 \pm 3.6 $ &  27 &  4.28 \\
BLG599 & $-5.4275$ & $-4.6916$ & $0.52 \pm 0.10$ & $ 3.9 \pm  0.7$ & $ 19.4 \pm   3.3$ & $31.4 \pm 4.2 $ &  41 &  7.80 \\
BLG600 & $-6.5036$ & $-5.2961$ & $0.44 \pm 0.12$ & $ 2.9 \pm  0.8$ & $  9.6 \pm   2.6$ & $35.1 \pm 8.7 $ &  18 &  5.29 \\
BLG603 & $-4.4401$ & $-2.7412$ & $1.21 \pm 0.12$ & $10.6 \pm  0.9$ & $ 66.8 \pm   5.4$ & $26.7 \pm 2.0 $ & 167 & 10.02 \\
BLG604 & $-5.5113$ & $-3.3505$ & $0.96 \pm 0.11$ & $ 7.0 \pm  0.8$ & $ 43.0 \pm   4.6$ & $32.0 \pm 2.6 $ & 100 &  9.80 \\
BLG605 & $-6.5850$ & $-3.9567$ & $0.77 \pm 0.17$ & $ 3.9 \pm  0.8$ & $ 14.7 \pm   3.0$ & $45.9 \pm 8.5 $ &  29 &  6.13 \\
BLG606 & $-7.6615$ & $-4.5597$ & $0.98 \pm 0.50$ & $ 3.8 \pm  1.2$ & $ 10.3 \pm   3.2$ & $60.3 \pm 24.7 $ &  11 &  4.48 \\
BLG609 & $-5.6069$ & $-2.0233$ & $1.24 \pm 0.15$ & $ 8.4 \pm  1.0$ & $ 36.7 \pm   4.2$ & $34.4 \pm 2.7 $ &  81 &  6.77 \\
BLG610 & $-6.6777$ & $-2.6334$ & $0.88 \pm 0.16$ & $ 6.2 \pm  1.3$ & $ 22.3 \pm   4.5$ & $32.8 \pm 5.9 $ &  38 &  5.78 \\
BLG611 & $ 0.3282$ & $ 2.8242$ & $1.52 \pm 0.15$ & $16.2 \pm  1.3$ & $ 70.0 \pm   5.7$ & $21.8 \pm 1.6 $ & 158 &  6.93 \\
BLG612 & $-1.0464$ & $ 3.4008$ & $1.19 \pm 0.15$ & $10.1 \pm  1.2$ & $ 48.2 \pm   5.5$ & $27.0 \pm 2.6 $ &  87 &  7.34 \\
BLG613 & $-2.0762$ & $ 2.7248$ & $1.31 \pm 0.14$ & $12.4 \pm  1.2$ & $ 62.7 \pm   6.3$ & $24.4 \pm 2.1 $ & 119 &  7.99 \\
BLG614 & $-2.2382$ & $ 4.0765$ & $0.78 \pm 0.14$ & $ 5.5 \pm  0.9$ & $ 21.9 \pm   3.6$ & $33.2 \pm 4.7 $ &  43 &  6.32 \\
BLG615 & $-3.2679$ & $ 3.3992$ & $1.09 \pm 0.16$ & $ 6.8 \pm  0.9$ & $ 35.9 \pm   4.4$ & $37.5 \pm 4.5 $ &  74 &  8.09 \\
BLG616 & $-3.1997$ & $ 5.6697$ & $0.41 \pm 0.12$ & $ 3.4 \pm  1.0$ & $ 11.2 \pm   3.3$ & $28.2 \pm 4.1 $ &  12 &  5.08 \\
BLG617 & $-4.2100$ & $ 4.9609$ & $0.70 \pm 0.14$ & $ 5.2 \pm  1.1$ & $ 20.9 \pm   4.4$ & $30.8 \pm 7.0 $ &  43 &  6.32 \\
BLG619 & $ 0.0082$ & $ 5.4341$ & $0.69 \pm 0.40$ & $ 3.3 \pm  2.0$ & $  9.4 \pm   5.6$ & $49.0 \pm 11.2 $ &   3 &  4.47 \\
BLG621 & $ 4.4323$ & $ 5.5773$ & $0.71 \pm 0.12$ & $ 6.8 \pm  1.2$ & $ 21.3 \pm   3.8$ & $24.5 \pm 3.3 $ &  44 &  5.15 \\
BLG622 & $ 3.3887$ & $ 4.9182$ & $0.75 \pm 0.17$ & $ 3.6 \pm  0.8$ & $ 10.6 \pm   2.3$ & $49.4 \pm 8.1 $ &  25 &  4.76 \\
BLG624 & $ 5.6025$ & $ 4.8747$ & $0.61 \pm 0.13$ & $ 4.3 \pm  0.9$ & $ 16.3 \pm   3.4$ & $33.6 \pm 6.5 $ &  31 &  6.02 \\
BLG625 & $ 4.5592$ & $ 4.2168$ & $0.78 \pm 0.15$ & $ 5.5 \pm  0.9$ & $ 22.9 \pm   3.5$ & $33.4 \pm 5.0 $ &  46 &  6.47 \\
BLG626 & $ 3.5176$ & $ 3.5577$ & $0.95 \pm 0.11$ & $ 8.3 \pm  1.0$ & $ 43.7 \pm   4.9$ & $26.8 \pm 2.4 $ &  92 &  8.32 \\
BLG629 & $ 7.8134$ & $ 4.8089$ & $0.53 \pm 0.17$ & $ 3.4 \pm  1.1$ & $  9.7 \pm   3.1$ & $36.7 \pm 7.8 $ &  11 &  4.56 \\
BLG630 & $ 6.7610$ & $ 4.1659$ & $0.73 \pm 0.17$ & $ 4.4 \pm  0.8$ & $ 13.8 \pm   2.6$ & $39.1 \pm 6.6 $ &  30 &  5.11 \\
BLG631 & $ 5.7162$ & $ 3.5117$ & $0.87 \pm 0.13$ & $ 7.1 \pm  1.0$ & $ 31.5 \pm   4.4$ & $28.7 \pm 3.5 $ &  61 &  7.09 \\
BLG632 & $ 4.6747$ & $ 2.8534$ & $1.11 \pm 0.12$ & $ 9.0 \pm  0.9$ & $ 55.1 \pm   5.1$ & $28.5 \pm 2.2 $ & 128 &  9.59 \\
BLG633 & $ 3.6341$ & $ 2.1945$ & $1.72 \pm 0.20$ & $13.1 \pm  1.4$ & $ 52.4 \pm   5.4$ & $30.4 \pm 2.8 $ & 106 &  6.30 \\
BLG636 & $ 7.9121$ & $ 3.4452$ & $0.64 \pm 0.17$ & $ 5.1 \pm  1.4$ & $ 13.8 \pm   3.9$ & $29.0 \pm 5.0 $ &  14 &  4.26 \\
BLG637 & $ 6.8600$ & $ 2.8041$ & $0.97 \pm 0.16$ & $ 5.2 \pm  0.9$ & $ 19.4 \pm   3.2$ & $43.0 \pm 4.7 $ &  40 &  5.84 \\
BLG638 & $ 5.8168$ & $ 2.1491$ & $0.87 \pm 0.13$ & $ 6.9 \pm  1.0$ & $ 35.5 \pm   4.9$ & $29.4 \pm 3.6 $ &  64 &  7.96 \\
BLG639 & $ 4.7762$ & $ 1.4906$ & $1.38 \pm 0.19$ & $10.4 \pm  1.3$ & $ 33.5 \pm   4.3$ & $30.9 \pm 3.0 $ &  70 &  5.08 \\
BLG641 & $ 7.9992$ & $ 2.0816$ & $0.82 \pm 0.16$ & $ 4.9 \pm  1.0$ & $ 16.5 \pm   3.2$ & $38.8 \pm 5.7 $ &  31 &  5.33 \\
BLG642 & $ 6.9487$ & $ 1.4394$ & $1.25 \pm 0.23$ & $ 8.7 \pm  1.5$ & $ 17.4 \pm   3.0$ & $33.5 \pm 3.6 $ &  36 &  3.22 \\
BLG643 & $ 5.9062$ & $ 0.7838$ & $1.17 \pm 0.33$ & $ 5.7 \pm  1.4$ & $  7.5 \pm   1.9$ & $47.6 \pm 9.0 $ &  17 &  2.11 \\
BLG644 & $ 4.2691$ & $-0.5348$ & -- & -- & -- & -- &   0 &  1.02 \\
BLG645 & $ 3.2039$ & $-1.1511$ & $2.02 \pm 0.32$ & $13.6 \pm  1.9$ & $ 29.2 \pm   4.0$ & $34.3 \pm 4.5 $ &  63 &  3.33 \\
BLG646 & $ 4.3401$ & $-1.8981$ & $1.12 \pm 0.17$ & $ 8.3 \pm  1.2$ & $ 26.3 \pm   3.7$ & $31.7 \pm 4.1 $ &  57 &  5.09 \\
BLG647 & $ 3.1204$ & $ 0.2092$ & -- & -- & -- & -- &   0 &  0.68 \\
BLG648 & $ 1.9634$ & $ 0.9419$ & $1.90 \pm 0.28$ & $18.3 \pm  2.4$ & $ 33.2 \pm   4.4$ & $24.0 \pm 2.8 $ &  67 &  2.86 \\
BLG649 & $ 0.8963$ & $ 0.3281$ & -- & -- & -- & -- &   0 &  0.93 \\
BLG650 & $-0.1665$ & $-0.2925$ & -- & -- & -- & -- &   0 &  0.91 \\
BLG651 & $-1.2329$ & $-0.9073$ & -- & -- & -- & -- &   0 &  1.56 \\
BLG652 & $ 1.8508$ & $ 2.3000$ & $1.52 \pm 0.17$ & $14.2 \pm  1.6$ & $ 60.0 \pm   6.9$ & $24.9 \pm 2.6 $ & 100 &  6.55 \\
BLG653 & $-0.8998$ & $ 2.0442$ & $2.29 \pm 0.24$ & $20.1 \pm  2.0$ & $ 67.7 \pm   6.5$ & $26.6 \pm 2.1 $ & 127 &  5.37 \\
BLG654 & $-1.9286$ & $ 1.3682$ & $1.84 \pm 0.20$ & $16.7 \pm  1.6$ & $ 53.0 \pm   5.3$ & $25.6 \pm 1.9 $ & 110 &  4.92 \\
BLG655 & $-2.9551$ & $ 0.6890$ & $1.07 \pm 0.62$ & $ 9.5 \pm  6.1$ & $ 12.7 \pm   8.2$ & $26.3 \pm 12.3 $ &   3 &  2.13 \\
BLG657 & $ 2.3179$ & $-7.2216$ & $0.32 \pm 0.13$ & $ 2.6 \pm  1.1$ & $  7.2 \pm   3.0$ & $27.2 \pm 4.0 $ &   6 &  4.43 \\
BLG659 & $-2.3030$ & $-1.5164$ & -- & -- & -- & -- &   0 &  1.05 \\
BLG660 & $-3.3695$ & $-2.1318$ & $2.59 \pm 0.34$ & $23.7 \pm  2.6$ & $ 81.8 \pm   9.1$ & $25.5 \pm 2.6 $ &  94 &  4.63 \\
BLG661 & $-4.5361$ & $-1.4138$ & $2.22 \pm 0.32$ & $15.2 \pm  2.2$ & $ 41.1 \pm   5.9$ & $34.0 \pm 5.0 $ &  80 &  4.22 \\
BLG662 & $-3.1047$ & $ 2.0479$ & $1.50 \pm 0.19$ & $12.2 \pm  1.4$ & $ 47.9 \pm   5.6$ & $28.6 \pm 2.8 $ &  87 &  6.16 \\
BLG663 & $ 6.5466$ & $-2.0595$ & $0.40 \pm 0.23$ & $ 3.7 \pm  2.3$ & $  5.1 \pm   3.1$ & $25.2 \pm 16.1 $ &   4 &  2.15 \\
BLG664 & $ 5.4117$ & $-1.2924$ & $0.51 \pm 0.22$ & $ 5.7 \pm  2.7$ & $  6.6 \pm   3.2$ & $21.0 \pm 8.5 $ &   6 &  1.85 \\
BLG665 & $ 1.2091$ & $ 4.8559$ & $0.41 \pm 0.41$ & $ 1.2 \pm  1.2$ & $  2.1 \pm   2.1$ & $79.6 \pm 0.0 $ &   1 &  2.72 \\
BLG666 & $ 0.1737$ & $ 4.1858$ & $1.57 \pm 0.94$ & $ 9.8 \pm  5.8$ & $ 12.7 \pm   7.5$ & $37.3 \pm 12.4 $ &   3 &  2.04 \\
BLG667 & $ 1.3523$ & $ 3.4996$ & $1.23 \pm 0.26$ & $11.3 \pm  2.5$ & $ 17.3 \pm   3.8$ & $25.3 \pm 5.1 $ &  27 &  2.41 \\
BLG668 & $-4.1383$ & $ 1.3792$ & $0.81 \pm 0.22$ & $ 8.8 \pm  2.9$ & $ 10.7 \pm   3.5$ & $21.3 \pm 5.7 $ &  15 &  1.93 \\
BLG670 & $-5.7392$ & $-0.6699$ & $1.70 \pm 0.55$ & $10.7 \pm  2.6$ & $ 11.1 \pm   2.8$ & $37.0 \pm 11.0 $ &  19 &  1.65 \\
BLG672 & $-7.9091$ & $-1.8529$ & $1.24 \pm 0.35$ & $ 5.6 \pm  1.4$ & $  8.3 \pm   2.1$ & $51.8 \pm 10.6 $ &  17 &  2.45 \\
BLG675 & $ 0.7819$ & $ 1.6875$ & $2.40 \pm 0.23$ & $26.5 \pm  2.3$ & $ 95.5 \pm   8.3$ & $21.0 \pm 1.8 $ & 160 &  5.64 \\
BLG676 & $-0.2799$ & $ 1.0642$ & $0.19 \pm 0.19$ & $ 3.9 \pm  3.9$ & $ 11.0 \pm  11.0$ & $10.8 \pm 10.8 $ &   1 &  4.58 \\
BLG677 & $-1.3467$ & $ 0.4490$ & -- & -- & -- & -- &   0 &  0.96 \\
BLG680 & $-7.7734$ & $-3.2045$ & $0.50 \pm 0.19$ & $ 5.0 \pm  2.0$ & $ 11.3 \pm   4.3$ & $23.4 \pm 6.6 $ &   8 &  3.57 \\
BLG683 & $ 3.0214$ & $ 1.5708$ & $3.63 \pm 0.88$ & $26.2 \pm  5.9$ & $ 73.0 \pm  16.3$ & $32.3 \pm 4.7 $ &  21 &  4.28 \\
BLG705 & $ 4.4295$ & $-11.4798$ & -- & -- & -- & -- &   0 &  1.10 \\
BLG706 & $ 5.4859$ & $-12.2582$ & $0.14 \pm 0.14$ & $ 3.4 \pm  3.4$ & $  2.0 \pm   2.0$ & $ 9.9 \pm 9.9 $ &   1 &  0.94 \\
BLG707 & $ 4.3744$ & $-12.8458$ & -- & -- & -- & -- &   0 &  0.81 \\
BLG708 & $ 5.4276$ & $-13.6408$ & -- & -- & -- & -- &   0 &  0.84 \\
BLG709 & $ 6.4709$ & $-14.4267$ & -- & -- & -- & -- &   0 &  0.61 \\
BLG710 & $ 5.3429$ & $-15.0054$ & $0.63 \pm 0.63$ & $ 2.6 \pm  2.6$ & $  1.0 \pm   1.0$ & $56.0 \pm 56.0 $ &   1 &  0.59 \\
BLG711 & $ 6.3893$ & $-15.8210$ & -- & -- & -- & -- &   0 &  0.48 \\
BLG714 & $ 2.4794$ & $ 2.8944$ & $1.24 \pm 0.20$ & $11.0 \pm  1.7$ & $ 41.6 \pm   6.6$ & $26.4 \pm 3.6 $ &  54 &  5.84 \\
BLG715 & $ 2.3472$ & $ 4.2575$ & $1.24 \pm 0.22$ & $10.2 \pm  1.9$ & $ 22.0 \pm   3.9$ & $28.0 \pm 3.1 $ &  36 &  3.42 \\
BLG717 & $ 7.6458$ & $-1.5033$ & $0.81 \pm 0.34$ & $ 3.9 \pm  1.6$ & $  5.1 \pm   2.1$ & $48.1 \pm 9.2 $ &   6 &  2.06 \\
\enddata
\tablecomments{OGLE-IV Galactic bulge fields with Galactic coordinates of the field center ($l$,$b$), optical depth $\tau_{\rm 300}$, microlensing event rate $\Gamma$, microlensing event rate per unit area $\Gamma_{\mathrm{deg}^2}$, average Einstein timescale $\langle\tE\rangle$, number of detected events $N_{\rm ev}$, and number of sources brighter than $I=21$ in millions $N_{\rm s}$. (The table is available in machine-readable form.)}
\end{deluxetable}

\clearpage

\bibliographystyle{aasjournal}
\bibliography{sample}

\begin{thebibliography}{}
\expandafter\ifx\csname natexlab\endcsname\relax\def\natexlab#1{#1}\fi

\bibitem[{{Afonso} {et~al.}(2003){Afonso}, {Albert}, {Alard}, {Andersen},
  {Ansari}, {Aubourg}, {Bareyre}, {Bauer}, {Beaulieu}, {Blanc}, {Bouquet},
  {Char}, {Charlot}, {Couchot}, {Coutures}, {Derue}, {Ferlet}, {Fouqu{\'e}},
  {Glicenstein}, {Goldman}, {Gould}, {Graff}, {Gros}, {Haissinski},
  {Hamadache}, {Hamilton}, {Hardin}, {de Kat}, {Kim}, {Lasserre}, {LeGuillou},
  {Lesquoy}, {Loup}, {Magneville}, {Mansoux}, {Marquette}, {Maurice}, {Maury},
  {Milsztajn}, {Moniez}, {Palanque-Delabrouille}, {Perdereau}, {Pr{\'e}vot},
  {Regnault}, {Rich}, {Spiro}, {Tisserand}, {Vidal-Madjar}, {Vigroux}, \&
  {Zylberajch}}]{afonso2003}
{Afonso}, C., {Albert}, J.~N., {Alard}, C., {et~al.} 2003, \aap, 404, 145

\bibitem[{{Alard} \& {Lupton}(1998)}]{alard1998}
{Alard}, C., \& {Lupton}, R.~H. 1998, \apj, 503, 325

\bibitem[{{Alcock} {et~al.}(1995){Alcock}, {Allsman}, {Axelrod}, {Bennett},
  {Cook}, {Freeman}, {Griest}, {Marshall}, {Perlmutter}, {Peterson}, {Pratt},
  {Quinn}, {Rodgers}, {Stubbs}, {Sutherland}, \& {MACHO
  Collaboration}}]{alcock1995}
{Alcock}, C., {Allsman}, R.~A., {Axelrod}, T.~S., {et~al.} 1995, \apj, 445, 133

\bibitem[{{Alcock} {et~al.}(1997){Alcock}, {Allsman}, {Alves}, {Axelrod},
  {Bennett}, {Cook}, {Freeman}, {Griest}, {Guern}, {Lehner}, {Marshall},
  {Park}, {Perlmutter}, {Peterson}, {Pratt}, {Quinn}, {Rodgers}, {Stubbs}, \&
  {Sutherland}}]{alcock1997b}
{Alcock}, C., {Allsman}, R.~A., {Alves}, D., {et~al.} 1997, \apj, 479, 119

\bibitem[{{Alcock} {et~al.}(2000){Alcock}, {Allsman}, {Alves}, {Axelrod},
  {Becker}, {Bennett}, {Cook}, {Drake}, {Freeman}, {Geha}, {Griest}, {Lehner},
  {Marshall}, {Minniti}, {Nelson}, {Peterson}, {Popowski}, {Pratt}, {Quinn},
  {Stubbs}, {Sutherland}, {Tomaney}, {Vandehei}, {Welch}, \& {MACHO
  Collaboration}}]{alcock2000}
{Alcock}, C., {Allsman}, R.~A., {Alves}, D.~R., {et~al.} 2000, \apj, 541, 734

\bibitem[{{Awiphan} {et~al.}(2016){Awiphan}, {Kerins}, \&
  {Robin}}]{awiphan2016}
{Awiphan}, S., {Kerins}, E., \& {Robin}, A.~C. 2016, \mnras, 456, 1666

\bibitem[{{Batista} {et~al.}(2011){Batista}, {Gould}, {Dieters}, {Dong},
  {Bond}, {Beaulieu}, {Maoz}, {Monard}, {Christie}, {McCormick}, {Albrow},
  {Horne}, {Tsapras}, {Burgdorf}, {Calchi Novati}, {Skottfelt}, {Caldwell},
  {Koz{\l}owski}, {Kubas}, {Gaudi}, {Han}, {Bennett}, {An}, {MOA
  Collaboration}, {Abe}, {Botzler}, {Douchin}, {Freeman}, {Fukui}, {Furusawa},
  {Hearnshaw}, {Hosaka}, {Itow}, {Kamiya}, {Kilmartin}, {Korpela}, {Lin},
  {Ling}, {Makita}, {Masuda}, {Matsubara}, {Miyake}, {Muraki}, {Nagaya},
  {Nishimoto}, {Ohnishi}, {Okumura}, {Perrott}, {Rattenbury}, {Saito},
  {Sullivan}, {Sumi}, {Sweatman}, {Tristram}, {von Seggern}, {Yock}, {PLANET
  Collaboration}, {Brillant}, {Calitz}, {Cassan}, {Cole}, {Cook}, {Coutures},
  {Dominis Prester}, {Donatowicz}, {Greenhill}, {Hoffman}, {Jablonski}, {Kane},
  {Kains}, {Marquette}, {Martin}, {Martioli}, {Meintjes}, {Menzies},
  {Pedretti}, {Pollard}, {Sahu}, {Vinter}, {Wambsganss}, {Watson}, {Williams},
  {Zub}, {FUN Collaboration}, {Allen}, {Bolt}, {Bos}, {DePoy}, {Drummond},
  {Eastman}, {Gal-Yam}, {Gorbikov}, {Higgins}, {Janczak}, {Kaspi}, {Lee},
  {Mallia}, {Maury}, {Monard}, {Moorhouse}, {Morgan}, {Natusch}, {Ofek},
  {Park}, {Pogge}, {Polishook}, {Santallo}, {Shporer}, {Spector}, {Thornley},
  {Yee}, {MiNDSTEp Consortium}, {Bozza}, {Browne}, {Dominik}, {Dreizler},
  {Finet}, {Glitrup}, {Grundahl}, {Harps{\o}e}, {Hessman}, {Hinse},
  {Hundertmark}, {J{\o}rgensen}, {Liebig}, {Maier}, {Mancini}, {Mathiasen},
  {Rahvar}, {Ricci}, {Scarpetta}, {Southworth}, {Surdej}, {Zimmer}, {RoboNet
  Collaboration}, {Allan}, {Bramich}, {Snodgrass}, {Steele}, \&
  {Street}}]{batista2011}
{Batista}, V., {Gould}, A., {Dieters}, S., {et~al.} 2011, \aap, 529, A102

\bibitem[{{Binney}(2018)}]{binney2018}
{Binney}, J. 2018, in IAU Symposium, Vol. 330, Astrometry and Astrophysics in
  the Gaia Sky, ed. A.~{Recio-Blanco}, P.~{de Laverny}, A.~G.~A. {Brown}, \&
  T.~{Prusti}, 111--118

\bibitem[{{Binney} {et~al.}(2000){Binney}, {Bissantz}, \&
  {Gerhard}}]{binney2000}
{Binney}, J., {Bissantz}, N., \& {Gerhard}, O. 2000, \apjl, 537, L99

\bibitem[{{Bissantz} \& {Gerhard}(2002)}]{bissantz2002}
{Bissantz}, N., \& {Gerhard}, O. 2002, \mnras, 330, 591

\bibitem[{{Blitz} \& {Spergel}(1991)}]{blitz1991}
{Blitz}, L., \& {Spergel}, D.~N. 1991, \apj, 379, 631

\bibitem[{{Brown} {et~al.}(2009){Brown}, {Sahu}, {Zoccali}, {Renzini},
  {Ferguson}, {Anderson}, {Smith}, {Bond}, {Minniti}, {Valenti}, {Casertano},
  {Livio}, {Panagia}, {Vanden Berg}, \& {Valenti}}]{brown2009}
{Brown}, T.~M., {Sahu}, K., {Zoccali}, M., {et~al.} 2009, \aj, 137, 3172

\bibitem[{{Brown} {et~al.}(2010){Brown}, {Sahu}, {Anderson}, {Tumlinson},
  {Valenti}, {Smith}, {Jeffery}, {Renzini}, {Zoccali}, {Ferguson},
  {VandenBerg}, {Bond}, {Casertano}, {Valenti}, {Minniti}, {Livio}, \&
  {Panagia}}]{brown2010}
{Brown}, T.~M., {Sahu}, K., {Anderson}, J., {et~al.} 2010, \apjl, 725, L19

\bibitem[{{Calchi Novati} {et~al.}(2008){Calchi Novati}, {de Luca}, {Jetzer},
  {Mancini}, \& {Scarpetta}}]{calchi_novati2008}
{Calchi Novati}, S., {de Luca}, F., {Jetzer}, P., {Mancini}, L., \&
  {Scarpetta}, G. 2008, \aap, 480, 723

\bibitem[{{Cseresnjes} \& {Alard}(2001)}]{alard2001}
{Cseresnjes}, P., \& {Alard}, C. 2001, \aap, 369, 778

\bibitem[{{Foreman-Mackey} {et~al.}(2013){Foreman-Mackey}, {Hogg}, {Lang}, \&
  {Goodman}}]{foreman2013}
{Foreman-Mackey}, D., {Hogg}, D.~W., {Lang}, D., \& {Goodman}, J. 2013, \pasp,
  125, 306

\bibitem[{{Fux}(1997)}]{fux1997}
{Fux}, R. 1997, \aap, 327, 983

\bibitem[{{Gaia Collaboration} {et~al.}(2018){Gaia Collaboration}, {Helmi},
  {van Leeuwen}, {McMillan}, {Massari}, {Antoja}, {Robin}, {Lindegren},
  {Bastian}, {Arenou}, {Babusiaux}, {Biermann}, {Breddels}, {Hobbs}, {Jordi},
  {Pancino}, {Reyl{\'e}}, {Veljanoski}, {Brown}, {Vallenari}, {Prusti}, {de
  Bruijne}, {Bailer-Jones}, {Evans}, {Eyer}, {Jansen}, {Klioner}, {Lammers},
  {Luri}, {Mignard}, {Panem}, {Pourbaix}, {Randich}, {Sartoretti}, {Siddiqui},
  {Soubiran}, {Walton}, {Cropper}, {Drimmel}, {Katz}, {Lattanzi}, {Bakker},
  {Cacciari}, {Casta{\~n}eda}, {Chaoul}, {Cheek}, {De Angeli}, \&
  {Fabricius}}]{helmi2018}
{Gaia Collaboration}, {Helmi}, A., {van Leeuwen}, F., {et~al.} 2018, \aap, 616,
  A12

\bibitem[{{Gonzalez} {et~al.}(2012){Gonzalez}, {Rejkuba}, {Zoccali}, {Valenti},
  {Minniti}, {Schultheis}, {Tobar}, \& {Chen}}]{gonzalez2012}
{Gonzalez}, O.~A., {Rejkuba}, M., {Zoccali}, M., {et~al.} 2012, \aap, 543, A13

\bibitem[{{Grenacher} {et~al.}(1999){Grenacher}, {Jetzer}, {Str{\"a}ssle}, \&
  {de Paolis}}]{grenacher1999}
{Grenacher}, L., {Jetzer}, P., {Str{\"a}ssle}, M., \& {de Paolis}, F. 1999,
  \aap, 351, 775

\bibitem[{{Griest} {et~al.}(1991){Griest}, {Alcock}, {Axelrod}, {Bennett},
  {Cook}, {Freeman}, {Park}, {Perlmutter}, {Peterson}, {Quinn}, {Rodgers},
  {Stubbs}, \& {MACHO Collaboration}}]{griest1991}
{Griest}, K., {Alcock}, C., {Axelrod}, T.~S., {et~al.} 1991, \apjl, 372, L79

\bibitem[{{Gyuk}(1999)}]{gyuk1999}
{Gyuk}, G. 1999, \apj, 510, 205

\bibitem[{{Hamadache} {et~al.}(2006){Hamadache}, {Le Guillou}, {Tisserand},
  {Afonso}, {Albert}, {Andersen}, {Ansari}, {Aubourg}, {Bareyre}, {Beaulieu},
  {Charlot}, {Coutures}, {Ferlet}, {Fouqu{\'e}}, {Glicenstein}, {Goldman},
  {Gould}, {Graff}, {Gros}, {Haissinski}, {de Kat}, {Lesquoy}, {Loup},
  {Magneville}, {Marquette}, {Maurice}, {Maury}, {Milsztajn}, {Moniez},
  {Palanque-Delabrouille}, {Perdereau}, {Rahal}, {Rich}, {Spiro},
  {Vidal-Madjar}, {Vigroux}, \& {Zylberajch}}]{hamadache2006}
{Hamadache}, C., {Le Guillou}, L., {Tisserand}, P., {et~al.} 2006, \aap, 454,
  185

\bibitem[{{Hamanowicz} {et~al.}(2016){Hamanowicz}, {Pietrukowicz}, {Udalski},
  {Mr{\'o}z}, {Soszy{\'n}ski}, {Szyma{\'n}ski}, {Skowron}, {Poleski},
  {Wyrzykowski}, {Koz{\l}owski}, {Pawlak}, \& {Ulaczyk}}]{hamanowicz2016}
{Hamanowicz}, A., {Pietrukowicz}, P., {Udalski}, A., {et~al.} 2016, \actaa, 66,
  197

\bibitem[{{Han} \& {Gould}(1995)}]{han1995_stat}
{Han}, C., \& {Gould}, A. 1995, \apj, 449, 521

\bibitem[{{Han} \& {Gould}(2003)}]{han_gould2003}
{Han}, C., \& {Gould}, A. 2003, \apj, 592, 172

\bibitem[{{Holtzman} {et~al.}(2006){Holtzman}, {Afonso}, \&
  {Dolphin}}]{holtzman2006}
{Holtzman}, J.~A., {Afonso}, C., \& {Dolphin}, A. 2006, \apjs, 166, 534

\bibitem[{{Holtzman} {et~al.}(1998){Holtzman}, {Watson}, {Baum}, {Grillmair},
  {Groth}, {Light}, {Lynds}, \& {O'Neil}}]{holtzman1998}
{Holtzman}, J.~A., {Watson}, A.~M., {Baum}, W.~A., {et~al.} 1998, \aj, 115,
  1946

\bibitem[{{Holtzman} {et~al.}(1995){Holtzman}, {Hester}, {Casertano},
  {Trauger}, {Watson}, {Ballester}, {Burrows}, {Clarke}, {Crisp}, {Evans},
  {Gallagher}, {Griffiths}, {Hoessel}, {Matthews}, {Mould}, {Scowen},
  {Stapelfeldt}, \& {Westphal}}]{holtzman1995}
{Holtzman}, J.~A., {Hester}, J.~J., {Casertano}, S., {et~al.} 1995, \pasp, 107,
  156

\bibitem[{{Ibata} {et~al.}(1994){Ibata}, {Gilmore}, \& {Irwin}}]{ibata1994}
{Ibata}, R.~A., {Gilmore}, G., \& {Irwin}, M.~J. 1994, \nat, 370, 194

\bibitem[{{Kerins} {et~al.}(2009){Kerins}, {Robin}, \& {Marshall}}]{kerins2009}
{Kerins}, E., {Robin}, A.~C., \& {Marshall}, D.~J. 2009, \mnras, 396, 1202

\bibitem[{{Kim} {et~al.}(2018{\natexlab{a}}){Kim}, {Kim}, {Hwang}, {Albrow},
  {Chung}, {Gould}, {Han}, {Jung}, {Ryu}, {Shin}, {Yee}, {Zhu}, {Cha}, {Kim},
  {Lee}, {Lee}, {Lee}, {Park}, {Pogge}, \& {KMTNet Collaboration}}]{kim2018}
{Kim}, D.-J., {Kim}, H.-W., {Hwang}, K.-H., {et~al.} 2018{\natexlab{a}}, \aj,
  155, 76

\bibitem[{{Kim} {et~al.}(2018{\natexlab{b}}){Kim}, {Hwang}, {Kim}, {Albrow},
  {Cha}, {Chung}, {Gould}, {Han}, {Jung}, {Kim}, {Lee}, {Lee}, {Lee}, {Park},
  {Pogge}, {Ryu}, {Shin}, {Shvartzvald}, {Yee}, {Zang}, {Zhu}, \& {KMTNet
  Collaboration}}]{kim2018_2}
{Kim}, H.-W., {Hwang}, K.-H., {Kim}, D.-J., {et~al.} 2018{\natexlab{b}}, \aj,
  155, 186

\bibitem[{{Kiraga} \& {Paczy{\'n}ski}(1994)}]{kiraga1994}
{Kiraga}, M., \& {Paczy{\'n}ski}, B. 1994, \apjl, 430, L101

\bibitem[{{MacKenty} {et~al.}(2010){MacKenty}, {Kimble}, {O'Connell}, \&
  {Townsend}}]{mackenty2010}
{MacKenty}, J.~W., {Kimble}, R.~A., {O'Connell}, R.~W., \& {Townsend}, J.~A.
  2010, in Society of Photo-Optical Instrumentation Engineers (SPIE) Conference
  Series, Vol. 7731, \procspie, 77310Z

\bibitem[{{Mao} \& {Paczy{\'n}ski}(1996)}]{mao1996}
{Mao}, S., \& {Paczy{\'n}ski}, B. 1996, \apj, 473, 57

\bibitem[{{Mao} {et~al.}(2002){Mao}, {Smith}, {Wo{\'z}niak}, {Udalski},
  {Szyma{\'n}ski}, {Kubiak}, {Pietrzy{\'n}ski}, {Soszy{\'n}ski}, \&
  {{\.Z}ebru{\'n}}}]{mao2002}
{Mao}, S., {Smith}, M.~C., {Wo{\'z}niak}, P., {et~al.} 2002, \mnras, 329, 349

\bibitem[{{Marshall} {et~al.}(2006){Marshall}, {Robin}, {Reyl{\'e}},
  {Schultheis}, \& {Picaud}}]{marshall2006}
{Marshall}, D.~J., {Robin}, A.~C., {Reyl{\'e}}, C., {Schultheis}, M., \&
  {Picaud}, S. 2006, \aap, 453, 635

\bibitem[{{Monahan}(2011)}]{monahan}
{Monahan}, J.~F. 2011, Numerical Methods of Statistics, 2nd edn. (Cambridge:
  Cambridge University Press)

\bibitem[{{Mr{\'o}z} {et~al.}(2017){Mr{\'o}z}, {Udalski}, {Skowron}, {Poleski},
  {Koz{\l}owski}, {Szyma{\'n}ski}, {Soszy{\'n}ski}, {Wyrzykowski},
  {Pietrukowicz}, {Ulaczyk}, {Skowron}, \& {Pawlak}}]{mroz2017}
{Mr{\'o}z}, P., {Udalski}, A., {Skowron}, J., {et~al.} 2017, \nat, 548, 183

\bibitem[{{Nataf} {et~al.}(2013){Nataf}, {Gould}, {Fouqu{\'e}}, {Gonzalez},
  {Johnson}, {Skowron}, {Udalski}, {Szyma{\'n}ski}, {Kubiak},
  {Pietrzy{\'n}ski}, {Soszy{\'n}ski}, {Ulaczyk}, {Wyrzykowski}, \&
  {Poleski}}]{nataf2013}
{Nataf}, D.~M., {Gould}, A., {Fouqu{\'e}}, P., {et~al.} 2013, \apj, 769, 88

\bibitem[{{Navarro} {et~al.}(2017){Navarro}, {Minniti}, \& {Contreras
  Ramos}}]{navarro2017}
{Navarro}, M.~G., {Minniti}, D., \& {Contreras Ramos}, R. 2017, \apjl, 851, L13

\bibitem[{{Navarro} {et~al.}(2018){Navarro}, {Minniti}, \&
  {Contreras-Ramos}}]{navarro2018}
{Navarro}, M.~G., {Minniti}, D., \& {Contreras-Ramos}, R. 2018, \apjl, 865, L5

\bibitem[{{Nikolaev} \& {Weinberg}(1997)}]{nikolaev1997}
{Nikolaev}, S., \& {Weinberg}, M.~D. 1997, \apj, 487, 885

\bibitem[{{Paczy{\'n}ski}(1991)}]{paczynski1991}
{Paczy{\'n}ski}, B. 1991, \apjl, 371, L63

\bibitem[{{Paczy{\'n}ski} {et~al.}(1994){Paczy{\'n}ski}, {Stanek}, {Udalski},
  {Szymanski}, {Kaluzny}, {Kubiak}, {Mateo}, \& {Krzeminski}}]{paczynski1994}
{Paczy{\'n}ski}, B., {Stanek}, K.~Z., {Udalski}, A., {et~al.} 1994, \apjl, 435,
  L113

\bibitem[{{Peale}(1998)}]{peale1998}
{Peale}, S.~J. 1998, \apj, 509, 177

\bibitem[{{Penny} {et~al.}(2019){Penny}, {Gaudi}, {Kerins}, {Rattenbury},
  {Mao}, {Robin}, \& {Calchi Novati}}]{penny2019}
{Penny}, M.~T., {Gaudi}, B.~S., {Kerins}, E., {et~al.} 2019, \apjs, 241, 3

\bibitem[{{Penny} {et~al.}(2013){Penny}, {Kerins}, {Rattenbury}, {Beaulieu},
  {Robin}, {Mao}, {Batista}, {Calchi Novati}, {Cassan}, {Fouqu{\'e}},
  {McDonald}, {Marquette}, {Tisserand}, \& {Zapatero Osorio}}]{penny2013}
{Penny}, M.~T., {Kerins}, E., {Rattenbury}, N., {et~al.} 2013, \mnras, 434, 2

\bibitem[{{Poindexter} {et~al.}(2005){Poindexter}, {Afonso}, {Bennett},
  {Glicenstein}, {Gould}, {Szyma{\'n}ski}, \& {Udalski}}]{poindexter2005}
{Poindexter}, S., {Afonso}, C., {Bennett}, D.~P., {et~al.} 2005, \apj, 633, 914

\bibitem[{{Popowski}(2002)}]{popowski2002}
{Popowski}, P. 2002, arXiv Astrophysics e-prints, astro-ph/0205044

\bibitem[{{Popowski} {et~al.}(2001){Popowski}, {Alcock}, {Allsman}, {Alves},
  {Axelrod}, {Becker}, {Bennett}, {Cook}, {Drake}, {Freeman}, {Geha}, {Griest},
  {Lehner}, {Marshall}, {Minniti}, {Nelson}, {Peterson}, {Pratt}, {Quinn},
  {Stubbs}, {Sutherland}, {Tomaney}, {Vandehei}, \& {Welch}}]{popowski2001}
{Popowski}, P., {Alcock}, C., {Allsman}, R.~A., {et~al.} 2001, in Astronomical
  Society of the Pacific Conference Series, Vol. 239, Microlensing 2000: A New
  Era of Microlensing Astrophysics, ed. J.~W. {Menzies} \& P.~D. {Sackett}, 244

\bibitem[{{Popowski} {et~al.}(2005){Popowski}, {Griest}, {Thomas}, {Cook},
  {Bennett}, {Becker}, {Alves}, {Minniti}, {Drake}, {Alcock}, {Allsman},
  {Axelrod}, {Freeman}, {Geha}, {Lehner}, {Marshall}, {Nelson}, {Peterson},
  {Quinn}, {Stubbs}, {Sutherland}, {Vandehei}, {Welch}, \& {MACHO
  Collaboration}}]{popowski2005}
{Popowski}, P., {Griest}, K., {Thomas}, C.~L., {et~al.} 2005, \apj, 631, 879

\bibitem[{{Robin} {et~al.}(2014){Robin}, {Reyl{\'e}}, {Fliri}, {Czekaj},
  {Robert}, \& {Martins}}]{robin2014}
{Robin}, A.~C., {Reyl{\'e}}, C., {Fliri}, J., {et~al.} 2014, \aap, 569, A13

\bibitem[{{Sevenster} {et~al.}(1999){Sevenster}, {Saha}, {Valls-Gabaud}, \&
  {Fux}}]{sevenster1999}
{Sevenster}, M., {Saha}, P., {Valls-Gabaud}, D., \& {Fux}, R. 1999, \mnras,
  307, 584

\bibitem[{{Shvartzvald} {et~al.}(2017){Shvartzvald}, {Bryden}, {Gould},
  {Henderson}, {Howell}, \& {Beichman}}]{ukirt2017}
{Shvartzvald}, Y., {Bryden}, G., {Gould}, A., {et~al.} 2017, \aj, 153, 61

\bibitem[{{Skowron} {et~al.}(2016){Skowron}, {Udalski}, {Koz{\l}owski},
  {Szyma{\'n}ski}, {Mr{\'o}z}, {Wyrzykowski}, {Poleski}, {Pietrukowicz},
  {Ulaczyk}, {Pawlak}, \& {Soszy{\'n}ski}}]{skowron2016}
{Skowron}, J., {Udalski}, A., {Koz{\l}owski}, S., {et~al.} 2016, \actaa, 66, 1

\bibitem[{{Smith} {et~al.}(2007){Smith}, {Wo{\'z}niak}, {Mao}, \&
  {Sumi}}]{smith2007}
{Smith}, M.~C., {Wo{\'z}niak}, P., {Mao}, S., \& {Sumi}, T. 2007, \mnras, 380,
  805

\bibitem[{{Spergel} {et~al.}(2015){Spergel}, {Gehrels}, {Baltay}, {Bennett},
  {Breckinridge}, {Donahue}, {Dressler}, {Gaudi}, {Greene}, {Guyon}, {Hirata},
  {Kalirai}, {Kasdin}, {Macintosh}, {Moos}, {Perlmutter}, {Postman},
  {Rauscher}, {Rhodes}, {Wang}, {Weinberg}, {Benford}, {Hudson}, {Jeong},
  {Mellier}, {Traub}, {Yamada}, {Capak}, {Colbert}, {Masters}, {Penny},
  {Savransky}, {Stern}, {Zimmerman}, {Barry}, {Bartusek}, {Carpenter}, {Cheng},
  {Content}, {Dekens}, {Demers}, {Grady}, {Jackson}, {Kuan}, {Kruk}, {Melton},
  {Nemati}, {Parvin}, {Poberezhskiy}, {Peddie}, {Ruffa}, {Wallace}, {Whipple},
  {Wollack}, \& {Zhao}}]{spergel2015}
{Spergel}, D., {Gehrels}, N., {Baltay}, C., {et~al.} 2015, arXiv e-prints,
  arXiv:1503.03757

\bibitem[{{Stanek} {et~al.}(1994){Stanek}, {Mateo}, {Udalski}, {Szymanski},
  {Kaluzny}, \& {Kubiak}}]{stanek1994}
{Stanek}, K.~Z., {Mateo}, M., {Udalski}, A., {et~al.} 1994, \apjl, 429, L73

\bibitem[{{Sumi} \& {Penny}(2016)}]{sumi_penny2016}
{Sumi}, T., \& {Penny}, M.~T. 2016, \apj, 827, 139

\bibitem[{{Sumi} {et~al.}(2003){Sumi}, {Abe}, {Bond}, {Dodd}, {Hearnshaw},
  {Honda}, {Honma}, {Kan-ya}, {Kilmartin}, {Masuda}, {Matsubara}, {Muraki},
  {Nakamura}, {Nishi}, {Noda}, {Ohnishi}, {Petterson}, {Rattenbury}, {Reid},
  {Saito}, {Saito}, {Sato}, {Sekiguchi}, {Skuljan}, {Sullivan}, {Takeuti},
  {Tristram}, {Wilkinson}, {Yanagisawa}, \& {Yock}}]{sumi2003}
{Sumi}, T., {Abe}, F., {Bond}, I.~A., {et~al.} 2003, \apj, 591, 204

\bibitem[{{Sumi} {et~al.}(2006){Sumi}, {Wo{\'z}niak}, {Udalski},
  {Szyma{\'n}ski}, {Kubiak}, {Pietrzy{\'n}ski}, {Soszy{\'n}ski},
  {{\.Z}ebru{\'n}}, {Szewczyk}, {Wyrzykowski}, \& {Paczy{\'n}ski}}]{sumi2006}
{Sumi}, T., {Wo{\'z}niak}, P.~R., {Udalski}, A., {et~al.} 2006, \apj, 636, 240

\bibitem[{{Sumi} {et~al.}(2011){Sumi}, {Kamiya}, {Bennett}, {Bond}, {Abe},
  {Botzler}, {Fukui}, {Furusawa}, {Hearnshaw}, {Itow}, {Kilmartin}, {Korpela},
  {Lin}, {Ling}, {Masuda}, {Matsubara}, {Miyake}, {Motomura}, {Muraki},
  {Nagaya}, {Nakamura}, {Ohnishi}, {Okumura}, {Perrott}, {Rattenbury}, {Saito},
  {Sako}, {Sullivan}, {Sweatman}, {Tristram}, {Udalski}, {Szyma{\'n}ski},
  {Kubiak}, {Pietrzy{\'n}ski}, {Poleski}, {Soszy{\'n}ski}, {Wyrzykowski},
  {Ulaczyk}, \& {Microlensing Observations in Astrophysics (MOA)
  Collaboration}}]{sumi2011}
{Sumi}, T., {Kamiya}, K., {Bennett}, D.~P., {et~al.} 2011, \nat, 473, 349

\bibitem[{{Sumi} {et~al.}(2013){Sumi}, {Bennett}, {Bond}, {Abe}, {Botzler},
  {Fukui}, {Furusawa}, {Itow}, {Ling}, {Masuda}, {Matsubara}, {Muraki},
  {Ohnishi}, {Rattenbury}, {Saito}, {Sullivan}, {Suzuki}, {Sweatman},
  {Tristram}, {Wada}, {Yock}, \& {MOA Collaboratoin}}]{sumi2013}
{Sumi}, T., {Bennett}, D.~P., {Bond}, I.~A., {et~al.} 2013, \apj, 778, 150

\bibitem[{{Udalski}(2003)}]{udalski2003}
{Udalski}, A. 2003, \actaa, 53, 291

\bibitem[{{Udalski} {et~al.}(2015){Udalski}, {Szyma{\'n}ski}, \&
  {Szyma{\'n}ski}}]{udalski2015}
{Udalski}, A., {Szyma{\'n}ski}, M.~K., \& {Szyma{\'n}ski}, G. 2015, \actaa, 65,
  1

\bibitem[{{Udalski} {et~al.}(1994){Udalski}, {Szyma\'nski}, {Stanek},
  {Kaluzny}, {Kubiak}, {Mateo}, {Krzeminski}, {Paczynski}, \&
  {Venkat}}]{udalski1994c}
{Udalski}, A., {Szyma\'nski}, M., {Stanek}, K.~Z., {et~al.} 1994, \actaa, 44,
  165

\bibitem[{{Wegg} {et~al.}(2016){Wegg}, {Gerhard}, \& {Portail}}]{wegg2016}
{Wegg}, C., {Gerhard}, O., \& {Portail}, M. 2016, \mnras, 463, 557

\bibitem[{{Wegg} {et~al.}(2017){Wegg}, {Gerhard}, \& {Portail}}]{wegg2017}
{Wegg}, C., {Gerhard}, O., \& {Portail}, M. 2017, \apjl, 843, L5

\bibitem[{{Wood} \& {Mao}(2005)}]{wood2005}
{Wood}, A., \& {Mao}, S. 2005, \mnras, 362, 945

\bibitem[{{Wo{\'z}niak} \& {Paczy{\'n}ski}(1997)}]{wozniak1997}
{Wo{\'z}niak}, P., \& {Paczy{\'n}ski}, B. 1997, \apj, 487, 55

\bibitem[{{Wo{\'z}niak}(2000)}]{wozniak2000}
{Wo{\'z}niak}, P.~R. 2000, \actaa, 50, 421

\bibitem[{{Wyrzykowski} {et~al.}(2015){Wyrzykowski}, {Rynkiewicz}, {Skowron},
  {Koz{\l}owski}, {Udalski}, {Szyma{\'n}ski}, {Kubiak}, {Soszy{\'n}ski},
  {Pietrzy{\'n}ski}, {Poleski}, {Pietrukowicz}, \& {Pawlak}}]{wyrzykowski2015}
{Wyrzykowski}, {\L}., {Rynkiewicz}, A.~E., {Skowron}, J., {et~al.} 2015, \apjs,
  216, 12

\bibitem[{{Wyrzykowski} {et~al.}(2016){Wyrzykowski}, {Kostrzewa-Rutkowska},
  {Skowron}, {Rybicki}, {Mr{\'o}z}, {Koz{\l}owski}, {Udalski}, {Szyma{\'n}ski},
  {Pietrzy{\'n}ski}, {Soszy{\'n}ski}, {Ulaczyk}, {Pietrukowicz}, {Poleski},
  {Pawlak}, {I{\l}kiewicz}, \& {Rattenbury}}]{wyrzykowski2016}
{Wyrzykowski}, {\L}., {Kostrzewa-Rutkowska}, Z., {Skowron}, J., {et~al.} 2016,
  \mnras, 458, 3012

\bibitem[{{Zhao} \& {Mao}(1996)}]{zhao1996}
{Zhao}, H., \& {Mao}, S. 1996, \mnras, 283, 1197

\bibitem[{{Zhao} {et~al.}(1995){Zhao}, {Spergel}, \& {Rich}}]{zhao1995}
{Zhao}, H., {Spergel}, D.~N., \& {Rich}, R.~M. 1995, \apjl, 440, L13

\end{thebibliography}

\end{document}